\documentclass[reprint, aps, amsmath, amssymb]{revtex4-2}

\usepackage{epsfig}
\usepackage{graphicx}   
\usepackage{dcolumn}    
\usepackage{bm}         
\usepackage{color}
\usepackage{multirow}

\usepackage{lmodern} 
\usepackage[bookmarks, colorlinks=false]{hyperref}
\usepackage{mathtools}
\hypersetup{colorlinks=true, allcolors=blue}
\usepackage[english]{babel}

\usepackage{siunitx}

\usepackage[caption=false]{subfig}

\begin{document}

\title{\texorpdfstring{Intense $\gamma$}{Gamma}-photon and high-energy electron production by neutron irradiation: effects of nuclear excitations on reactor materials}

\author{Luca Reali}
\email{Luca.Reali@ukaea.uk}
\affiliation{CCFE, United Kingdom Atomic Energy Authority, Culham Science Centre, Oxfordshire OX14 3DB, UK}

\author{Mark R. Gilbert}
\email{Mark.Gilbert@ukaea.uk}
\affiliation{CCFE, United Kingdom Atomic Energy Authority, Culham Science Centre, Oxfordshire OX14 3DB, UK}

\author{Max Boleininger}
\email{Max.Boleininger@ukaea.uk}
\affiliation{CCFE, United Kingdom Atomic Energy Authority, Culham Science Centre, Oxfordshire OX14 3DB, UK}

\author{Sergei L. Dudarev}
\email{Sergei.Dudarev@ukaea.uk}
\affiliation{CCFE, United Kingdom Atomic Energy Authority, Culham Science Centre, Oxfordshire OX14 3DB, UK}
\affiliation{Department of Physics and Thomas Young Centre, Imperial College London,
South Kensington Campus, London SW7 2AZ, United Kingdom}

\begin{abstract}
Effects of neutron irradiation on materials are often interpreted in terms of atomic recoils, initiated by neutron impacts and producing crystal lattice defects. In addition, there is a remarkable two-step process, strongly pronounced in the medium-weight and heavy elements. This process involves the generation of energetic $\gamma$ photons in nonelastic collisions of neutrons with atomic nuclei, achieved \emph{via} capture and inelastic reactions. Subsequently, high-energy electrons are excited through the scattering of $\gamma$ photons by the atomic electrons. We derive and validate equations enabling a fast and robust evaluation of photon and electron fluxes produced by the neutrons in the bulk of materials. The two-step $n-\gamma -e$ scattering creates a non-equilibrium dynamically fluctuating steady-state population of high-energy electrons, with the spectra of photon and electron energies extending well into the MeV range. This stimulates vacancy diffusion through electron-triggered atomic recoils, primarily involving vacancy-impurity dissociation, even if thermal activation is ineffective.
Tungsten converts the energy of fusion or fission neutrons into a flux of $\gamma$-radiation at the conversion efficiency approaching 99\%, with implications for structural materials, superconductors and insulators, as well as phenomena like corrosion, and helium and hydrogen isotope retention. 

\end{abstract}

\maketitle

\section{Introduction}

Neutron irradiation alters the microstructure of a material by producing atomic scale defects. These defects interact and coalesce, forming extended dislocation networks and voids. This ultimately results in macroscopic swelling and dimensional changes, which have to be considered in the engineering design of reactor components alongside thermomechanical and electromagnetic loads. Populations of defects act as microscopic sources for the macroscopic fields of stress and strain \cite{Reali2022}. These populations evolve, driven by the combined effects of stochastic generation of defects by neutron and ion impacts, elastic interaction between the defects, thermal migration and athermal relaxation. The resulting pattern of microstructural evolution depends on the irradiation dose rate, temperature, and stress. 

The goal of nuclear fission or fusion power generation is to exploit the energy of the particles released during the nuclear reactions, particularly the fission fragments and neutrons from the deuterium-tritium fusion reactions, and it is this energy that is ultimately deposited in the coolant, structural materials or the tritium-generating fusion reactor blanket. One way this energy deposition occurs is through elastic collisions of neutrons with the atomic nuclei. Primary knock-on atoms (PKA) initiate displacement cascades that melt the material on atomic length and time scales, leading to what is known as neutron heating. Another energy deposition mechanism involves inelastic collisions of neutrons with the nuclei, either in the form of neutron capture reactions at low neutron energies or nuclear break-up reactions at higher (MeV) energies. These and other nuclear reactions excite the internal degrees of freedom of atomic nuclei and give rise to the generation of energetic photons through the subsequent nuclear de-excitation. This process of absorption of these energetic photons is known as \(\gamma\)-heating. It has long been realised that  $\gamma$-heating depends strongly on the atomic weight of an element and can be the dominant heating process, more intense than heating through atomic recoils, for example by a factor of ten in niobium \cite{Abdou1975}. It is also known that $\gamma$-photons can produce high concentration of radiation defects in covalent and ionic crystals even if they do not displace atoms from their lattice sites \cite{Lushchik1977}.

Exposure to neutrons is characterised by the magnitude of the total neutron flux and its energy spectrum. The \(\gamma\)-photon flux, and in turn the high-energy electron flux that the \(\gamma\)-photons produce in a material, also have their own characteristic energies. The energy distribution of all the three types of particles determine the likelihood of various scattering events via the energy dependence of the corresponding cross sections.

Electron irradiation on its own has been extensively used in the studies of production and evolution of defects in transmission electron microscopy (TEM) \cite{Maury1978,Kiritani1976,Arakawa2020, Reimer1984, Griffiths1994}. Below, we show that in the nuclear environments --- whether fission or fusion --- both neutron and electron irradiation are {\it always} present simultaneously. Notably, there are no $\gamma$-photons and high-energy electrons generated during ion irradiation, a fact that is not usually highlighted when comparing neutron and ion irradiation experiments \cite{Was2014,WasZinkle2020}.

Interaction between high-energy electrons and atoms in a crystal can generate new defects {\it and} drive their subsequent motion \cite{Vajda1977,Kiritani1976,Arakawa2020}. The maximum recoil energy $E_{\textnormal{R}}^{\textnormal{max}}$ that an atom of mass $M$ can receive in a collision with an electron of kinetic energy $E_{\textnormal{el}}$ and mass $m$ is  \cite{Dicarlo1971, Vajda1977}
\begin{equation}\label{E_R_max}
    E_{\textnormal{R}}^{\textnormal{max}} = \frac{2(E_{\textnormal{el}}+2mc^2)}{Mc^2}E_{\textnormal{el}},
\end{equation}
where $c$ is the speed of light. For instance, if $E_{\textnormal{el}}=3$~MeV then in tungsten $E_{\textnormal{R}}^{\textnormal{max}}=140.8$~eV, which is nearly twice the threshold displacement energy (TDE) \cite{Xu2003, Nordlund2018}, the minimum energy required to generate a stable vacancy-interstitial pair. If $E_{\textnormal{el}}=500$~keV, then  $E_{\textnormal{R}}^{\textnormal{max}}=8.9$~eV, which is about five times the vacancy migration energy in pure elemental W \cite{Nguyen2006}. Vacancy migration in W, driven by electron impacts, is observed under a 500~keV electron beam in a transmission electron microscope even at cryogenic temperatures \cite{Arakawa2020}. Similarly, athermal vacancy diffusion was reported in Zr illuminated by a 800~keV electron beam at 150 K \cite{Griffiths1994}, and in Pb exposed to a beam of 390~keV electrons at 58 K \cite{Urban1974}. The accelerated diffusion of vacancies stimulates rapid microstructural evolution, involving the shrinkage of interstitial-type dislocation loops \cite{Arakawa2020} and growth of vacancy-type loops \cite{Griffiths1994}. Electron irradiation also stimulates the motion of defect clusters at low temperatures in a variety of alloys and steels \cite{Satoh2014} and even in covalent materials like SiC \cite{Jiang2016}.

Athermal vacancy migration, induced by electron impacts, was modelled using molecular dynamics (MD) by Xu {\it et al.} in W \cite{Xu2003} and Satoh {\it et al.} in Fe and Cu \cite{Satoh2017}. They found that the minimum recoil energy for displacing a vacancy was directionally anisotropic, and strongly varied as a function of distance from a vacancy. Satoh {\it et al.} \cite{Satoh2017} quantified the probability of a driven vacancy hop as a function of recoil energy $E_{\textnormal{R}}$, the distance between an atom impacted by an electron and a vacancy, and temperature. Even at high $E_{\textnormal{R}}$, approaching 20-100 times the vacancy migration energy, the probability of an electron impact on a nearest neighbour atom resulting in a vacancy hop remains close to 0.3-0.5. This probability is higher at 300~K than at 20~K, particularly at low $E_{\textnormal{R}}$.

Below, we detail the generation of \(\gamma\)-photons in a reactor environment and its contribution to nuclear heating in various materials. We explain how the photon spectra are derived from the neutron spectra, and how the high-energy electron spectra are related to the $\gamma$-photon spectra. Then, we simulate and analyse the effect of collisions between the  high-energy electrons and atoms in a lattice, and quantify a relationship between atomic recoils and vacancy diffusion. Finally, we combine all the above steps and evaluate the rates of neutron-$\gamma$-electron-stimulated vacancy migration in irradiated materials.

\section{ \texorpdfstring{$\gamma$}{Gamma}-photons produced by exposure to fission and fusion neutrons}\label{sec:spectra}

There are various sources of \(\gamma\)-photons (or \(\gamma\)-rays or $\gamma$-radiation) in a nuclear reactor environment. In addition to \(\gamma\)-photons, which are often produced by nuclei decay as well as directly (promptly) during nuclear reactions, it is also possible to produce X-rays or, equivalently, the R\"{o}ntgen photons, which are are commonly generated through the de-excitation of atomic electrons~\cite{knoll}. These two forms of electromagnetic radiation are fundamentally similar and there is no universally agreed differentiation between X-rays and $\gamma$-photons. X-rays are often defined as photons with energies below 100~keV, a value that, as we shall see later, is conveniently close to the cut-off energy threshold below which the energy of the electrons produced by the absorption of photons is too low to influence the dynamics of atoms. The \(\gamma\)-photons have the energies in the range from 100~keV to 10~MeV~\cite{krane}. In this study, we investigate the \(\gamma\)-photons produced by the relaxation of atomic nuclei from their excited high energy configurations.

\begin{figure*}[t]
\subfloat[]{%
 \includegraphics[width=\columnwidth]{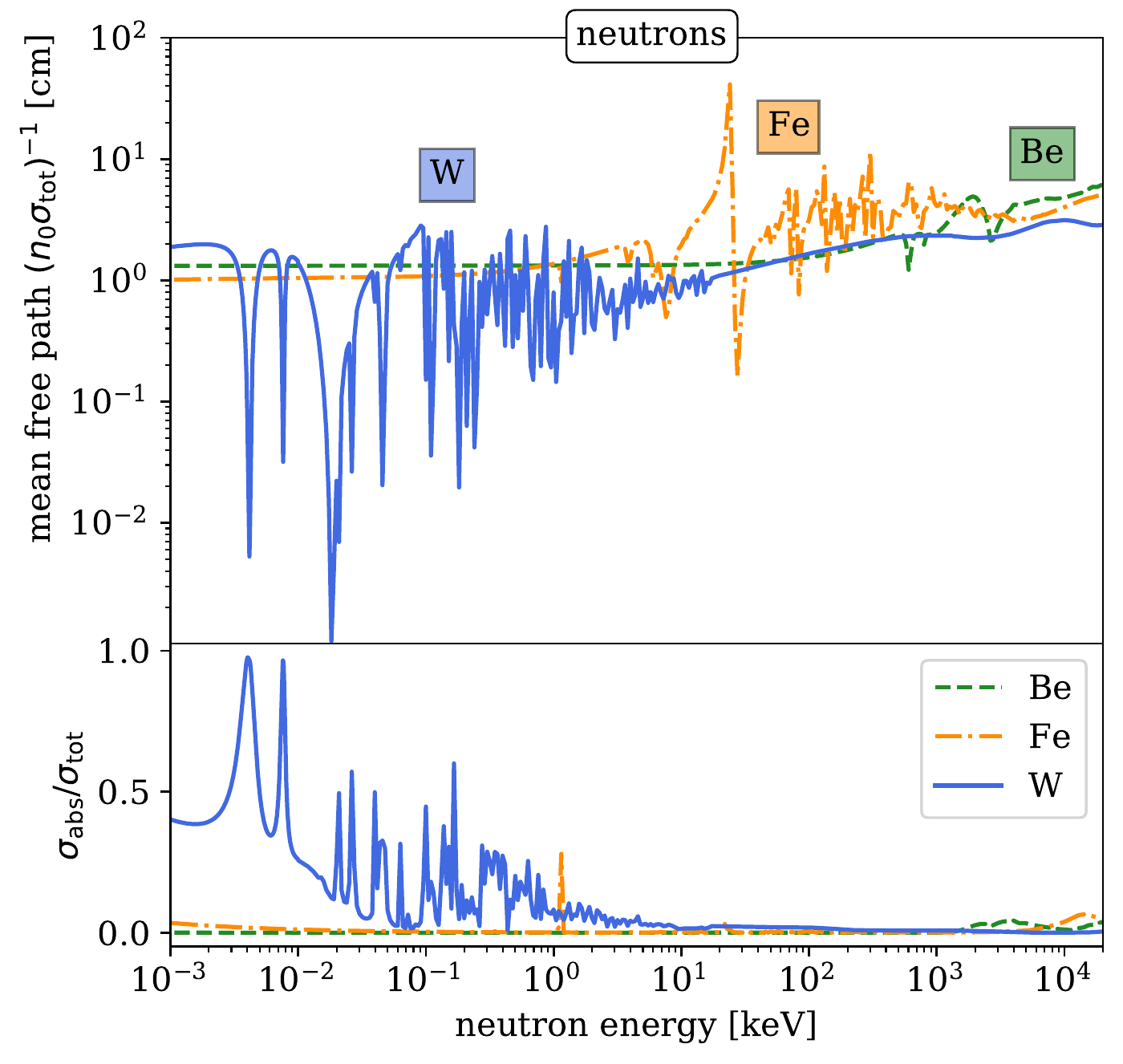}%
}\hfill
\subfloat[]{%
 \includegraphics[width=\columnwidth]{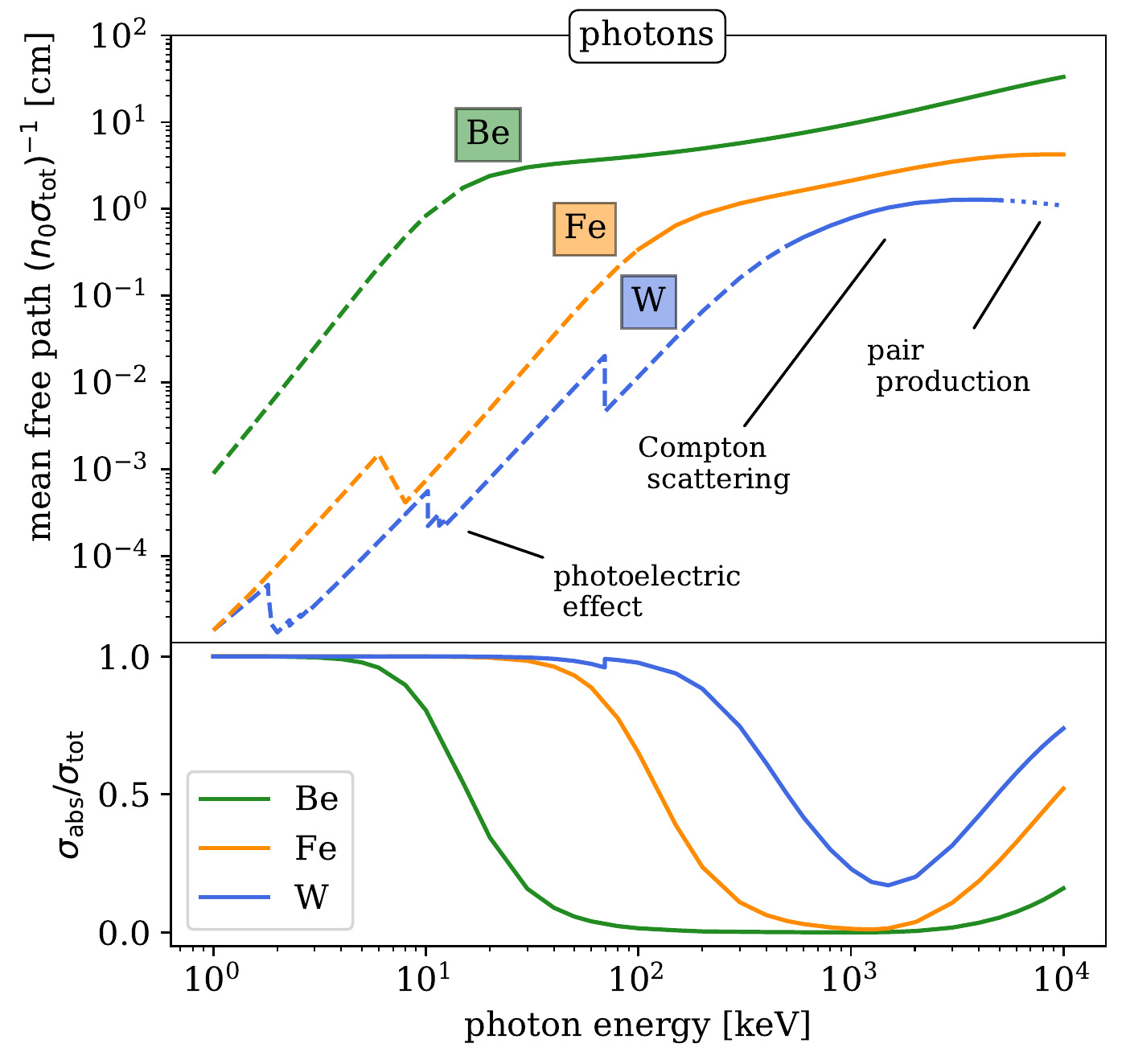}%
}\hfill
    \caption{Mean free paths of neutrons (a) and photons (b) in Be, Fe and W as a function of energy, defined as the reciprocal of atomic density times the total cross section. The effect of atomic number on the position and character of nuclear resonances on the energy axis is clearly visible. In (b), the dominant scattering processes are highlighted. At $\sim$~MeV energies, neutrons likely travel between 2 to 4~cm before experiencing mostly elastic scatterings; $\gamma$-photons in W and Fe likely travel for 1 to 2~cm, before undergoing a Compton scattering event. The lower panels show the ratio between the absorption and total cross sections, highlighting that neutrons have a much lower likelihood of being absorbed than photons, despite their mean free paths being comparable.}
    \label{fig:mean_free_path} 
\end{figure*}

In fusion, the energy released in D-T reactions is carried by neutrons and $\alpha$-particles, formed in these events. Other types of fusion reactions are also possible, but the international fusion research is focused primarily on the D-T route. \(\gamma\)-photons with energies of 16.75 MeV and 13.5 MeV are occasionally produced in D-T reactions, and this involves the formation of a \(^5\)He particle instead of the dominant \(\alpha\)-particle (\(^4\)He). The branching ratio for the \(^5\)He channel is very low compared to the main channel. Inertial confinement fusion and accelerator-based experiments show that the branching ratio for the \(\gamma\)-photon production directly from D-T reactions is in the range from \(10^{-4}\) to \(5\times10^{-5}\)~\cite{jeet2021}. Thus, this flux of \(\gamma\)-photons originating directly from the D-T reactions in the fusion plasma is negligible. 

Meanwhile, the \(\alpha\)-particles produced in the D-T reactions do not have the sufficient energy to induce any of the nuclear reactions in the plasma-facing materials that could lead to \(\gamma\)-photon emission. The threshold minimum energies for reactions such as (\(\alpha\),n), in other words  the \(\alpha\)-absorption followed by neutron emission, which could lead to excited isomeric states and thus \(\gamma\)-photons production analysed below, are typically above the maximum 3.5~MeV energy of $\alpha$-particles produced in D-T collisions, even for metals lighter than tungsten~\cite{matsunobu2002,aydin2016,yildiz2017}. The direct absorption reactions (\(\alpha\),\(\gamma\)) resulting in the photon production, are extremely rare~\cite{janis}. Therefore, the only significant source of photons in a fusion reactor are the neutrons produced by the D-T reactions themselves, or those created by neutron multiplications, interacting with the nuclei in the reactor materials. 

In fission, the situation is more complex. Only a relatively small proportion of the $\sim$200~MeV energy released in the fission of \(^{235}\)U is carried by neutrons -- an average of around 5~MeV per fission event, whereas the vast majority (\(\sim\) 80\%) is carried by the fission fragments themselves~\cite{krane}. Additionally, prompt \(\gamma\)-photons, emitted within \(10^{-14}\)~s of a fission event, typically carry around 8~MeV, while the decay of fission fragments releases around 19~MeV via \(\beta\) decays and 7~MeV through delayed \(\gamma\)-decay. The exact energy of individual \(\gamma\)-photons and \(\beta\) particles depends on the nature of the fission fragments, which vary from one fission event to another, and their decays. However, the overall intensity of $\gamma$-photon production is significant; for example, around 8 prompt \(\gamma\)-photons are produced per \(^{235}\)U fission~\cite{oberstedt2013}, and there are of the order of \(3\times10^{19}\) fission events per second per GW of fission power, assuming 200~MeV of energy release per a fission event, leading to \(\sim2.4\times10^{20}\) \(\gamma\)-photons per GW per second.

Such a large flux of photons could be problematic to structural components, were it not for the relatively short penetration range of \(\gamma\)-rays compared to neutrons. For example, the \(\gamma\)-photons produced by the decay of \(^{60}\)Co have energies in the range from 1.1 to 1.4~MeV. Their absorption mean free path in stainless steel is 1.6~cm and they would lose more than 90\% of their intensity within 3~cm~\cite{Buyuk2015}. Hence, despite the large \(\gamma\)-flux in the fuel channels themselves and for experimental material samples placed in core locations, fission does not generate an appreciable external flux of $\gamma$-photons. This results in the same conclusion for fission as for fusion above, namely that it can be assumed, as we will in the remainder of this paper, that the dominant and the only source of photons in a bulk structural material in a nuclear reactor is that originating from within the material itself due to the nuclear reactions triggered by the incident neutrons.

Fig.~\ref{fig:mean_free_path} shows the mean free paths of neutrons and photons in Be, Fe and W. This defines the length scale between the subsequent collision events. The penetration depth of the two types of particles also depends on the probability of them being absorbed or scattered/re-emitted, see the lower panels in Fig.~\ref{fig:mean_free_path}. The relative probability of absorption can be estimated from the ratio between the absorption ($\sigma_{\textnormal{abs}}$) and total ($\sigma_{\textnormal{tot}}$) cross sections. The values were taken from the TENDL-2021 \cite{t21} and the XCOM \cite{Berger1987} libraries for neutrons and photons, respectively. For $\sigma_{\textnormal{abs}}$ of photons, we considered the photoelectric effect and pair production. We find that in the $\gtrsim$~keV energy range, the mean free paths of neutrons and photons are comparable, but the photons are far more likely to be absorbed.

\begin{figure}[t]
  \includegraphics[width=\columnwidth]{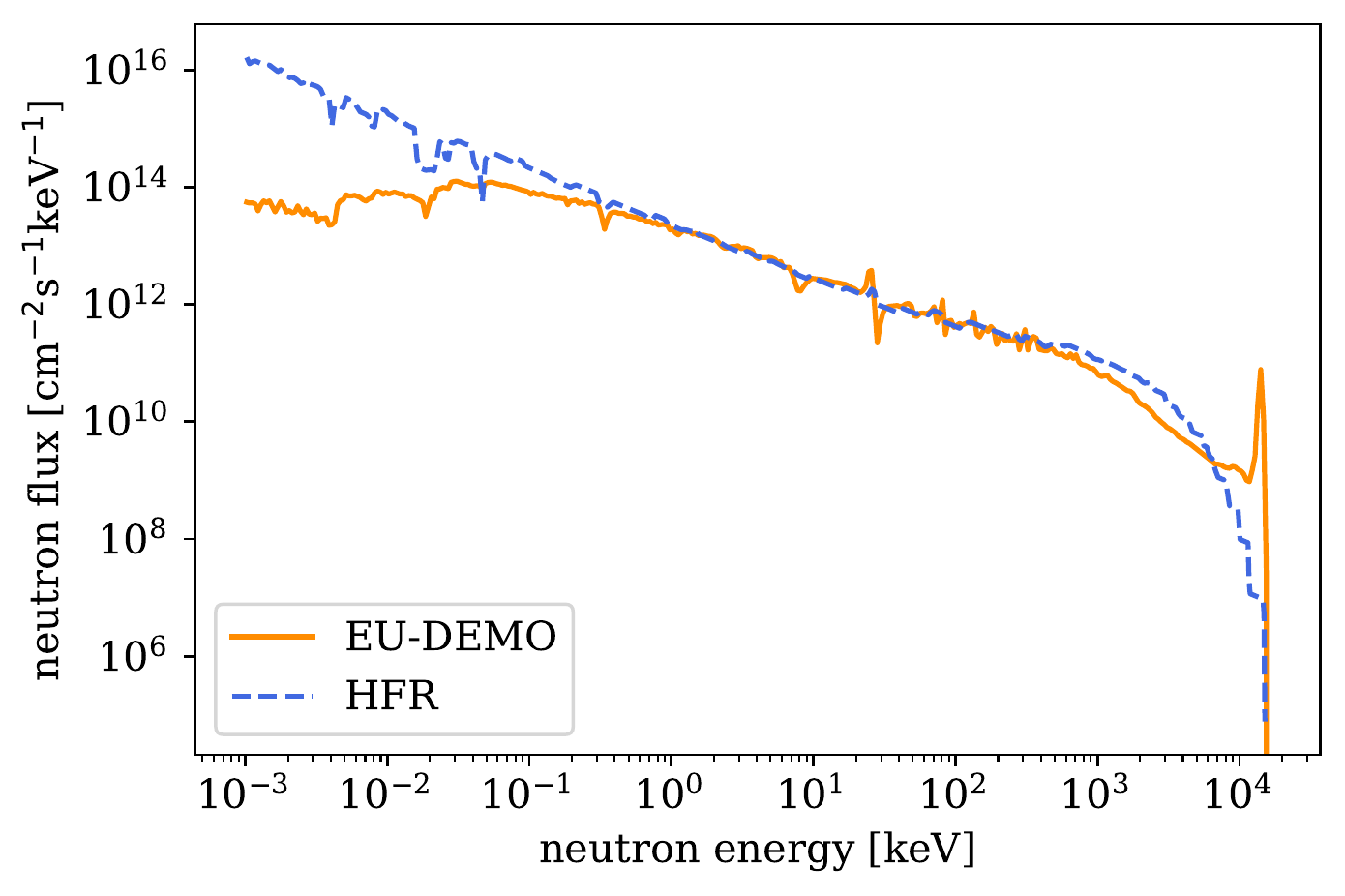}%
    \caption{HFR (fission) \cite{Gilbert2017} and EU-DEMO first wall (fusion) \cite{Gilbert2017b} energy-differential neutron spectra. These spectra were calculated through stochastic Monte Carlo simulations by MCNP~\cite{MCNPusersmanual}, and are the volume-averaged flux as a function of energy (originally in tallied histogram format, but converted to differential fluxes here by dividing by bin width). They are obtained by integrating (over angle) the energy-dependent angular flux of neutrons in a region to give the density of particles, regardless of trajectory, at a point. Over a distance element d$s$ this density can be thought of as the track length density; MCNP estimates the average flux by summing track lengths, which is a reliable approach in well-populated statistics~\cite{MCNPusersmanual}. The resulting flux per source neutron are multiplied by the reactor power (converted to number of neutrons) to give the flux-per-second values shown.}
    \label{fig:neutron_spectra} 
\end{figure}

Unlike photons, high-energy neutrons undergo multiple scattering and propagate through materials over much longer distances typically of the order of 10~cm and, to a first approximation, the high-energy neutron flux can be considered constant in the bulk of materials on atomic length scales; near interfaces between materials, neutron fluxes can vary more rapidly~\cite{Gilbert_2017}, but we do not consider that case here.
Fig.~\ref{fig:neutron_spectra} compares the neutron spectra predicted for materials in the first wall, plasma-facing environment of a fusion reactor to that expected at typical core locations in the High-Flux Reactor (HFR) fission experimental facility in Petten, Netherlands. These spectra were obtained using the Monte Carlo N-Particle (MCNP) transport code~\cite{MCNPusersmanual,MCNPrelasenotes} and account for the reactor geometries -- in the case of HFR for a digital model of the physical reactor, and for fusion using a conceptual digital design of DEMO, the next-step demonstration fusion reactor, see~\cite{Federici_2017,Gilbert2017b} for details. The HFR spectrum was calculated specifically for a W irradiation experiment, and this took into account the local experimental environment as opposed to assuming a generic spectrum often quoted for experimental facilities. Using this refined spectrum, analysis shows that transmutation burn-up calculations with the FISPACT-II~\cite{subletetalnds2017} inventory code accurately predict the composition evolution of W samples placed in this environment~\cite{Gilbert_2017,LLOYD2022}.

\begin{figure*}[t]
\subfloat[]{%
  \includegraphics[width=\columnwidth]{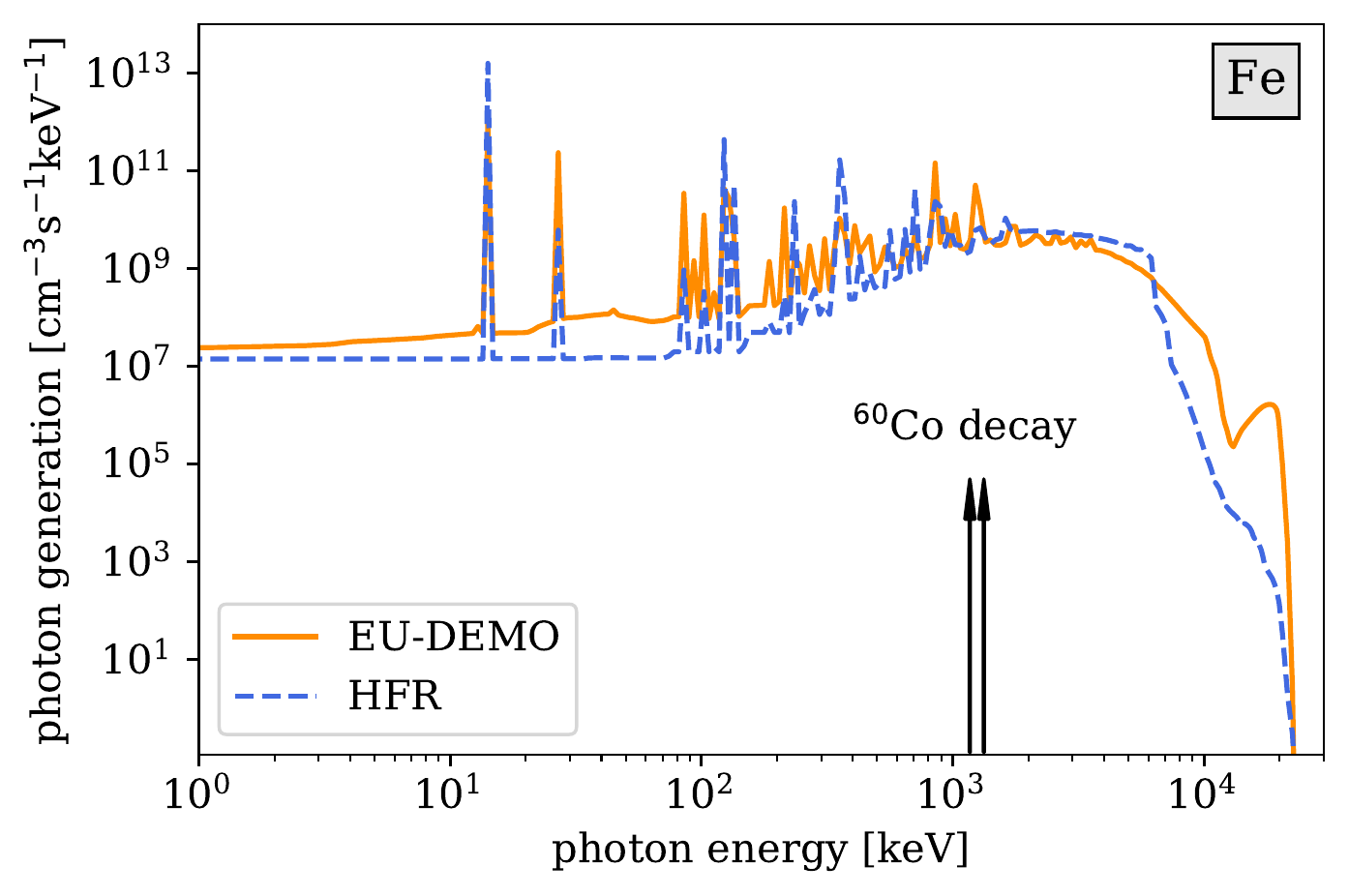}%
}\hfill
\subfloat[]{%
  \includegraphics[width=\columnwidth]{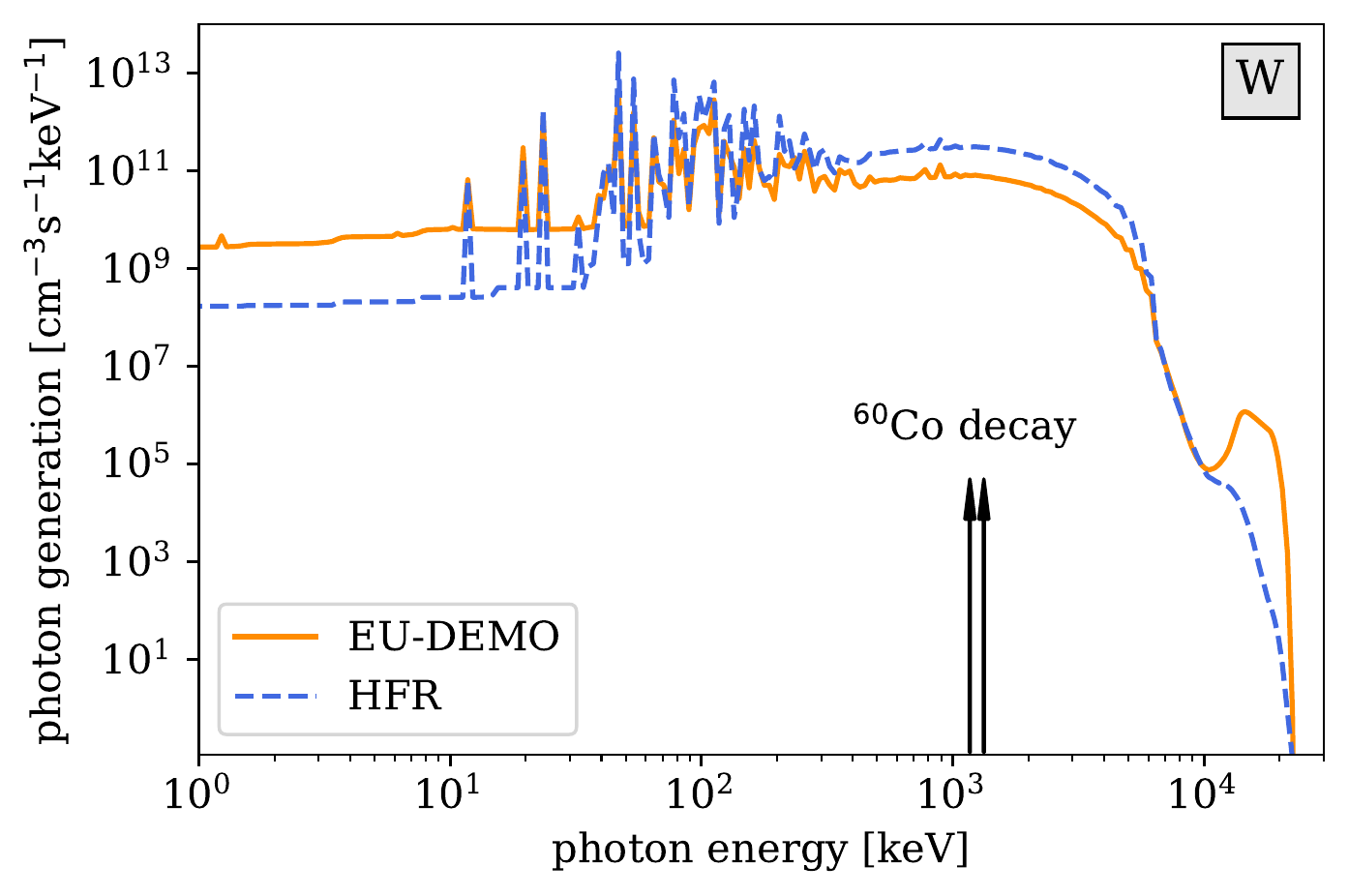}%
}\hfill
    \caption{Source photon distributions in iron (a) and tungsten (b) computed using the SPECTRA-PKA code for the neutron spectra shown in Fig. \ref{fig:neutron_spectra}.
    $\gamma$-photon intensities in beryllium are negligible and omitted, see Table~\ref{tab:heating}. The curves give the energy-resolved values of the source term $Q_{\textnormal{ph}}(E)$, defined by Eq. (\ref{eq:flux_source_definitions}) and referring to the generation of $\gamma$-photons per unit volume.  The energies of the \(\gamma\)-photons emitted during the decay of $^{60}$Co \textemdash  the two sharp spectral lines at 1.17~MeV and 1.33~MeV \textemdash are shown for comparison.}
    \label{fig:photon_source} 
\end{figure*}

Photons are generated by various nuclear reactions involving neutrons, such as inelastic scattering, neutron multiplication (n,2n) or neutron capture, followed by direct \(\gamma\) emission (n,\(\gamma\)). The fundamental origin of $\gamma$-photon emission in all the cases is the same: when a neutron interaction leaves a daughter reaction product or nuclide in an excited  state, often referred to as an isomeric state, or a metastable state if it is sufficiently long-lived, photons are emitted to release energy and allow the excited nuclei to transition through various energy levels to its lowest energy ground state. A negligibly small amount of energy is also taken by the recoiling nucleus as it decays~\cite{lilley}. While this de-excited state may also be unstable on longer time scales, producing further \(\gamma\)-photons as part of \(\beta\)-decay or \(\alpha\)-decay, these ``delayed'' photons are comparatively rare compared to the ``prompt'' \(\gamma\)-photons considered in the present work. The exact definition of prompt photons is not universally accepted, but in the context of fission reactions, prompt \(\gamma\)-photons are considered to be emitted within 10~fs of the initial reaction event~\cite{oberstedt2013,mihalczo2004}. However, the minimum half-life for an excited nuclide to be called ``isomeric'', and thus not an emitter of prompt \(\gamma\)-photons upon decay, can be anywhere from 1~ps to 1~\textmu s, and is often taken to be around 1~ns~\cite{krane,walker2007}. Such metastable nuclides may persist for much longer -- sometimes even for 100s or 1000s of years -- but, again, these infrequent strongly delayed \(\gamma\)-photons are not the subject of analysis here. 

Instead of using a transport code to calculate the prompt photon fluxes, which is often done in complex nuclear geometries to evaluate the dose rates outside of shielding to assess the safety of maintenance operations, we can instead use an approach independent of geometry, to calculate the instantaneous source term of photons in a region exposed to a known flux and energy spectrum of neutrons. SPECTRA-PKA~\cite{Gilbert2015,spectrapkaavailable} is a code developed to calculate the atomic displacement source terms due to neutron irradiation, but it can also be used to provide the equivalent \(\gamma\)-photon source terms using the available nuclear library data of \(\gamma\)-photon emission cross sections as a function of the incident neutron energy. 

In what follows, we treat the photon sources and account for the attenuation of photons in materials while at the same time evaluating the flux of high-energy electrons generated as photons undergo interactions with {\it atomic} electrons. This approach, starting from a neutron spectrum for a given geometry, through a \(\gamma\)-photon source term and then \(\gamma\)-photon attenuation and electron interactions in a material, is self-consistent and avoids having to disentangle the local geometry  attenuation from the pure source spectrum of \(\gamma\)-photons.

Before using SPECTRA-PKA to define the flux and energy spectrum of these local, prompt \(\gamma\)-photons, we first evaluate the neutron-induced energy deposition, often called heating. To perform these calculations, we use the FISPACT-II code, which can access the KERMA (Kinetic Energy Released per unit MAss) cross sections, expressed in barns-eV units and included in nuclear reaction data libraries. Nuclear heating is an integral measure of the energy transferred to the material by neutron irradiation. By analysing the constituent contributions to this heating, we can understand the relative significance and microscopic effects resulting from the nuclear heating caused by the $\gamma$-photons. 

Table~\ref{tab:heating} summarizes FISPACT-II nuclear heating calculations for W, Fe and Be, assuming exposure to the two neutron spectra shown in Fig.~\ref{fig:neutron_spectra}, and hence enabling the comparison of fusion (DEMO) and fission (HFR) irradiation environments. The total nuclear heating, given in W/g units, includes the energy deposited as a result of elastic, inelastic and non-elastic (break-up) nuclear reactions initiated by neutrons, as well as any locally deposited energy from emitted secondary particles, including \(\gamma\)-photons. The table also details the absolute and relative contributions to the total heating due to the energy deposited by photons. The KERMA cross sections used here were evaluated for the TENDL-2017~\cite{tendl2017} nuclear data library by the NJOY processing code~\cite{macfarlanekahler2010,Macfarlane2017}, assuming local absorption of the prompt photons and thus not considering their transport and attenuation. 

These FISPACT-II results represent the instantaneous heating rate at time \(t=0\), before any transmutation has taken place. The most striking observation from these calculations is that for W and to a lesser extent for Fe, the two technologically most significant fusion materials \cite{Rieth2013,Rieth2021,Terentyev2022},  \(\gamma\)-photon heating entirely dominates the energy deposited in the material exposed to neutron irradiation under both fission and fusion conditions. 

This result can be understood by considering that in high atomic mass elements like W there is a greater scope for nuclear excited states because there are more possible configurations of the nucleus in which to store energy. When the nucleon number, i.e. the total number of protons and neutrons, is high there is a significant and often dominant probability that the nuclear reaction energy will be trapped in an excited isomeric state and subsequently converted into \(\gamma\)-photons during isomeric transitions. In contrast, in simpler low nucleon number materials like Be, there are fewer isomeric possibilities and so the energy is more readily released directly in the form of kinetic energy of the daughter products of nuclear reactions.

The results summarised in Table~\ref{tab:heating} demonstrate how significant it is to consider the production of \(\gamma\)-photons in W and Fe, the two main fusion and fission nuclear materials \cite{Rieth2013,Rieth2021,Terentyev2022}. In them, the energy released during nuclear reactions is dominated by the photons and not by atomic recoils. Table~\ref{tab:heating} shows that tungsten acts as a highly efficient converter of the energy of fusion or fission neutrons into electromagnetic $\gamma$-radiation, with the conversion efficiency approaching 99\%.

\begin{table}[t]
\caption{Nuclear heating, detailing the contribution of $\gamma$-photon emission to the energy deposited in materials exposed to fusion and fission neutrons. }
\label{tab:heating}
\begin{ruledtabular}  
\begin{tabular}{l|ccc|cccc}
         Material & \multicolumn{6}{c}{Heating (W/g)}\\
         &\multicolumn{3}{c}{DEMO}&\multicolumn{3}{c}{HFR} \\
         & Total & Photon & Fraction & Total & Photon & Fraction \\
         W & 2.73 & 2.65 & 97.0\% & 10.48 & 10.36 & 98.9\% \\
         Fe & 1.68 & 1.19 & 70.8\% & 2.00 & 1.63 & 81.5\% \\
         Be & 3.65 & $3\times 10^{-3}$ & 0.1\% & 2.15 & $8\times 10^{-7}$ & 4$\times 10^{-5}$\% \\

  	\end{tabular}
\end{ruledtabular}
\end{table}

\begin{figure*}[t]
\subfloat[]{%
  \includegraphics[width=\columnwidth]{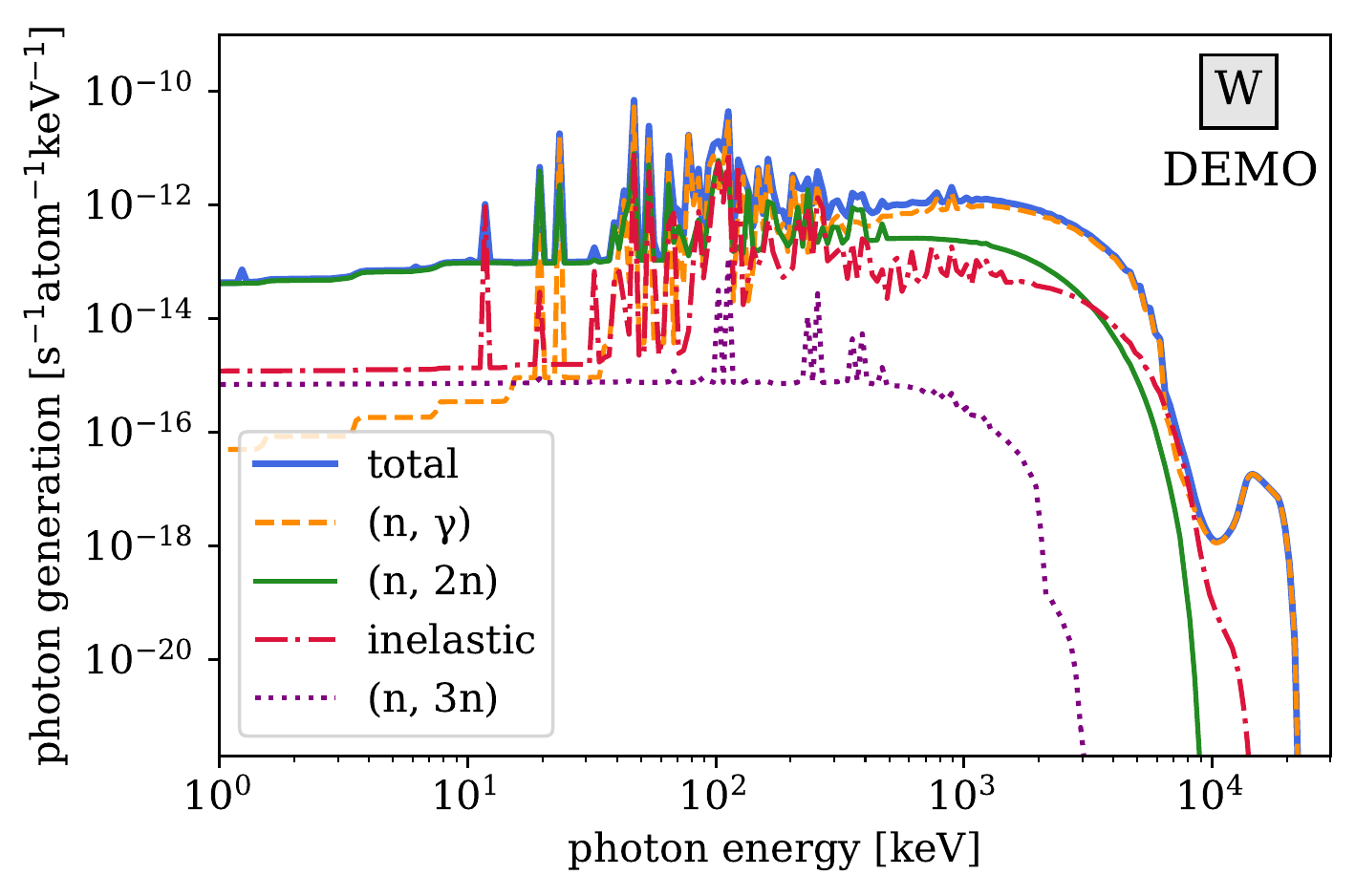}%
}\hfill
\subfloat[]{%
  \includegraphics[width=\columnwidth]{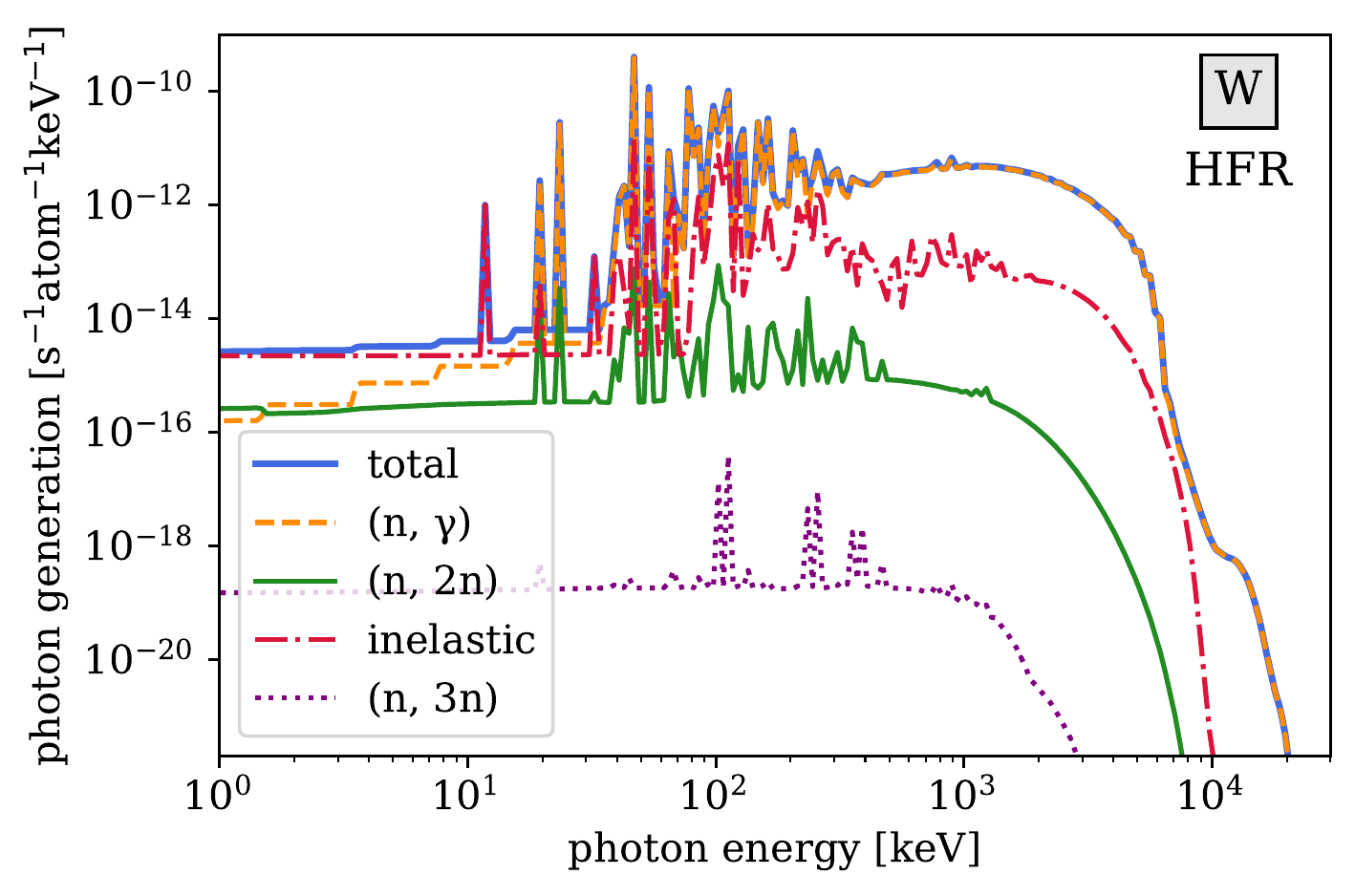}%
}\hfill
    \caption{Various nuclear reaction channel contributions to the $\gamma$-photon generation in W. Left panel: spectra of photons  generated by fusion neutrons, right panel: spectra of photons produced by fission neutrons. Fusion and fission neutron spectra are taken from Fig. \ref{fig:neutron_spectra}. The curves give the energy-resolved values of the source term $Q_{\textnormal{ph}}(E)$, defined by Eq. (\ref{eq:flux_source_definitions}) and referring to the generation of $\gamma$-photons by an individual atomic nucleus.  }
    \label{fig:photon_source_channels} 
\end{figure*}

Using the neutron spectra shown in Fig.~\ref{fig:neutron_spectra} as input for SPECTRA-PKA calculations, we find the \(\gamma\)-photon flux distributions for Fe and W shown in Fig.~\ref{fig:photon_source}. Beryllium is omitted from this figure as it generates a negligible flux of \(\gamma\)-photons, in agreement with the data given in Table~\ref{tab:heating}. These distributions will be used later in the paper to evaluate the flux of high-energy electrons produced by the $\gamma$-photons in the material. The spectral lines of the 1.17~MeV and 1.33~MeV \(\gamma\)-photons emitted by $^{60}$Co, an important high-intensity \(\gamma\)-source for medical and experimental applications, are also shown in the figure, highlighting that the photon generation in DEMO and HFR is near its maximum intensity at similar energies, and moreover it extends to photons of about one order of magnitude higher energy. At this point it is instructive to highlight the fundamental origin of spectra of $\gamma$-photons. To illustrate the principle, in Fig.~\ref{fig:photon_source_channels} we plot the dominant channel contributions for W under fusion (DEMO) and fission (HFR) conditions. In both cases we see an overwhelming dominance of the neutron capture (n,\(\gamma\)) reactions, particularly at higher \(>\)MeV \(\gamma\)-photon energies. Note that in both Fig.~\ref{fig:photon_source} and Fig.~\ref{fig:photon_source_channels} the \(\gamma\)-spectra appear as continuous distributions, while in reality \(\gamma\)-photons are emitted with discrete energy lines (i.e. associated with the discrete energy levels of excited nuclei). However, the spectra appear continuous here due to the energy-binning necessary in the simulations and the multiple discrete lines in each bin.

In what follows, we assume that the \(\gamma\)-photons emitted by the excited nuclei, and the high-energy electrons that these $\gamma$-photons produce in a material, are generated isotropically and their flux has no directional dependence. The neutron flux in a nuclear reactor inevitably has some angular anisotropy because, for example, the 14.1~MeV neutrons produced in a fusion plasma initially travel radially away from the plasma, whereas the lower energy neutrons originating from multiple scattering events in the surrounding structure have a strong back-scattered component. Any directional dependence of the neutron flux is lost at low energies due to multiple scattering. Similarly, in fission, there is a directional dependence of the fast (above 1 MeV) neutron flux emitted from the fuel pins, which impinges on the surrounding structural materials, but the high thermal component shown in Fig.~\ref{fig:neutron_spectra} is directionally more isotropic. 

The prompt \(\gamma\)-photons are generated in materials primarily from the neutron capture (n, $\gamma$) or inelastic scattering reactions, and can have angular anisotropy associated with the initial direction of the neutron momentum transferred to the compound nucleus~\cite{Motz1970}, as observed in inelastic scattering experiments on iron~\cite{beyer2018}. $\gamma$-radiation intensities have been found to be correlated with the direction of motion of fission fragments~\cite{graff1965}. However, since neutron capture and inelastic scattering dominate the generation of \(\gamma\)-photons in structural materials, as illustrated in Fig.~\ref{fig:photon_source_channels}, and since these reactions typically occur at MeV energies and below, especially in the case of neutron capture, most of the directional dependence on the neutron flux will have been lost at the point of \(\gamma\)-photon production due to the multiple scattering of neutrons. It is therefore reasonable to assume the angular isotropy of the flux of $\gamma$-photons. The flux of electrons $\phi_{\textnormal{el}}({\bf n},E)$, produced by the scattering of $\gamma$-quanta, is also isotropic, and in the absence of directional variables is solely a function of the kinetic energy of the electrons.

\section{Energy spectra of \texorpdfstring{$\gamma$}{gamma}-photons and high-energy electrons}\label{sec:scatterings}

A quantity central to the treatment of rates of scattering, atomic recoils, and reactions in solids, is the flux of high-energy particles initiating the respective scattering and reaction events. This flux, denoted by $\phi ({\bf n}, {\bf r}, E)$, equals the number of particles or photons with kinetic energy $E$ that cross a unit area in the direction of unit vector ${\bf n}$ per unit time in the vicinity of point ${\bf r}$. This flux can be computed by multiplying the number density $f({\bf n}, {\bf r}, E)$ of particles with kinetic energy $E$, and moving in direction ${\bf n}$, by their velocity $v$.   

For example, if $\phi_{\textnormal{el}}({\bf n}, {\bf r}, E)$ is the flux of high energy electrons, the rate of scattering of electrons by an atom located at ${\bf r}$ is
\begin{equation}
    \frac{\text{d}^2 \sigma ({\bf n},E \rightarrow {\bf n}',E')}{\text{d}o' \text{d}E'}
        \phi_{\textnormal{el}}({\bf n}, {\bf r}, E), \label{collision_rate_differential}
\end{equation}
where $\text{d}^2\sigma /\text{d}o'dE'$ is the differential cross section of scattering of electrons into an element of solid angle $do'$ corresponding to direction ${\bf n}'$, and an energy interval $dE'$. If the flux of electrons is mono-energetic then integrating (\ref{collision_rate_differential}) over all the directions of scattering ${\bf n}'$ and energies $E'$, as well over the directions of incidence ${\bf n}$, we find the total rate of collisions of electrons with an atom situated at ${\bf r}$, namely
\begin{equation}
    \nu _{\textnormal{tot}}(E,{\bf r})=\sigma _{\textnormal{tot}}(E)\int \text{d}o\, \phi_{\textnormal{el}}({\bf n}, {\bf r}, E) . \label{collision_rate_total}
\end{equation}
The total cross section of scattering
\begin{equation}
    \sigma _{\textnormal{tot}}(E)=\int \text{d}o' \int \text{d}E'\, {\text{d}^2 \sigma ({\bf n},E \rightarrow {\bf n}',E') \over \text{d}o' \text{d}E'}
\end{equation}
is independent of ${\bf n}$ due to the rotational invariance of the process of scattering, requiring that the differential cross section depends only on the angle between vectors ${\bf n}$ and ${\bf n}'$ \cite{GolbergerWatson,Newton1966}. 

In the treatment of threshold atomic recoil events below, we explore quantities similar to (\ref{collision_rate_total}), and often involving integration over a selected range of solid angles corresponding to a reaction, for example a hop of an atom from an occupied to a vacant lattice site. Such a hop occurs only if the direction of the recoil is favourable and is able to initiate a transition along a trajectory that crosses the energy barrier for the reaction. 

The flux of neutrons, electrons or photons  satisfies the Boltzmann transport equation \cite{Kalashnikov1985,Vassiliev}, extensively used in the theory of radiative transfer \cite{Chandrasekhar1960}, namely
\begin{equation}
    {\bf n}{\partial \over \partial {\bf r}}\phi({\bf n},{\bf r},E)=I_{\textnormal{coll}}[\phi({\bf n},{\bf r},E)]+Q({\bf n},{\bf r},E). \label{Boltzmann_transport_equation}
\end{equation}
In this equation, $I_{\textnormal{coll}}[\phi({\bf n},{\bf r},E)]$ is the so-called collision term that describes the effect of scattering by atoms or nuclei on the propagation of particles through the material, and $Q({\bf n},{\bf r},E)$ is the source term accounting for the generation of particles. For example, in the Boltzmann transport equation for high-energy electrons,  $Q({\bf n},{\bf r},E)$ describes the generation of electrons by the $\gamma$-photons, whereas in the transport equation for $\gamma$-photons the source term refers to the generation of photons by the relaxation of excited states of atomic nuclei \cite{Yang2021}.  

For the $\gamma$-photons generated by neutrons, the source term has the form
\begin{eqnarray}
&&Q_{\textnormal{ph}}({\bf n}_{\gamma},{\bf r}, E_{\gamma}) \nonumber \\
&=&n_0\int \text{d}o' \text{d}o \int \text{d}E' \text{d}E {\text{d}^2 \sigma _{\textnormal{n}\gamma}({\bf n}',E' \rightarrow {\bf n},E; {\bf n}_{\gamma},E_{\gamma}) \over \text{d}o' \text{d}E'} \nonumber \\
&&\times \phi _{\textnormal{n}}({\bf n}', {\bf r}, E'),
\label{eq:photon_source}
\nonumber \\
\end{eqnarray}
where $n_0$ is the number density of atomic nuclei, $\phi _{\textnormal{n}}({\bf n}', {\bf r}, E')$ is the flux of neutrons interacting with the nuclei, and 
\begin{equation}
   {\text{d}^2 \sigma _{\textnormal{n}\gamma}({\bf n}',E' \rightarrow {\bf n},E; {\bf n}_{\gamma},E_{\gamma}) \over \text{d}o' \text{d}E'} 
\end{equation}
is the differential cross section of scattering of a neutron by an atomic nucleus, with the neutron changing its direction of propagation and energy from $({\bf n}',E')$ to $({\bf n},E)$, accompanied by the production of a $\gamma$-photon with energy $E_{\gamma}$ travelling in the direction ${\bf n}_{\gamma}$. The integration of (\ref{eq:photon_source}) over directions ${\bf n}_{\gamma}$ and energies $E_{\gamma}$ gives the total number of $\gamma$-photons generated in a unit volume of the material per unit time. This quantity can also be computed numerically. For example, the curves shown in Figs.~\ref{fig:photon_source} and \ref{fig:photon_source_channels} were computed using Monte Carlo MCNP simulations. They are equivalent to integrating Eq. (\ref{eq:photon_source}) over directions ${\bf n}_{\gamma}$ but not energies $E_{\gamma}$, yielding the source term in a form differential with respect to the energy of photons $E_{\gamma}$. 

Eq.~\eqref{Boltzmann_transport_equation} itself in some cases can be solved analytically \cite{Chandrasekhar1960,Kalashnikov1985}, but more often its solutions are found numerically using Monte Carlo methods \cite{Berger1963,Vassiliev}, see Ref. \cite{Booth2012} for more detail. 

The collision term in the right-hand side of the Boltzmann transport equation (\ref{Boltzmann_transport_equation}) has the form \cite{Kalashnikov1985,Vassiliev}
\begin{eqnarray}
    &&I_{\textnormal{coll}}[\phi({\bf n},{\bf r},E)]=-n_0\sigma _{\textnormal{tot}}(E) \phi({\bf n}, {\bf r}, E)\nonumber \\
    &+&n_0\int \text{d}o' \int \text{d}E' {\text{d}^2 \sigma ({\bf n}',E' \rightarrow {\bf n},E) \over \text{d}o' \text{d}E'} \phi({\bf n}', {\bf r}, E'),\label{collision_term}
\end{eqnarray}
where $n_0$ is the number density of atoms or, equivalently, atomic nuclei in the material, and integration over $\text{d}o'$ and $\text{d}E'$ is performed over the solid angle and energy of particles scattered by these atoms or nuclei. The first, negative, part of the collision term in (\ref{collision_term}) describes the differential rate of loss of flux from an element of phase space $({\bf n},E)$, whereas the second part of the collision term accounts for the rate of scattering {\it into} this element of phase space. If absorption is the dominant channel of scattering, only the first, negative, term needs to be retained in the right-hand side of  (\ref{collision_term}).

When the above equations are applied to the treatment of transport of high-energy electrons, velocity ${\bf v}=v{\bf n}$ of an electron is related to its momentum ${\bf p}$ through the relativistic kinematic formula
${\bf p}=m{\bf v}/\sqrt{1-{\bf v}^2/c^2}$. The kinetic energy of an electron is $E= c\sqrt {p^2+m^2c^2} - mc^2$ \cite{ClassicalTheoryOfFields}, where $p=|{\bf p}|$ and $c$ is the speed of light. 

\subsection{Photons: iterative solution of the transport equation}

If the source of $\gamma$-photons is directionally isotropic and the rate of spatial variation of the field of photons is negligible in comparison with all the other length scales involved, we can neglect the derivative with respect to spatial coordinates in the left-hand side of (\ref{Boltzmann_transport_equation}), and arrive at
\begin{equation}
    I_{\textnormal{coll}}[\phi_{\textnormal{ph}}(E)]+Q_{\textnormal{ph}}(E)=0. \label{Boltzmann_transport_equation1}
\end{equation}
The angularly isotropic photon flux and the source term, shown in Figs.~\ref{fig:photon_source} and
\ref{fig:photon_source_channels}, are  
\begin{eqnarray}
    \phi _{\textnormal{ph}}({\bf n}, E)&=&{1\over 4 \pi} \phi _{\textnormal{ph}}(E),\nonumber \\
     Q _{\textnormal{ph}}({\bf n}, E)&=&{1\over 4 \pi} Q _{\textnormal{ph}}(E). \label{eq:flux_source_definitions}
\end{eqnarray}
In this approximation, the collision term (\ref{collision_term}) transforms to
\begin{eqnarray}
    &&I_{\textnormal{coll}}[\phi_{\textnormal{ph}}(E)]=-n_0\sigma _{\textnormal{tot}}(E) \phi_{\textnormal{ph}}(E)\nonumber \\
    &+&n_0\int \text{d}E' \phi_{\textnormal{ph}}(E') \int \text{d}o' {\text{d}^2 \sigma ({\bf n}',E' \rightarrow {\bf n},E) \over \text{d}o' \text{d}E'}.\label{collision_term1}
\end{eqnarray}
By denoting the kernel of the integral over the solid angle $do'$ in (\ref{collision_term1}) by
\begin{equation}
    K(E,E') =\int \text{d}o' {\text{d}^2 \sigma ({\bf n}',E' \rightarrow {\bf n},E) \over \text{d}o' \text{d}E'},\label{kernel}
\end{equation}
we arrive at a closed equation for the energy spectrum of $\gamma$-photons 
\begin{eqnarray}
    && Q_{\textnormal{ph}}(E)-n_0\sigma _{\textnormal{tot}}(E) \phi_{\textnormal{ph}}(E) \nonumber \\
    &+& n_0\int \text{d}E'\, K(E,E')\phi_{\textnormal{ph}}(E')=0. \label{Boltzmann_transport_equation2}
\end{eqnarray}
This equation can be readily solved by iteration. This involves representing the energy spectrum in the form of a series, where each term refers to the number of scattering events undergone by an energetic $\gamma$-photon in a material
\begin{equation}
    \phi_{\textnormal{ph}}(E)=\phi_{\textnormal{ph}}^{(0)}(E)+\phi_{\textnormal{ph}}^{(1)}(E)+\phi_{\textnormal{ph}}^{(2)}(E)+ \dots \label{eq:photon_series}
\end{equation}
The zero-order term in the series describes the flux of photons directly emitted by the atomic nuclei
\begin{equation}\label{eq:phi_ph_0}
    \phi_{\textnormal{ph}}^{(0)}(E)={Q_{\textnormal{ph}}(E) \over n_0\sigma _{\textnormal{tot}}(E)},
\end{equation}
and the subsequent terms
\begin{equation}\label{eq:phi_ph_i}
    \phi_{\textnormal{ph}}^{(i)}(E)={1\over \sigma _{\textnormal{tot}}(E)}\int \text{d}E'\, K(E,E')\phi^{(i-1)}(E')
\end{equation}
refer to the contributions to the energy spectrum of $\gamma$-photons from trajectories involving $i$ events of Compton scattering of photons by conduction and inner-shell atomic electrons. Expression for the kernel $K(E,E')$ computed using (\ref{kernel}) has the form
\begin{align} \label{eq:kernel_photons}
K(E, E')=    \begin{cases}
        \frac{\pi r_{\textnormal{c}}^2mc^2}{E'^2}\bigg[\frac{E}{E'}+\frac{E'}{E}-1\\ +\left(\frac{mc^2}{E'}-\frac{mc^2}{E}+1\right)^2\bigg], & \parbox[t]{.6\textwidth}{ $\frac{E'}{1+ \frac{2E'}{mc^2}} < E < E'$}\\
        0, & \text{otherwise}
    \end{cases}
\end{align}
where $r_\textnormal{c}=2.8179$~fm is the classical electron radius. The above formula results from inserting into Eq.~\eqref{kernel} the Klein-Nishina cross section
\begin{equation}\label{eq:KN_cross_section}
    \frac{\text{d}\sigma}{\text{d}o}=\frac{r_{\textnormal{c}}^2}{2} \left(\frac{E'}{E}\right)^2\left[ \frac{E'}{E} +  \frac{E}{E'}-\sin^2\theta\right],
\end{equation}
where $\theta$ is the photon scattering angle.

The photon generation term $Q_{\textnormal{ph}}(E)$ in Eqs.~\eqref{eq:photon_source} and ~\eqref{eq:phi_ph_0}, expressed either in volumetric units cm$^{-3}$s$^{-1}$eV$^{-1}$ in Fig. \ref{fig:photon_source}, or in atomic units  s$^{-1}$atom$^{-1}$eV$^{-1}$ in Fig. \ref{fig:photon_source_channels}, is obtained by folding the (n, $\gamma$) cross section matrices with the energy-spectrum-resolved neutron flux. It is calculated internally in transport codes such as MCNP, and can also be explicitly obtained by inputting the cross sections generated by NJOY \cite{Macfarlane2017} into the processing code SPECTRA-PKA \cite{Gilbert2015}. The cross section in the denominator of Eq.~\eqref{eq:phi_ph_0} and Eq.~\eqref{eq:phi_ph_i} depends on the photon energy. For photons in the energy range from about 10 keV to 10 MeV, $\sigma_{\textnormal{tot}}(E)$ is dominated by photoelectric absorption (PE, the lower end of the energy spectrum), the Compton scattering (CS, the middle part of the spectrum), or by the electron-positron pair production (PP, the upper end of the spectrum). The total cross section can be written as a sum
\begin{equation}\label{eq:sigma_total}
    \sigma_{\textnormal{tot}}(E) = \sigma_{\textnormal{PE}}(E)+\sigma_{\textnormal{CS}}(E)+\sigma_{\textnormal{PP}}(E),
\end{equation}
where the numerical values for individual terms are available from tables \cite{Hubbell1975} or databases \cite{Berger1987}. Of the three processes included in (\ref{eq:sigma_total}), only the Compton scattering does not lead to the absorption of photons, and hence is the only process that contributes to the scattering kernel~\eqref{kernel}. 

Eq.~\eqref{kernel} assumes that electrons in a material act as independent centres of scattering of photons, regardless of them being free or bound in the inner electronic shells of atoms. This approximation is justified if we are interested in the photon energies higher than approximately 100~keV, see Appendix~\ref{app:kernel}. Fig.~\ref{fig:mean_free_path} shows plots of the mean distance between photon scattering events $[n_0\sigma_{\textnormal{tot}}(E)]^{-1}$ that, in the limit where absorption is dominant, determine the characteristic attenuation distance of a photon flux emitted by a localised source. 

The inverse total scattering cross section is the most important scaling factor in the representation of the photon flux as a series in the number of scattering events~\eqref{eq:phi_ph_0} and \eqref{eq:phi_ph_i}. In the interval of energies where the Compton scattering dominates, photons may lose nearly half of their energy in a single scattering event, depending on the angle between directions ${\bf n}$ and ${\bf n}'$. We also note that the above calculations assume that the $\gamma$-photons are generated in the bulk of the material and that the local flux of photons is proportional to the local flux of neutrons. If in addition there is flux of photons from an external source, it will provide an extra contribution to the various scattering events, including the generation of high-energy electrons considered in the next section. 

\subsection{High-energy electrons: the continuous slowing down approximation}

The PE, CS and PP processes, giving rise to the absorption or scattering of $\gamma$-photons, result in the generation of high energy electrons. Similarly to how nonelastic nuclear reactions give rise to the generation of $\gamma$-photons described by the source term $Q_{\textnormal{ph}}(E)$ in (\ref{eq:photon_source}), the PE, CS and PP processes are responsible for the \emph{electron} generation term $Q_{\textnormal{el}}(E)$ in Eq.~\eqref{Boltzmann_transport_equation}. This phenomenon of production of high-energy electrons by $\gamma$-photons is well recognised, for example the authors of Ref. \cite{Moll1997} note the equivalence of $\gamma$-photon and electron irradiation. The source term describing scattering of $\gamma$-photons by atoms serves as a starting point in a calculation of the flux of high-energy electrons.

If the flux of electrons is directionally isotropic and its variation on the spatial scale of the problem is negligible, we write 
\begin{equation}
    \phi_{\textnormal{el}}({\bf n},E)={1\over 4\pi}\phi_{\textnormal{el}}(E). \label{isotropic_flux}
\end{equation}
The only variable characterising the flux of electrons is their energy $E$, and now the collision term (\ref{collision_term}) describes electron energy losses, which can be treated in the continuous slowing down approximation \cite{Vassiliev,Kalashnikov1985,Dudarev1995}. In this approximation, assuming that the flux of electrons is independent of spatial coordinates, the Boltzmann transport equation acquires the form 
\begin{equation}
    0={\partial \over \partial E}[\overline \varepsilon (E)\phi_{\textnormal{el}}(E)]+Q_{\textnormal{el}}(E),\label{Boltzmann_stopping}
\end{equation}
where $\overline {\varepsilon }(E)$ is the average rate of energy losses of an electron with energy $E$, and $Q_{\textnormal{el}}(E)$ is the rate of generation of electrons with energy $E$ by the $\gamma$-photons. The average rate of energy losses per unit distance travelled $\overline {\varepsilon }(E)$ is related to the range $R(E)$ of electrons in the material {\it via} \cite{Bethe1932,Kalashnikov1985,Vassiliev,Grimes2017,ElGhossain2017}
\begin{equation}
    R(E)=\int \limits _{0} ^{E} {\text{d}E' \over \overline \varepsilon (E')}.\label{range}
\end{equation}
Equation (\ref{Boltzmann_stopping}) can be readily solved, and its solution has the form
\begin{equation}\label{eq:el_flux}
    \phi_{\textnormal{el}}(E)={1\over {\overline \varepsilon (E)}} \int \limits _E^{\infty}Q_{\textnormal{el}}(E')\text{d}E'.
\end{equation}
This analytical expression describes a directionally isotropic distribution of high energy electrons (\ref{isotropic_flux}) that energetic $\gamma$-photons generate in the bulk of the material. In practice the upper limit of integration in (\ref{eq:el_flux}) is determined by the energy span of the source function $Q_{\textnormal{el}}(E)$, whereas numerical values of function $\overline \varepsilon (E)$ are available in literature  \cite{Berger1999}. The flux of electrons 
$\phi_{\textnormal{el}}(E)$ can also be computed numerically using Monte Carlo simulations \cite{Berger1963,Vassiliev}, and below we show that results of such simulations compare favourably with the analytical result \eqref{eq:el_flux}.

The electron generation term has the form
\begin{equation}
    Q_{\textnormal{el}}(E) = Q_{\textnormal{PE}}(E) + Q_{\textnormal{CS}}(E)  + Q_{\textnormal{PP}}(E).
\end{equation}
At high energies well above the keV range, we neglect the electron binding energy effects and write
\begin{equation}
    Q_{\textnormal{PE}}(E) = n_0\sigma_{\textnormal{PE}}(E)\phi_{\textnormal{ph}}(E).
\end{equation}

The Compton scattering contribution can be evaluated by noting that the energy balance dictates that the electron energy equals the difference between the photon energy before and after the event, $E=E_{\textnormal{ph}}-E'_{\textnormal{ph}}$. Therefore it follows from Eq.~\eqref{eq:kernel_photons} that the energy-differential Compton cross section involving an electron receiving recoil energy $E$ is
\begin{align}\label{eq:KN_energy_diff}
\frac{\text{d}\sigma}{\text{d}E}=    \begin{cases}
        \frac{\pi r_{\textnormal{c}}^2mc^2}{E_{\textnormal{ph}}^2}\bigg[\frac{E_{\textnormal{ph}}}{E_{\textnormal{ph}}-E}-\frac{E}{E_{\textnormal{ph}}}\\ +\left(\frac{mc^2}{E_{\textnormal{ph}}}-\frac{mc^2}{E_{\textnormal{ph}}-E}+1\right)^2\bigg], & \parbox[t]{.6\textwidth}{$0 < E < \frac{2E_{\textnormal{ph}}^2}{2E_{\textnormal{ph}}+mc^2}$}\\
        0. & \text{otherwise}
    \end{cases}
\end{align}
If the number density of electrons is $n_{\textnormal{el}}$, the resulting electron generation term is
\begin{equation}\label{eq:Q_CS}
    Q_{\textnormal{CS}}(E) = n_{\textnormal{el}}\int\phi_{\textnormal{ph}}(E_{\textnormal{ph}})\frac{\text{d}\sigma}{\text{d}E}(E_{\textnormal{ph}}, E)\text{d}E_{\textnormal{ph}}.
\end{equation}

If the photon flux is represented by a discrete $N$-point set on a grid of photon energies $E_i$, we can write the energy-differential flux of electrons in the form of a discrete sum of Dirac delta functions
\begin{equation}
     \phi_{\textnormal{ph}}(E_{\textnormal{ph}})=\sum_{i=1}^N\Phi_{\textnormal{ph}}^{(i)}\delta(E_{\textnormal{ph}}-E_i).
\end{equation}
This then simplifies Eq.~\eqref{eq:Q_CS} to
\begin{equation}
     Q_{\textnormal{CS}}(E) = n_{\textnormal{el}}\sum_{i=1}^N\Phi_{\textnormal{ph}}^{(i)}\frac{\text{d}\sigma}{\text{d}E}(E_i, E).
\end{equation}

Finally, high-energy electrons can be generated through the production of electron-positron pairs, resulting from the conversion of photons with energy greater than $2mc^2\sim1.022$~MeV into electrons and positrons. The sum of kinetic energies of an electron and a positron is then $E_{\textnormal{ph}}-2mc^2$. Assuming that this energy is split equally between the two particles, we find that $E=\frac{1}{2}(E_{\textnormal{ph}}-2mc^2)$ and
\begin{equation}
    Q_{\textnormal{PP}}(E) = n_0\sigma_{\textnormal{PP}}(E_{\textnormal{ph}})\phi_{\textnormal{ph}}(E_{\textnormal{ph}}).
\end{equation}

\begin{figure}
  \includegraphics[width=0.49\textwidth]{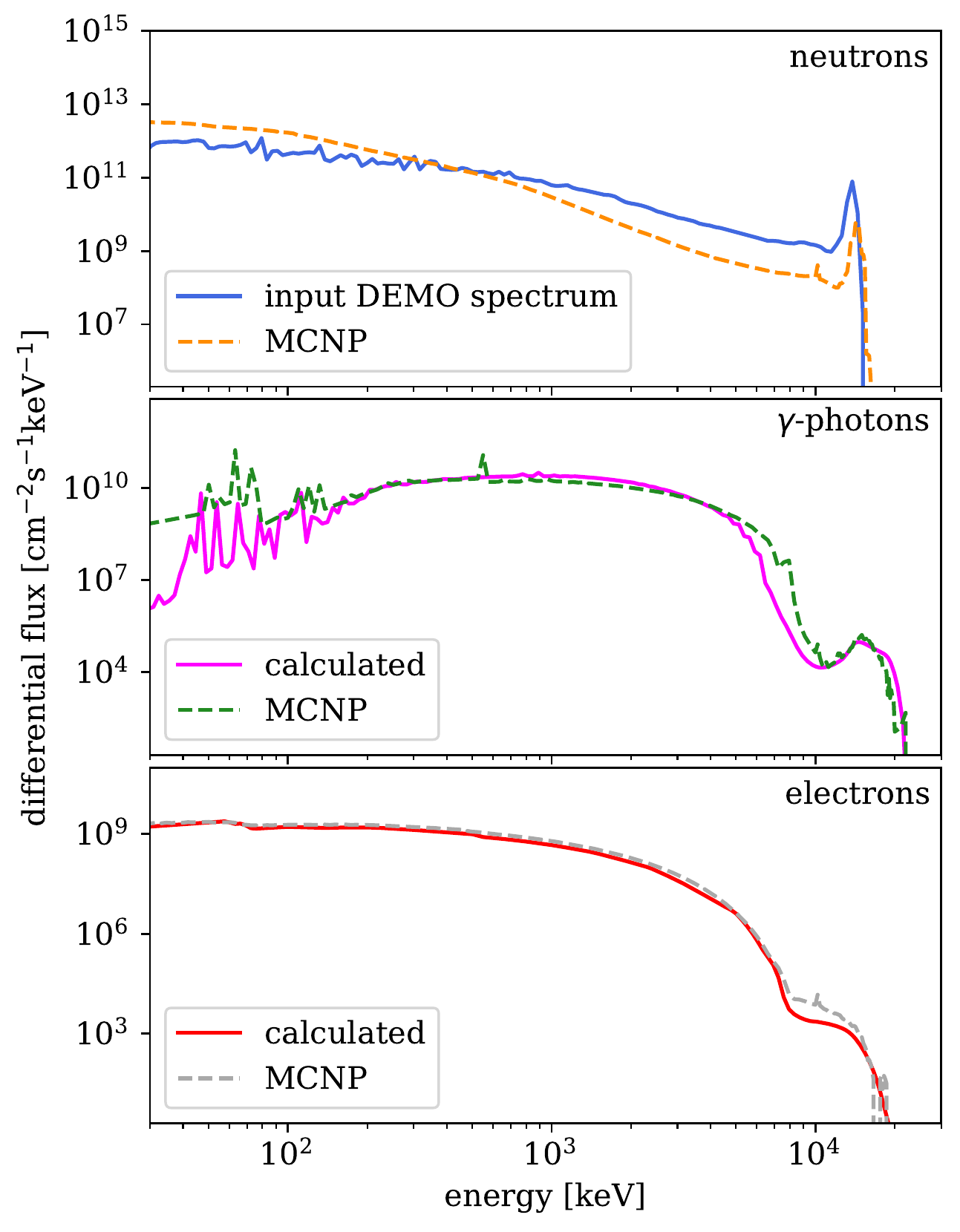}%
    \caption{Top: input DEMO spectrum shown together with the neutron spectrum it generates in  MCNP under periodic boundary conditions. Central and bottom panels: MCNP photon and electron spectra compared to the  $\phi_{\textnormal{ph}}(E)$ and $\phi_{\textnormal{el}}(E)$ spectra computed using equations given in the text. The abscissa is the same for the three panels. Note that the two neutron spectra are different as one is representative of a thin first-wall W layer in a mixed-material DEMO reactor and the other is the result of that same spectrum being propagated through pure bulk W. The photon and electron spectra are consistent with the MCNP bulk W simulation and thus the equivalence between the MCNP and calculated curves validates the methodology for the latter.}
    \label{fig:MCNP_validation} 
\end{figure}

Monte Carlo MCNP simulations were used for validating the approximations involved in Eq.~\eqref{eq:phi_ph_0} and Eq.~\eqref{eq:phi_ph_i} for $\gamma$-photons and Eq.~\eqref{eq:el_flux} for high-energy electrons. The curves computed independently using the above equations and MCNP simulations are shown in Fig.~\ref{fig:MCNP_validation}. The DEMO spectrum of neutrons from Fig.~\ref{fig:neutron_spectra} was used as a spatially homogeneous incident flux of neutrons in a $20\times20\times20$~cm$^3$ cube of tungsten, with periodic boundary conditions (PBCs) applied along $x$ and $y$, and reflecting planes bounding the cell in $z$. Neutrons, photons and electrons were tallied in the box to produce the MCNP results shown in Fig.~\ref{fig:MCNP_validation}. Using tallies on a grid of 8000 {1\,cm\(^3\)} voxels, we confirmed that there was no detectable spatial variation of the spectra and thus the results are representative of an effectively infinite bulk sample of W. The total neutron flux of the input source spectrum was $5.04\times10^{14}$~cm$^{-2}$s$^{-1}$. The neutron flux calculated by MCNP was normalised to the same value. This resulting tallied neutron spectrum is slightly different from the input spectrum, shown in the top panel of the figure because, as opposed to the current simulation, the input neutron spectrum was produced in simulation assuming a thin W layer in the first wall of a full-reactor DEMO involving also other materials. Because of the high computational cost of simulating the photon and particularly electron transport in MCNP, it was not feasible to perform the $n-\gamma-e$ transport simulation using the full DEMO geometry. The MCNP-generated neutron flux was used for evaluating $\phi_{\textnormal{ph}}(E)$ through Eqs.~\eqref{eq:photon_series}, \eqref{eq:phi_ph_0} and \eqref{eq:phi_ph_i}, and then for comparing the outcome to the MCNP \(\gamma\)-photons result. Similarly, the MCNP-generated photon flux represented input for high-energy electron flux calculations, which was then compared to the MCNP result. It is worth noting that the MCNP solution with full photon and electron transport for this relatively simple system (single cube with no complex surfaces or long transport paths) and a relatively modest 10\(^8\) source neutrons took approximately 2 days using 8 processors, whereas the calculation according to the treatment proposed in this section required of the order of seconds on a laptop.

\begin{figure*}[t]
\subfloat[]{%
  \includegraphics[width=\columnwidth]{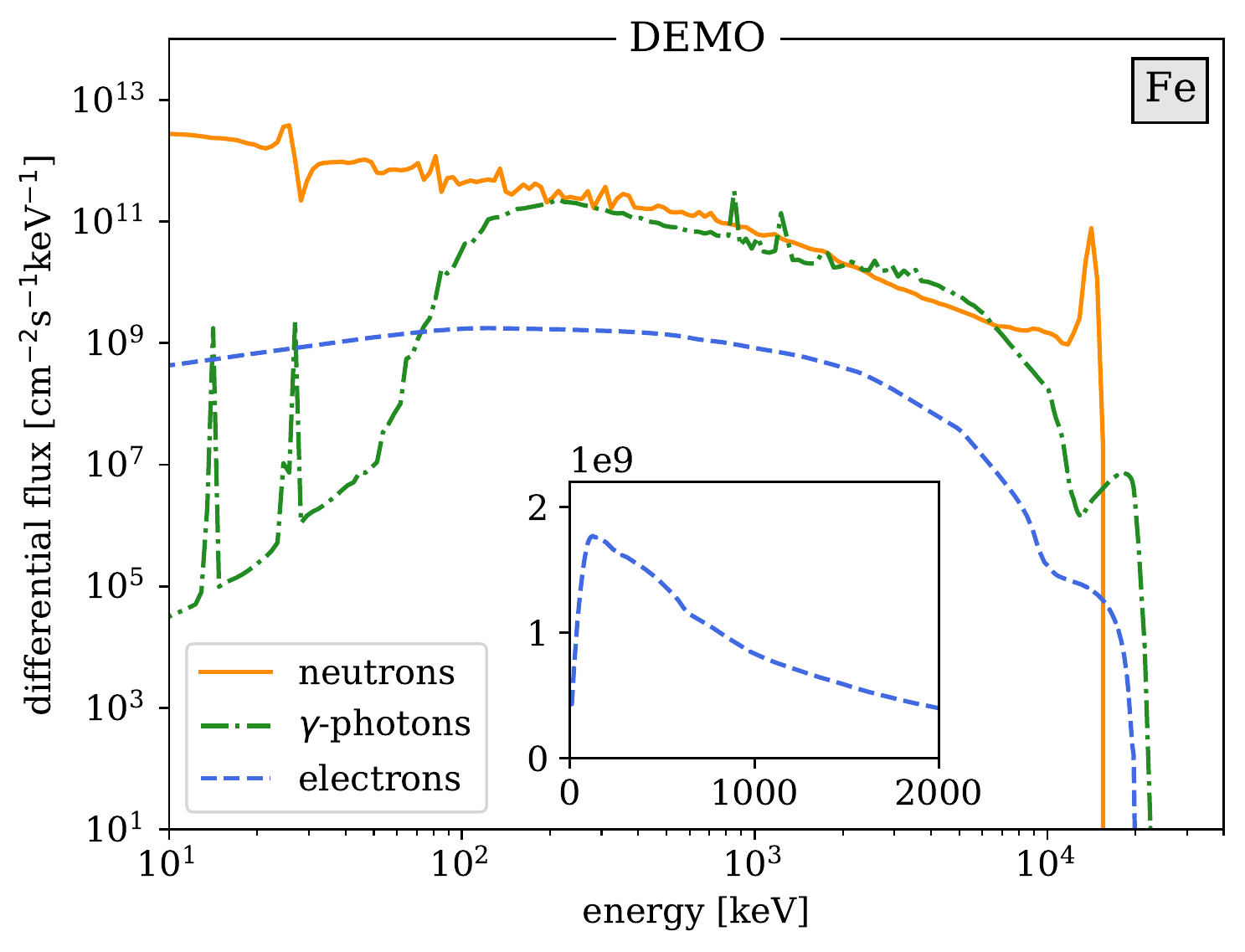}%
}\hfill
\subfloat[]{%
  \includegraphics[width=\columnwidth]{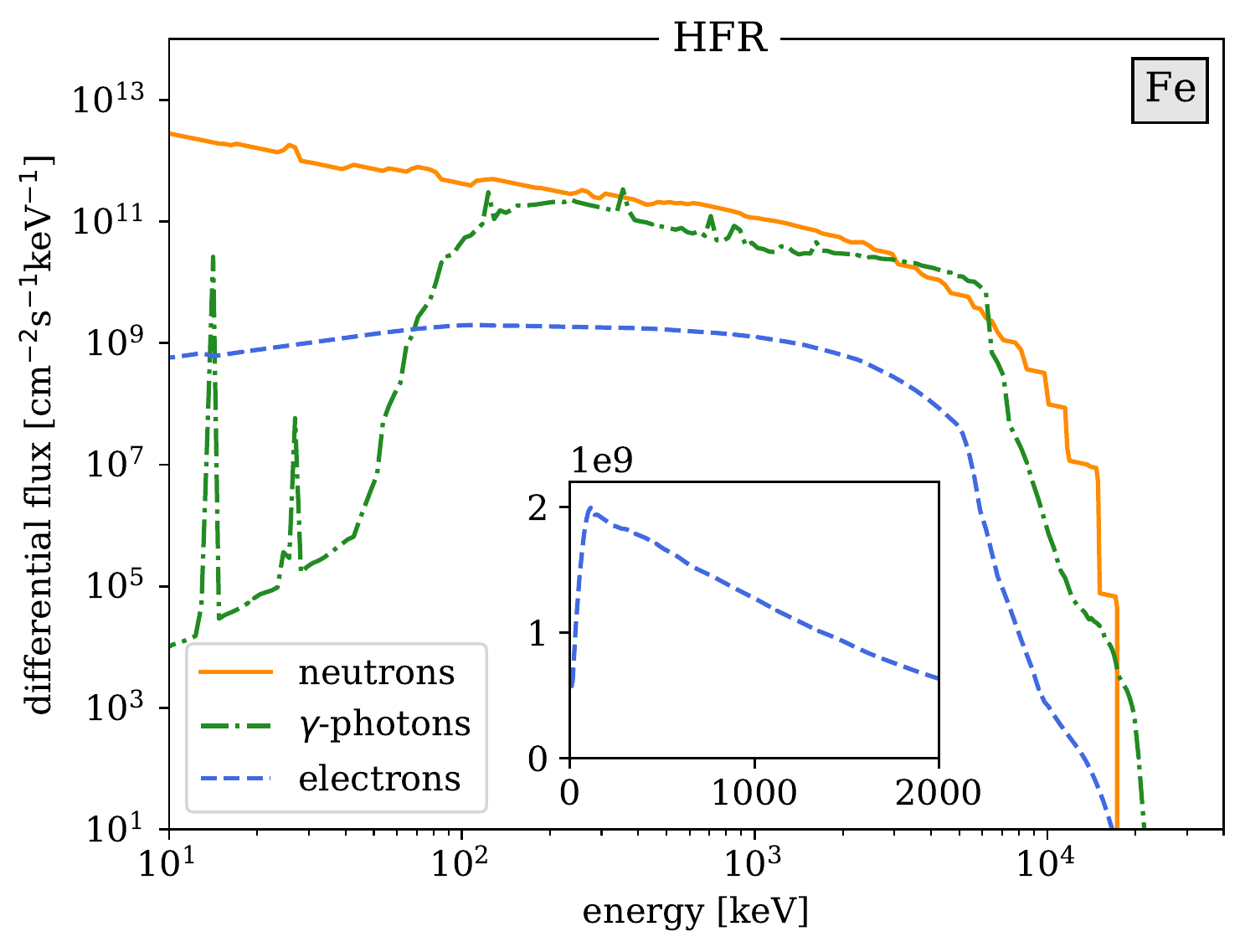}%
}\hfill
\subfloat[]{%
  \includegraphics[width=\columnwidth]{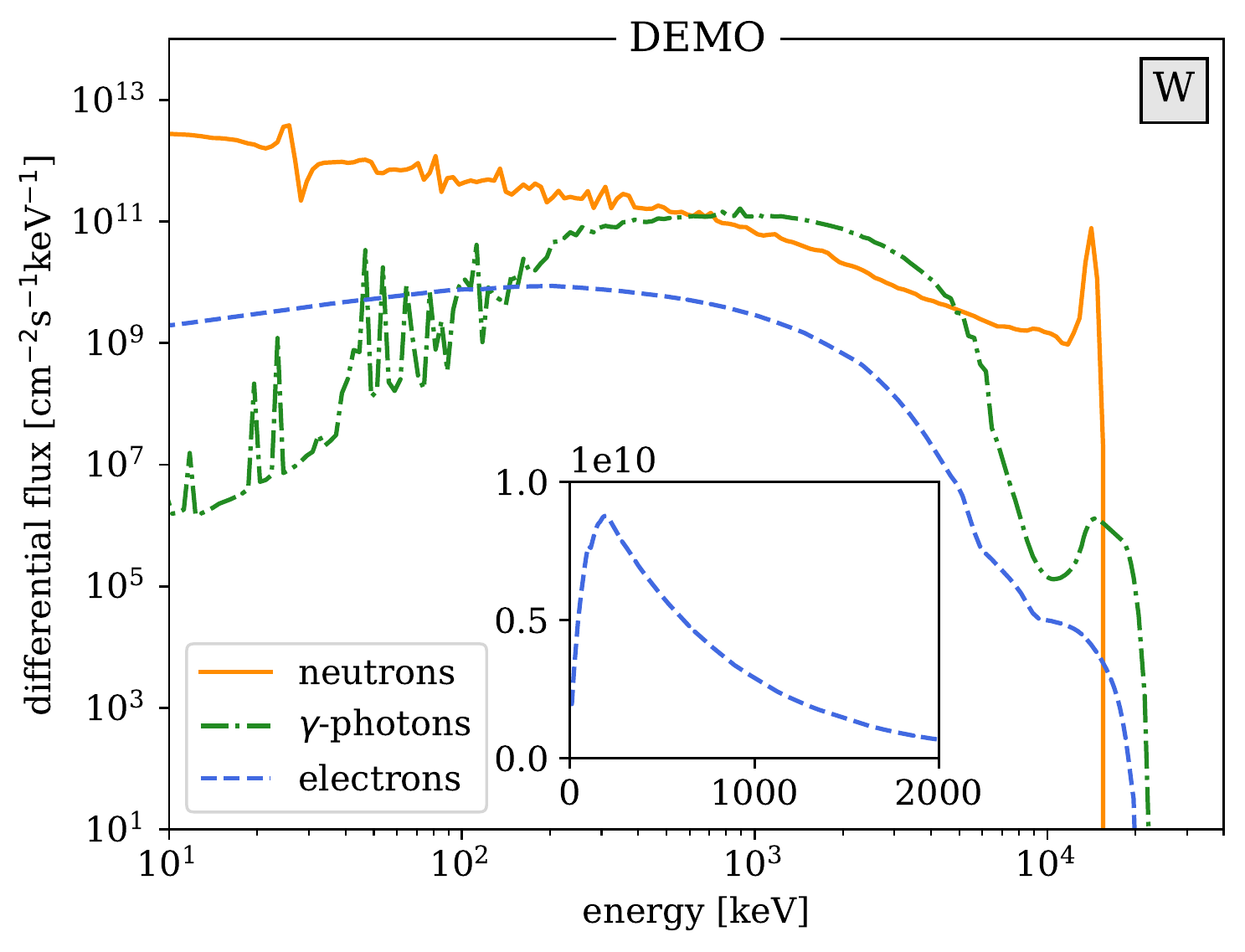}%
}\hfill
\subfloat[]{%
  \includegraphics[width=\columnwidth]{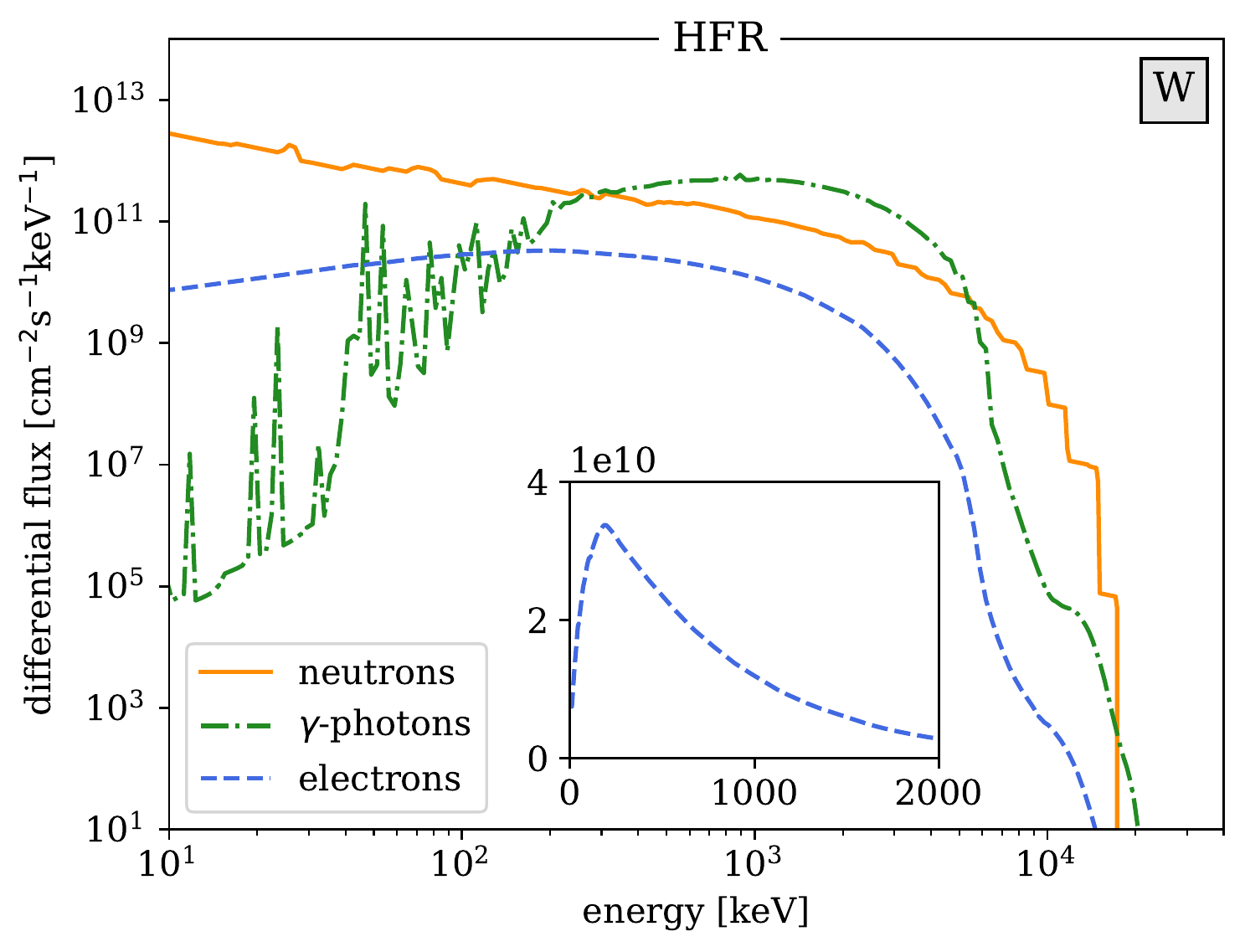}%
}\hfill
    \caption{Energy-resolved differential neutron, $\gamma$-photon and electron spectra in the bulk of neutron-irradiated Fe (a, b) and W (c, d) corresponding to fusion (DEMO) and fission (HFR) conditions. Insets show the energy-differential electron fluxes on a linear scale. These are all maximum at approximately 200~keV, but even at $\sim2$~MeV they still retain $\sim10$~\% of the peak intensity. Photon and electron spectra were calculated according to the theory presented in Sec.~\ref{sec:scatterings}.}
    \label{fig:spectra_DEMO_HFR} 
\end{figure*}

Neutron flux used for evaluating the generation rate of $\gamma$-photons and high-energy electrons in Fig.~\ref{fig:MCNP_validation} refers to what were to be expected if scattering occurred in pure W. In Fig.~\ref{fig:spectra_DEMO_HFR} and Table~\ref{tab:fluxes} we plot the curves and provide numerical data for the photon and electron spectra developing in bulk Fe and W exposed to the neutron spectra shown in Fig.~\ref{fig:neutron_spectra}, calculated as explained in Sec.~\ref{sec:scatterings}. In both fission and fusion scenarios the photon and electron spectra are qualitatively similar. The intensity of photon fluxes is comparable to the neutron fluxes. Notably, the neutrons generate $\gamma$-photon spectra with the characteristic energy of $\sim1-1.5$~MeV, and electrons spectra with the characteristic energy of $\sim0.5-1$~MeV, which are skewed towards high energies. 

The characteristic scale of electron energies that we see in Fig.~\ref{fig:spectra_DEMO_HFR}
is the same as the energy range of electrons used in transmission electron microscope experiments \cite{Arakawa2020} where it was found that the flux of electrons was able to drive microstructural evolution of the material exposed to an electron beam. In W, the photon and electron fluxes computed for the HFR input spectrum are about 4 times higher than those computed for the DEMO neutron spectrum.

\begin{table*}[t]
\caption{Total neutron ($\Phi_{\textnormal{n}}$), $\gamma$-photon ($\Phi_{\textnormal{ph}}$) and electron ($\Phi_{\textnormal{el}}$) fluxes in W and Fe and in DEMO and HFR conditions, calculated as presented in Sec.~\ref{sec:scatterings}, and associated median neutron energy (MNE), median photon energy (MPE) and median electron energy (MEE), calculated as the energy below and above which lie the two halves of the total flux.}
\label{tab:fluxes}
\begin{ruledtabular}  
\begin{tabular}{lcccccccccc}
 & $\Phi_{\textnormal{n}}$ & MNE & $\Phi_{\textnormal{ph}}$ in W & MPE in W & $\Phi_{\textnormal{ph}}$ in Fe & MPE in Fe &  $\Phi_{\textnormal{el}}$ in W & MEE in W & $\Phi_{\textnormal{el}}$ in Fe & MEE in Fe \\
 & [cm$^{-2}$s$^{-1}$] & [keV] & [cm$^{-2}$s$^{-1}$] & [keV] & [cm$^{-2}$s$^{-1}$] & [keV] & [cm$^{-2}$s$^{-1}$] & [keV] & [cm$^{-2}$s$^{-1}$] & [keV] \\
DEMO & $5.04\times10^{14}$ & 295 & $2.74\times10^{14}$ & 1,410 & $1.78\times10^{14}$ & 891 & $7.57\times10^{12}$ & 589 & $2.36\times10^{12}$ & 934 \\
HFR & $6.83\times10^{14}$ & 129 & $1.09\times10^{15}$ & 1,480 & $2.04\times10^{14}$ & 1,230 & $3.03\times10^{13}$ & 589 & $3.22\times10^{12}$ & 1,070 \\
\end{tabular}
\end{ruledtabular}
\end{table*}

\section{Atomic recoils produced by high-energy electrons}
In the treatment of scattering of energetic particles by atoms in a material, involving either the Boltzmann transport equation \cite{Kalashnikov1985,Vassiliev} or Monte Carlo simulations \cite{Berger1963}, the fact that scattering involves not only the change of momenta of the incident particles but also atomic recoils, is often not recognised. Experimental observations and theoretical analysis \cite{Boersch1967,Fujikawa2006,Winkelmann2011,Arakawa2020} show that even the relatively low energy electrons produce atomic recoils with energies in the electron-Volt range, comparable with the energy barrier for vacancy migration in metals \cite{Dudarev2013}. 

High-energy atomic recoils can be formally treated as multi-phonon excitation events, whereas at low energies it is often sufficient to retain only the lowest order single-phonon terms when computing the scattering structure factor \cite{Dudarev1993}. Irrespectively of the recoil energy, electron-atom recoils can be treated as electron-phonon interaction events, resulting in the eventual dissipation of the energy of high-energy electrons into the heat bath of thermal atomic vibrations. 

The treatment of relativistic collision kinematics involving an electron with mass $m$ and an atom with mass $M$ gives a relation between the kinetic energy of the recoil atom $E_{\textnormal{R}}$ and the angle of scattering $\theta $, defined in the centre of mass frame \cite{ClassicalTheoryOfFields} 
\begin{equation}
    E_{\textnormal{R}}={Mc^2E_{\textnormal{el}}(E_{\textnormal{el}}+2mc^2)\over (m+M)^2c^4+2Mc^2E_{\textnormal{el}}}(1-\cos \theta),\label{eq:recoil_energy}
\end{equation}
where $E_{\textnormal{el}}$ is the kinetic energy of the fast electron, ${E_{\textnormal{el}}=mc^2/\sqrt{1-v^2/c^2}-mc^2}$, and $\cos \theta = ({\bf n}\cdot {\bf n}')$. 
In the limit $m\ll M$,  Eq.~\eqref{eq:recoil_energy} can be simplified as \cite{Reimer1984,Was2016,Yang2021}
\begin{equation}
    E_{\textnormal{R}}={E_{\textnormal{el}}(E_{\textnormal{el}}+2mc^2)\over Mc^2}(1-\cos \theta).\label{eq:recoil_energy_simplified}
\end{equation}
The maximum  amount of energy is transferred to an atom in a collision where the electron is scattered exactly backwards $\theta= \pi$. Taking $\cos \theta =-1$, from Eq.~\eqref{eq:recoil_energy_simplified} we find
\begin{equation}
    E_{\textnormal{R}}^{\textnormal{max}}(E_{\textnormal{el}})=2E_{\textnormal{el}}{E_{\textnormal{el}}+2mc^2 \over Mc^2},\label{eq:recoil_energy_maximum}
\end{equation}
which agrees with Eq. (\ref{E_R_max}).
Estimates based on this formula suggest that backscattering of electrons with kinetic energies in the MeV range can readily generate atomic recoils with energies many times the magnitude of the potential barrier for vacancy migration in metals. 

The characteristic scale of angle $\theta$ in \eqref{eq:recoil_energy} depends on the differential cross section of scattering. Elastic scattering of relativistic electrons by atoms is well described by the screened Coulomb Rutherford cross section \cite{Reimer1984,Boschini2013, Dudarev1995}
\begin{equation}
    \left(\frac{\text{d}\sigma}{\text{d}o}\right)= \left(\frac{Ze^2}{4\pi\varepsilon_0 m c^2}\right)^2\left(\frac{1-\beta^2}{\beta^4}\right)\frac{1}{(1+\kappa-\cos\theta)^2} \label{eq:screened_Rutherford} ,
\end{equation}
where $\beta ^2=v^2/c^2=1-(1+E_{\textnormal{el}}/mc^2)^{-2}$, $Z$ is the atomic number, and $\kappa$ is the screening parameter \cite{Dudarev1995} inversely proportional to the effective size of the atom,  
\begin{equation}
    \kappa= {\hbar ^2\over 2p^2a^2_{\textnormal{TF}}},\: \kappa \ll 1
\end{equation}
where $p$ is the relativistic momentum of the incident electron ${p=mv/\sqrt{1-v^2/c^2}}$, and ${a_{\textnormal{TF}}=0.885a_{\textnormal{B}}/Z^{1/3}}$ is the Thomas-Fermi atomic radius. Here, we use the atomic system of units where the Bohr radius is ${a_{\textnormal{B}}={\hbar ^2 /me^2}=0.52918}$~\AA$\;$ and $e^2/a_{\textnormal{B}}=27.2116$ eV.

The original Rutherford formula does not treat the effect of screening of the electrostatic potential of the nucleus nor the effects of electron spin. The Mott cross section takes the latter into account by solving the Dirac rather than the Schr{\"o}dinger equation \cite{Rez1982}. The Mott cross section is usually expressed in terms of the Rutherford cross section as
\begin{equation}\label{eq:Mott_CS}
     \left(\frac{\text{d}\sigma_{\textnormal{M}}}{\text{d}o}\right)= \left(\frac{\text{d}\sigma_{\textnormal{R}}}{\text{d}o}\right)R_{\textnormal{M}},
\end{equation}
where factor $R_M$ for elements with $Z>20$ is found numerically \cite{Reimer1984}. Lijian {\it et al.} \cite{Lijian1995} proposed a polynomial interpolation for the unscreened Mott cross section
\begin{equation}
    R_{\textnormal{M}} = \sum_{j=0}^{4}a_j(Z, \beta)(1-\cos\theta)^{j/2}
\end{equation}
where
\begin{equation}
    a_j(Z, \beta) = \sum_{k=1}^{6}b_{k,j}(Z) (\beta-\bar{\beta})^{k-1}.
\end{equation}
and $\bar{\beta}=0.7181287$. The numerical values of the coefficients $b_{k,j}$ can be found in tables \cite{Boschini2013, Lijian1995}.

Since the scattering potential is radially symmetric, in the expression for an element of the solid angle $\text{d}o$ it is sufficient to retain only the polar component ${\text{d}o=2\pi\sin\theta\text{d}\theta}$, where $\theta$ is the angle of scattering of the electron. In the treatment of  generation and induced motion of lattice defects, it is convenient to work with the kinetic energy of atomic recoils, related to $\theta$ by 
\begin{equation}\label{E_R}
    E_{\textnormal{R}}(\theta) = E_{\textnormal{R}}^{\textnormal{max}}\sin^2\frac{\theta}{2},
\end{equation}
which follows from Eqs.~\eqref{eq:recoil_energy_simplified} and \eqref{eq:recoil_energy_maximum}. Using the chain rule, we define a quantity
\begin{equation}\label{eq:dS_dEr}
    \left(\frac{\text{d}\sigma}{\text{d}E_{\textnormal{R}}}\right)=\left[2\pi\sin\theta \left(\frac{\text{d}\sigma}{\text{d}o}\right)\right]_{\theta=\theta(E_{\textnormal{R}})} \left|\frac{\text{d}\theta}{\text{d}E_{\textnormal{R}}}\right|,
\end{equation}
where
$$
\theta(E_{\textnormal{R}})=2\sin^{-1}\sqrt{\frac{E_{\textnormal{R}}}{E_{\textnormal{R}}^{\textnormal{max}}}},
$$
$$
\left|\frac{\text{d} \theta}{\text{d} E_{\textnormal{R}}}\right|=\frac{1}{\sqrt{E_{\textnormal{R}}(E_{\textnormal{R}}^{\textnormal{max}}-E_{\textnormal{R}})}}
$$

Similarly to how $ \left({\text{d}\sigma}/{\text{d}\theta}\right)$ gives the probability of an electron being scattered by angle $\theta$, $\left({\text{d}\sigma}/{\text{d}E_{\textnormal{R}}}\right)$ is proportional to the probability of a target atom receiving recoil energy $E_{\textnormal{R}}$.

\begin{figure}[t]
\includegraphics[width=\columnwidth]{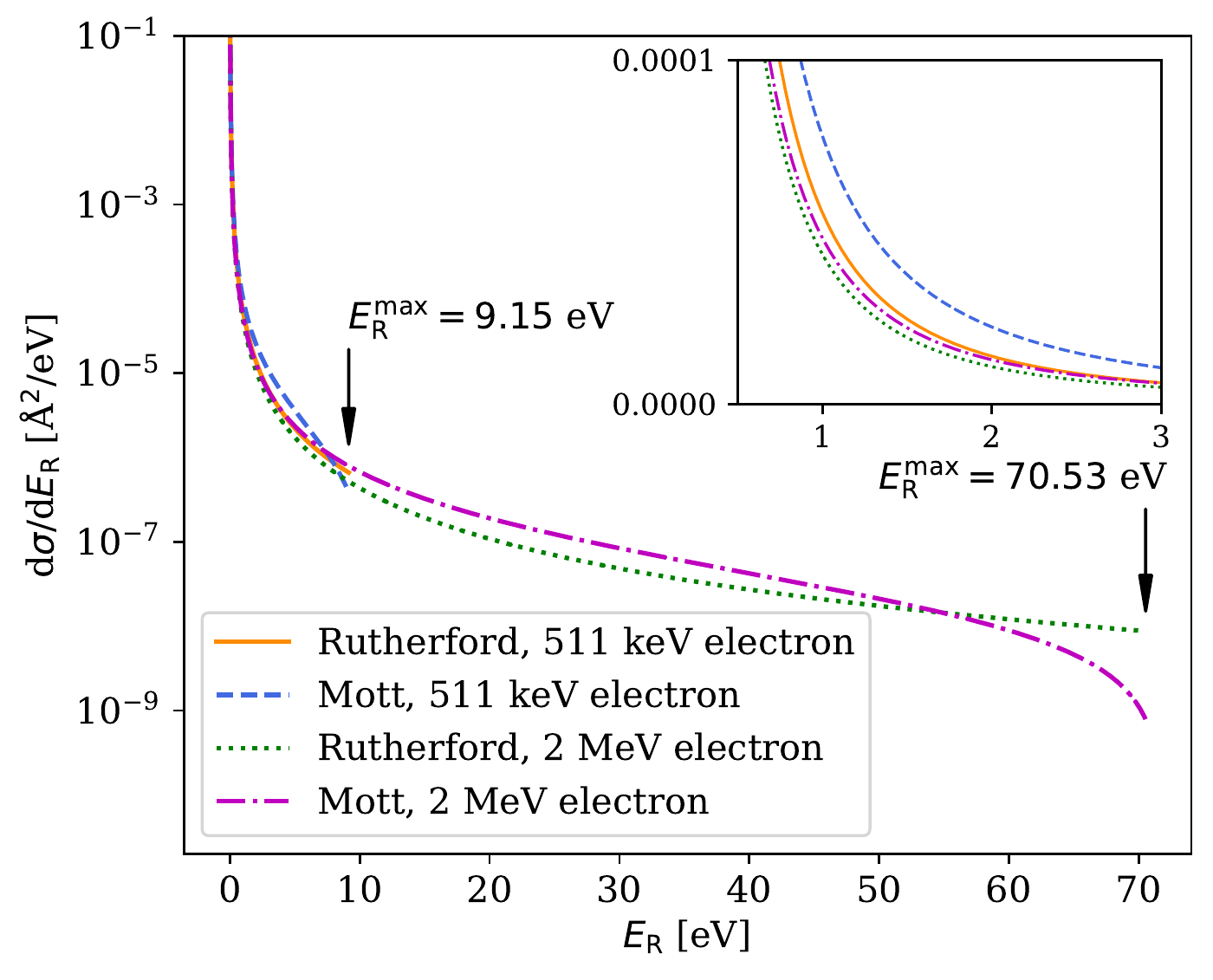}%
\caption{Differential cross sections for a tungsten atom to receive recoil energy $E_{\textnormal{R}}$ after a collision with an electron with kinetic of 511~keV or 2~MeV, respectively. The screened relativistic Rutherford cross section is compared to the Mott cross section. Both terminate at the maximum recoil energy given by Eq.~\eqref{E_R_max}. The inset shows a magnified view of the curve with the vertical axis shown on a linear scale.}
\label{fig:dS_dE}
\end{figure}

Using Eqs.~\eqref{eq:screened_Rutherford} and \eqref{eq:Mott_CS}, and noting that
$$
\frac{\sin\theta(E_{\textnormal{R}})}{\sqrt{E_{\textnormal{R}}(E_{\textnormal{R}}^{\textnormal{max}}-E_{\textnormal{R}})}}=\frac{2}{E_{\textnormal{R}}^{\textnormal{max}}},
$$
we simplify Eq.~\eqref{eq:dS_dEr} as
\begin{equation}\label{eq:dS_dEr_simplified}
    \left(\frac{\text{d}\sigma}{\text{d}E_{\textnormal{R}}}\right)=
    2\pi\left(\frac{Ze^2}{4\pi\varepsilon_0 m c^2}\right)^2\left(\frac{1-\beta^2}{\beta^4}\right)\frac{2E_{\textnormal{R}}^{\textnormal{max}}}{(\kappa E_{\textnormal{R}}^{\textnormal{max}}+2E_{\textnormal{R}})^2}
\end{equation}
if scattering is described by the screened relativistic Rutherford cross section. Alternatively, 
\begin{equation}\label{eq:dS_dEr_simplified_Mott}
    \left(\frac{\text{d}\sigma}{\text{d}E_{\textnormal{R}}}\right)=
    2\pi R_M\left(\frac{Ze^2}{4\pi\varepsilon_0 m c^2}\right)^2\left(\frac{1-\beta^2}{\beta^4}\right)\frac{E_{\textnormal{R}}^{\textnormal{max}}}{2E_{\textnormal{R}}^2}
\end{equation}
if scattering is described by the Mott formula. In Fig.~\ref{fig:dS_dE} we compare the two expressions at two different electron energies. The Mott cross section is more accurate, but computing it requires a large number of coefficients; the Rutherford cross section may be more appropriate since the error that it introduces is small given all the other approximations involved in the analysis.

To validate the method, we now apply Eq.~\eqref{eq:dS_dEr} to the evaluation of the rate at which high-energy electrons generate Frenkel pairs (FP) once the transferred recoil energy $E_{\textnormal{R}}$ surpasses a certain threshold atomic  displacement energy barrier \cite{Maury1978,Vajda1977}. This amounts to estimating the number of stable defects $N_{\textnormal{d}}(E_{\textnormal{R}})$ remaining after a collision event. Using the data derived from MD simulations, Yang and Olsson \cite{Yang2021} modified the arc-dpa (athermal recombination-corrected displacement per atom) model \cite{Nordlund2018} and found that the number of stable defects increases linearly as a function of $E_{\textnormal{R}}$ above a certain minimum energy $E_{\textnormal{d}}^{\textnormal{min}}$. A ``defect production differential cross section'' can be defined by combining the energy-differential cross section given by Eq.~\eqref{eq:dS_dEr} and the Yang-Olsson expression
\begin{equation}\label{eq:arcdpa}
N_{\textnormal{d}}(E_{\textnormal{R}})=
\begin{cases}
    0 , & E_{\textnormal{R}} < E_{\textnormal{d}}^{\textnormal{min}} \\
    \frac{0.8E_{\textnormal{R}}}{2E_{\textnormal{d}}^{\textnormal{avr}}} , & E_{\textnormal{d}}^{\textnormal{min}} \leq E_{\textnormal{R}} < \frac{2E_{\textnormal{d}}^{\textnormal{avr}}}{0.8} \\
    \frac{0.8E_{\textnormal{R}}}{2E_{\textnormal{d}}^{\textnormal{avr}}}\xi(E_{\textnormal{R}}) , &  E_{\textnormal{R}} \geq \frac{2E_{\textnormal{d}}^{\textnormal{avr}}}{0.8}
\end{cases}
\end{equation}
where
$$
\xi(E_{\textnormal{R}}) = (1-c_{\textnormal{arcdpa}})\left(\frac{E_{\textnormal{R}}}{2E_{\textnormal{d}}^{\textnormal{avr}}/0.8}\right)^{b_{\textnormal{arcdpa}}}+c_{\textnormal{arcdpa}}.
$$
We then obtain the number of stable defects as a function of recoil energy. The four parameters in Eq.~\eqref{eq:arcdpa} can be obtained by MD simulations, and thus depend on the chosen interatomic potential. For the W potential used in this work, after Ref.~\cite{Mason2017}, we obtained the values: ${E_{\textnormal{d}}^{\textnormal{min}}=47}$~eV, ${E_{\textnormal{d}}^{\textnormal{avr}}=106}$~eV, ${b_{\textnormal{arcdpa}}=-0.80}$, ${c_{\textnormal{arcdpa}}=0.23}$. We note that in Eq.~\eqref{eq:arcdpa} we assume that the recoil energy is equal to the damage energy. We are, in other words, neglecting electronic losses because during recoils of energy 10–100 eV only about 5\% of the energy is lost to electronic stopping \cite{Yang2021}. 

Since in a single collision any amount of energy ${0<E_{\textnormal{R}}<E_{\textnormal{R}}^{\textnormal{max}}}$ can be transferred, the defect production cross section for a single atom is given by the integral
\begin{equation}\label{eq:sigmaFP}
\sigma_{FP}=\int \limits_{E_{\textnormal{d}}^{\textnormal{min}}}^{E_{\textnormal{R}}^{\textnormal{max}}}\left(\frac{\text{d}\sigma}{\text{d}E_{\textnormal{R}}}\right)
N_{\textnormal{d}}(E_{\textnormal{R}}) \ \text{d} E_{\textnormal{R}}.
\end{equation}
The Frenkel pair production rate per unit volume is proportional to $\sigma_{\textnormal{FP}}$, to the electron flux $\phi_{\textnormal{el}}$ and the target atomic density $n_0$,
$$
\propto \phi_{\textnormal{el}}n_0\sigma_{\textnormal{FP}}.
$$
This appears to suggest that the concentration of Frenkel pairs increases linearly with time, which is approximately valid only at the very early stages of irradiation, where the defects are isolated and do not coalesce or recombine because of elastic interactions, and only if the temperature of the material is sufficiently low so that the thermal diffusion of defects is not activated.

\begin{figure}[t]
\includegraphics[width=\columnwidth]{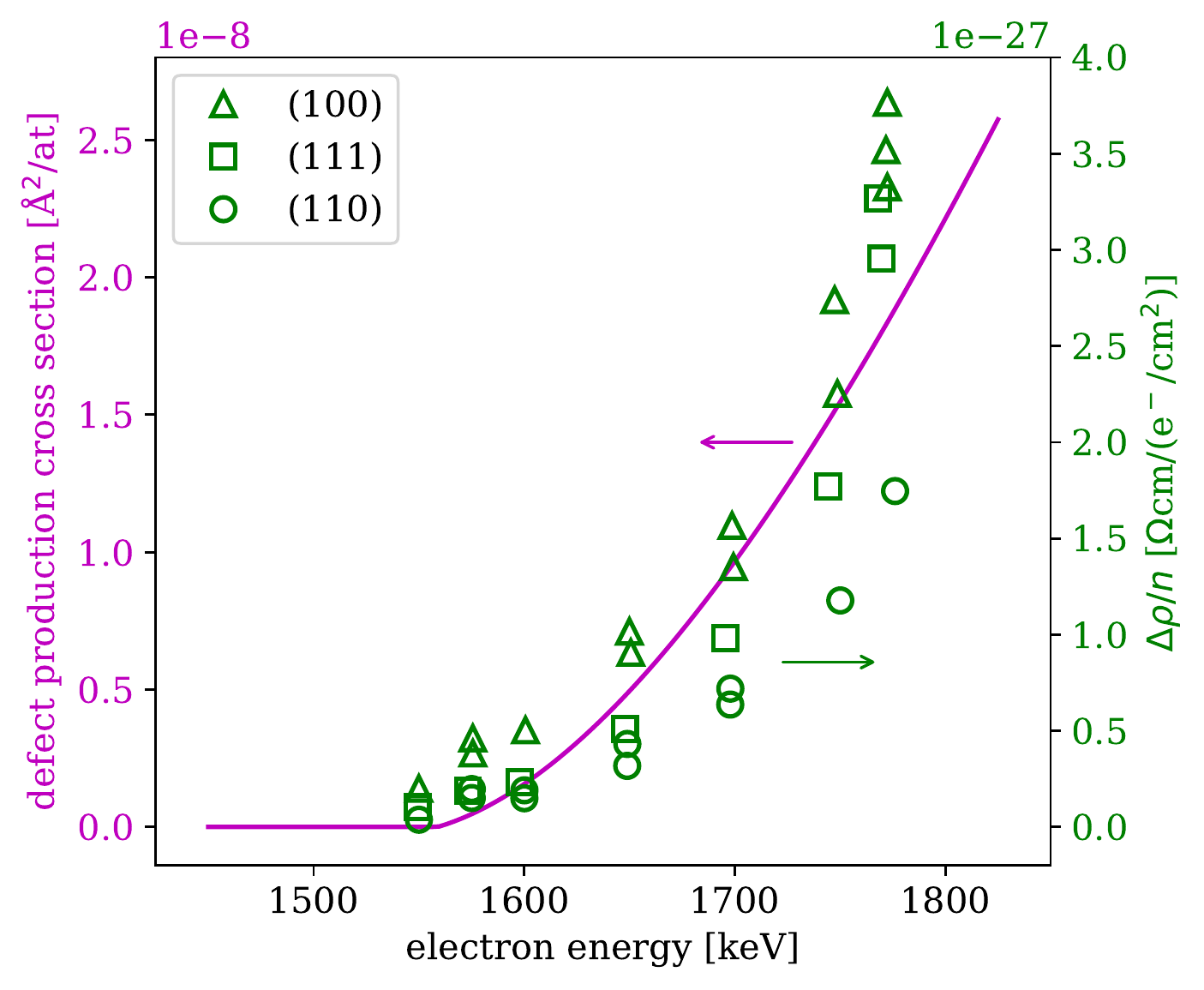}%
\caption{If the maximum recoil energy given by Eq.~\eqref{E_R_max} exceeds the threshold displacement energy, electrons start creating Frenkel pairs that in turn notably increase the electric resistivity of the material \cite{Maury1978}. In this figure, we compare the solid line, providing an estimate for the effective cross section of production of Frenkel pairs in W, see text for detail, with experimental results by Maury {\it et al.} \cite{Maury1978}.}
\label{fig:FP_rho}
\end{figure}

Maury {\it et al.} \cite{Maury1978} measured the variation of electrical resistivity due to the formation of Frenkel pairs in electron-irradiated W at low cryogenic temperatures below 7~K. Assuming that the resistivity increase is proportional to the concentration of Frenkel pairs, in Fig.~\ref{fig:FP_rho} we compare the data by Maury {\it et al.} with the results derived from Eq.~\eqref{eq:sigmaFP}, finding good agreement for the onset of damage \textemdash\ corresponding in our model to a change of slope at $\sim1560$~keV \textemdash\ as well as for the shape of the curve with respect to the experimental points.

\section{Molecular dynamics simulations}\label{sec:MD}
\begin{figure}[t]
\includegraphics[width=\columnwidth]{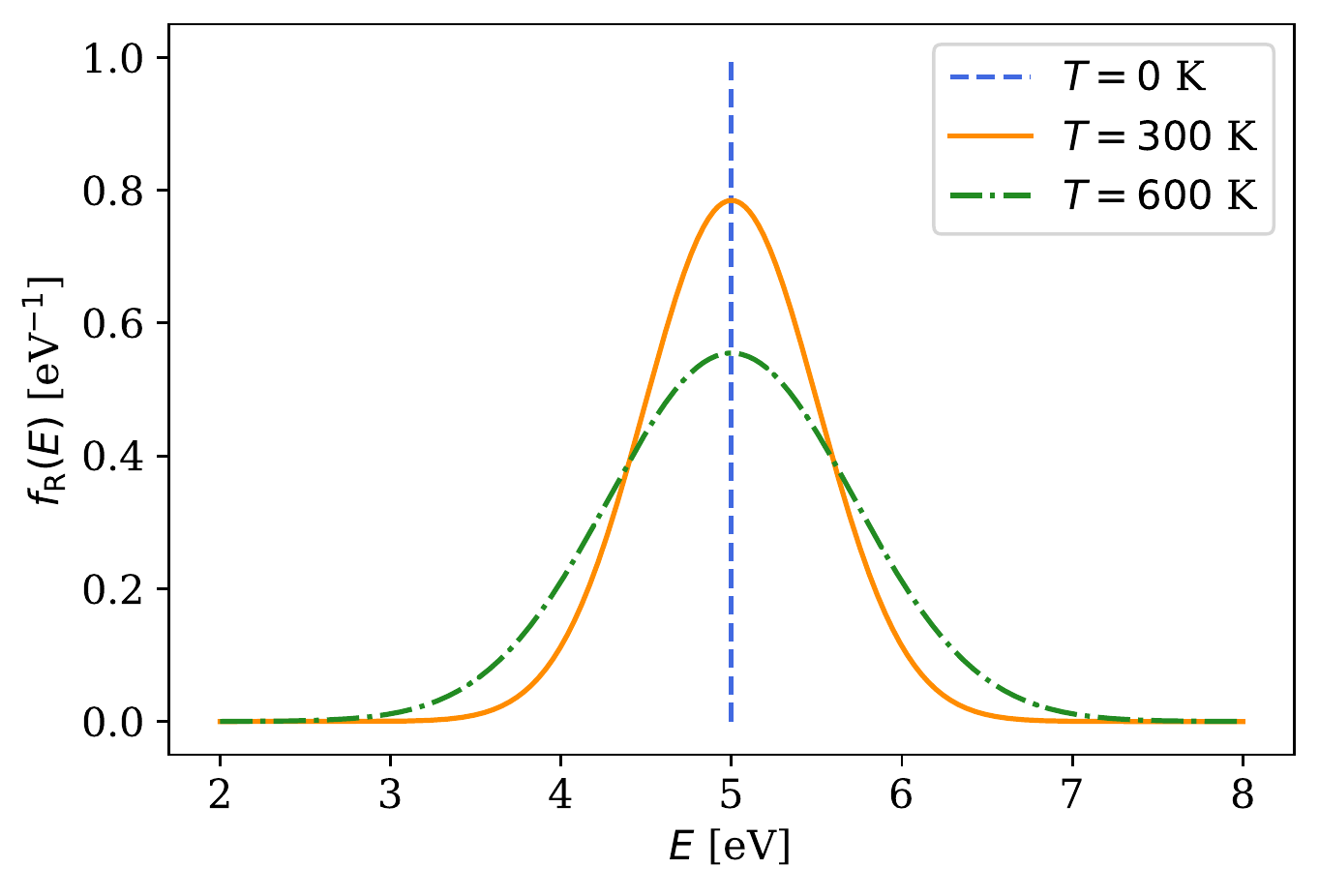}%
\caption{Probability distribution for the energy of recoils $x$ computed from Eq.~\eqref{eq:Gauss_kTEr} assuming that electrons collide with an atom undergoing thermal vibrations at a lattice site. The recoil energy transferred in a collision would be exactly equal to $E_{\textnormal{R}}=5$~eV if the atom were at rest. Even if the thermal energy is as low as about 0.05~eV, the broadening of the recoil spectrum is close to 2~eV.}
\label{fig:Doppler}
\end{figure}

Molecular dynamics simulations of W and Fe were performed using LAMMPS \cite{Plimpton1995} and empirical potentials developed by Mason {\it et al.} \cite{Mason2017} and Gordon {\it et al.} \cite{Gordon2011}, respectively. A single vacancy was created at the centre of a bcc simulation cell containing 1024 atoms fully relaxed under periodic boundary conditions. The structure was thermalised at a target temperature using a Langevin thermostat with the damping constant of 15.7~ps for W and 0.84~ps for Fe \cite{Mason2015}. Electron collisions were simulated by adding a randomly oriented momentum vector $\mathbf{P}$, corresponding to a given amount of energy transferred to the atom by an electron, to the instantaneous thermal momentum of the atom. The magnitude of $P$ was chosen in the way that the energy that the struck atom would have gained in the absence of thermal motion was
$$
E_{\textnormal{R}} = \frac{P^2}{2M}.
$$
In what follows, we shall demonstrate how thermal vibrations of atoms influence the spectrum of recoil energies caused by high-energy electrons. To see why thermal motion is relevant, let us find the distribution of recoil energies $f_{\textnormal{R}}(E)$ of atoms that, if at rest, would recoil with $E_{\textnormal{R}}$. By adding the transferred momentum to the thermal momentum, see Refs. \cite{Boersch1967,Fujikawa2006} and Appendix~\ref{app:Doppler} for detail, we find that the distribution of recoil energies is well approximated by a Gaussian centred at $E_{\textnormal{R}}$ and with variance ${\sigma^2=2(k_{\textnormal{B}} T) E_{\textnormal{R}}}$, namely
\begin{equation}\label{eq:Gauss_kTEr}
    f_{\textnormal{R}}(E)=\frac{1}{\sqrt{4\pi k_{\textnormal{B}}TE_{\textnormal{R}}}}\exp \left [-\frac{(E-E_{\textnormal{R}})^2}{4 k_{\textnormal{B}}TE_{\textnormal{R}}}\right].
\end{equation}
In the above equation, $k_{\textnormal{B}}$ is the Boltzmann constant. This expression is valid in the limit $k_{\textnormal{B}} T \ll E_{\textnormal{R}}$, applicable to the treatment of collisions of atoms with high-energy electrons considered here. We note that although the thermal energy is very small compared to the recoil energy, it has a major effect on the shape of spectrum of atomic recoils \cite{Boersch1967,Fujikawa2006}. This remarkable manifestation of the Doppler effect is illustrated in Fig.~\ref{fig:Doppler}.

Simulations were carried out for temperatures $T=0$, 300, 600, and 900 K in W, and 0, 150, 300, and 450 K in Fe. In relation to Fe, we note that two other potentials, by Mendelev {\it et al.} \cite{Mendelev2003} and Malerba {\it et al.} \cite{Malerba2010}, frequently produce a spurious ``split vacancy'' defect in near-threshold collisions, see Appendix~\ref{app:Fe_pot}. This configuration was very rarely found in the simulations performed using the potential by Gordon {\it et al.} \cite{Gordon2011}.

We note that although the treatment of collisions with $\sim$~MeV electrons requires using relativistic mechanics \cite{ClassicalTheoryOfFields}, the velocities of atoms receiving recoils are in the non-relativistic $10^3$ m/s range. 50,000 collisions were simulated at each recoil energy over the energy interval extending up to 10~eV. Depending on the relative orientation of thermal and transferred momenta, the energy of the recoil atom after a collision can be higher or lower than $E_{\textnormal{R}}$, as shown in Fig.~\ref{fig:Doppler}. This is significant, as an electron may collide with an atom moving thermally towards a vacancy, away from it, or at an angle. The total duration of the simulation less than 1~\textmu s is such that pure thermally activated hops have a negligible probability. Even at 900~K, a vacancy hops in tungsten on the timescale of milliseconds, and hence MD simulations performed in this study describe only the mode of diffusion resulting from atomic recoils generated by high-energy electron impacts.

Since collisions with electrons delivering recoil energies in the eV range are quite rare, we simulated them starting each time from a different thermalised configuration. A computational loop was set up as follows: (i) a simulation was run at a constant temperature for 0.310~ps for W and for 0.220~ps for Fe; (ii) the resulting atomic configuration was saved at the end of each run; (iii) a randomly oriented momentum $\mathbf{P}$ was added to the thermal momentum of an atom in the neighbourhood of a vacancy; (iv) MD simulation was run for 1~ns to give the vacancy a chance to hop to a neighbouring site; (v) simulation was restarted from a configuration saved at step (ii). The time between samplings of  initial thermalised configurations was about 12 times the inverse Debye frequency ${\nu_{\textnormal{D}}=k_{\textnormal{B}} T_{\textnormal{D}}/\hbar}$, where $T_{\textnormal{D}}$ is the Debye temperature, equal to 310~K for W and 410~K for Fe \cite{Ashcroft1976}, giving $\nu_{\textnormal{D}}=4.06\times 10^{13}$~s$^{-1}$ and $\nu_{\textnormal{D}}=5.37\times 10^{13}$~s$^{-1}$, respectively. The sampling frequency ensured that a broad spectrum of initial conditions was explored. As expected, we found that the $x$, $y$ and $z$ components of initial velocities of atoms followed three independent Maxwell-Boltzmann distributions corresponding to the thermalisation temperature.

\begin{figure}[t]
\subfloat[]{%
  {\includegraphics[width=\columnwidth]{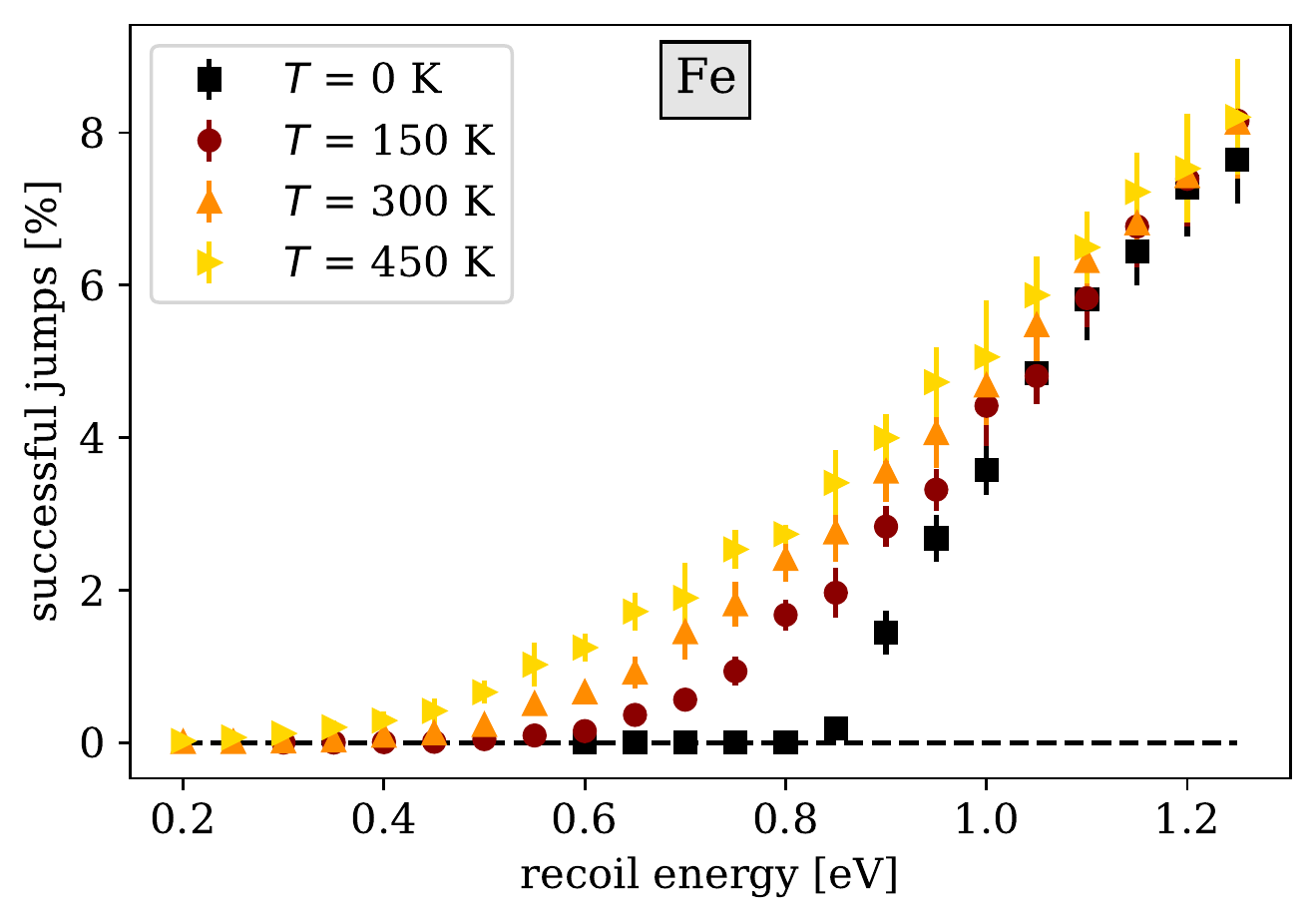}}%
}\hfill
\subfloat[]{%
  {\includegraphics[width=\columnwidth]{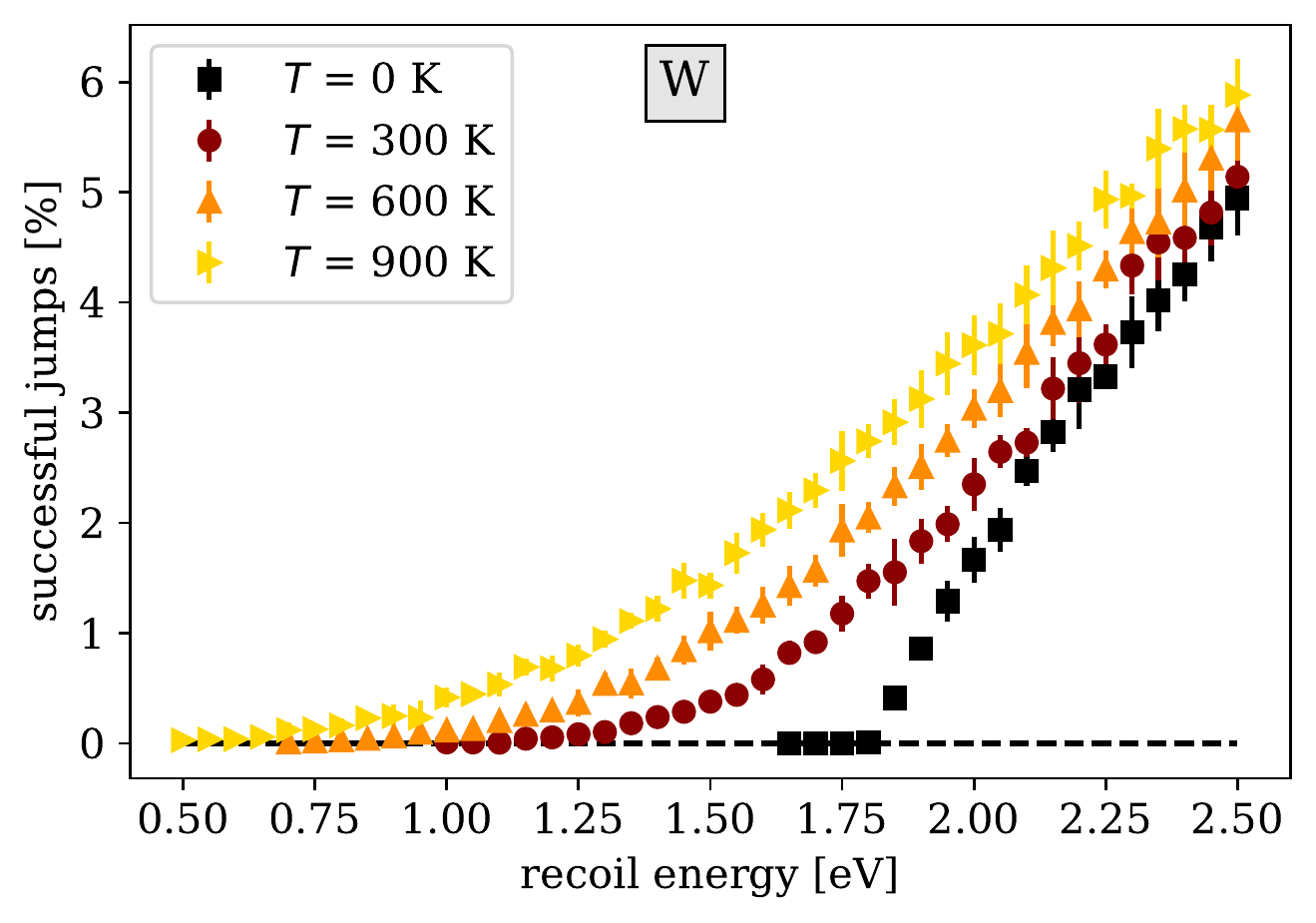}}%
}\hfill
\caption{Fraction of atomic recoils resulting in a vacancy hop increases with the amount of energy transferred in a collision, as illustrated  by the plots for Fe (a) and W (b). At 0 K, there is a sharp threshold for the reaction. At a finite temperature, thresholds are smooth and vacancy hops occur even at very low recoil energies.}
\label{fig:jump_frequency}
\end{figure}

By recording whether the vacancy had moved after each kick, we found the fraction of successful hops as a function of recoil energy at different temperatures, shown in Fig.~\ref{fig:jump_frequency}. A reader will appreciate that, owing to the Doppler effect of Eq.~\eqref{eq:Gauss_kTEr}, vacancies in Fe or W at a temperature of about 15-25\% of the melting point have a finite probability of hopping even when the neighbouring atom recoils with up to about half of the minimum energy that is required to initiate a vacancy hop at 0~K. However, the effect of temperature is no longer significant if the recoil energy is above approximately twice this minimum energy.

\section{Vacancy diffusion driven by direct impacts of high-energy electrons on atoms} \label{sec:D_el}
We now evaluate the effect of high-energy electron collisions with atoms in a material on vacancy diffusion. We start by evaluating the contribution to diffusion from direct impacts of electrons on atoms. Although, as we show below, this direct contribution to diffusion is rarely dominant, the approach itself is generic, and we subsequently extend it to treat cases where the electron-stimulated contribution to the mobility of defects is many orders of magnitude greater than the effect of thermal activation.

In a pure material, the part of the vacancy diffusion coefficient $D$ associated with thermal activation equals \cite{Allnatt1993,Ma2019,diffusion} 
\begin{equation}\label{eq:th_D}
    D_{\textnormal{th}}= \frac{N_{\textnormal{1NN}}}{6}d^2\nu_{\textnormal{D}}\exp\left[-\frac{E_{\textnormal{a}}}{k_{\textnormal{B}}T}\right],
\end{equation}
where $N_{\textnormal{1NN}}$ is the number of first nearest neighbour (1NN) atoms, $d$ is the vacancy hopping distance (if $a$ is the lattice parameter then $d=a\sqrt{3}/2$ for bcc and $d=a\sqrt{2}/2$ for fcc crystals), $\nu_{\textnormal{D}}$ is taken as the thermal attempt frequency, and $E_{\textnormal{a}}$ is the vacancy migration energy.


In the presence of impurities, the thermal part of the diffusion coefficient can be estimated following Refs.~\cite{Fu2008, Liu2014}. For example, in Fe and W containing carbon at a concentration exceeding the concentration of vacancies, we find  
\begin{equation}\label{eq:D_th_C}
    D_{\textnormal{th}}^{\textnormal{C}}= D_{\textnormal{th}}\frac{[V]}{[V_{\textnormal{tot}}]},
\end{equation}
where $[V]$ is the vacancy concentration and ${[V_{\textnormal{tot}}]=[V]+[VC]+[VC_2]+...}$ is the total concentration of isolated vacancies and carbon-vacancy clusters VC$_n$. A single vacancy in Fe and W can trap between one and four C atoms, with the VC$_2$ cluster being the most stable \cite{Fu2008,Kabir2010,Liu2014}. The part played by other impurities like nitrogen and oxygen can be as significant as that of carbon \cite{Schuler2015}. Using the binding energies $E_n^{\textnormal{b}}$ calculated using density functional theory for Fe \cite{Fu2008} and for W \cite{Liu2014} we find 
$$
[VC_n]=[V][C]^n\exp\left(\frac{E_n^{\textnormal{b}}}{k_{\textnormal{B}}T}\right)
$$
resulting in
\begin{equation}
    D_{\textnormal{th}}^{\textnormal{C}}= \frac{D_{\textnormal{th}}}{1+\sum _{n = 1}^4 [C]^n\exp\displaystyle \left(\frac{E_n^{\textnormal{b}}}{k_{\textnormal{B}}T}\right) }.
\end{equation}
In these equations $[C]$ is the carbon concentration in solid solution. In polycrystalline W, some of the carbon atoms segregate to grain boundaries \cite{Murphy2009}, and do not therefore pin the vacancies.

If atoms are bombarded by high-energy electrons, diffusion is accelerated through a variety of processes, involving for example a direct transfer of the kinematic momentum from electrons to atoms. A collision of a high-energy electron with an atom near a vacancy can stimulate an atomic hop or, equivalently, a vacancy hop to an adjacent lattice site. The threshold energy for a hop depends on the direction of the impact with respect to the lattice, and on the distance between the impacted atom and the vacancy. 

To estimate the contribution of direct electron impacts  to vacancy diffusion, we identify the atoms around a vacancy that participate in the hopping events through recoils requiring the least amount of energy to initiate a successful vacancy hop. We simulated, at $T=0$~K, the atomic processes initiated by electron impacts using the interatomic potentials for tungsten by Mason \cite{Mason2017} and Marinica \cite{Marinica2013}, and the interatomic potential for iron by Gordon \cite{Gordon2011}. We considered initial impacts in the $\langle111\rangle$, $\langle100\rangle$ and $\langle110\rangle$ directions involving the first, second and third nearest neighbours of a vacancy, and also atoms further away from it.

\begin{table}[t]
\caption{Minimum recoil energies resulting in a vacancy jump if the recoil is precisely towards the vacant site. The values, given in eV units, were computed for $T=0$ using two different potentials for W and one for Fe. The atoms receiving the recoils were assumed to be in the first, second, and third NN position in relation to the vacancy. Recoil energy thresholds are visualised for the Mason potential for bcc W \cite{Mason2017}.}
\label{tab:min_E}
\begin{ruledtabular}  
\begin{tabular}{lcccc}
     & & W \cite{Mason2017} & W \cite{Marinica2013} & Fe \cite{Gordon2011} \\\hline
    Direction & distance  & \multicolumn{3}{c}{minimum energy [eV]}  \\\hline
    \multirow{7}{*}{$\langle111\rangle$} & 1$^{\text{st}}$ & 1.80 & 2.23 & 0.83 \\
     & 2$^{\text{nd}}$ & 2.19 & 3.72 & 1.25 \\
     & 3$^{\text{rd}}$ & 3.03 & 5.81 &  1.99 \\
     & 4$^{\text{th}}$ & 3.97 & 8.08 &  2.87 \\
     & 5$^{\text{th}}$ & 4.99 & 10.3 &  3.78 \\
     & 6$^{\text{th}}$ & 6.06 & 12.5 &  4.70 \\
     & 7$^{\text{th}}$ & 7.19 & 14.3 &  5.59 \\\hline
    \multirow{3}{*}{$\langle100\rangle$} & 1$^{\text{st}}$ & 10.1 & 10.1 & 4.76 \\
     & 2$^{\text{nd}}$ & 19.6 & 19.1 & 8.46 \\
     & 3$^{\text{rd}}$ & 30.1 & 28.2 &  12.7 \\\hline
    $\langle110\rangle$ & 1$^{\text{st}}$ & 43.3 & 32.9 & 17.0 \\ \hline
    &&& \\
    \multicolumn{5}{c}{\includegraphics[width=0.35\textwidth]{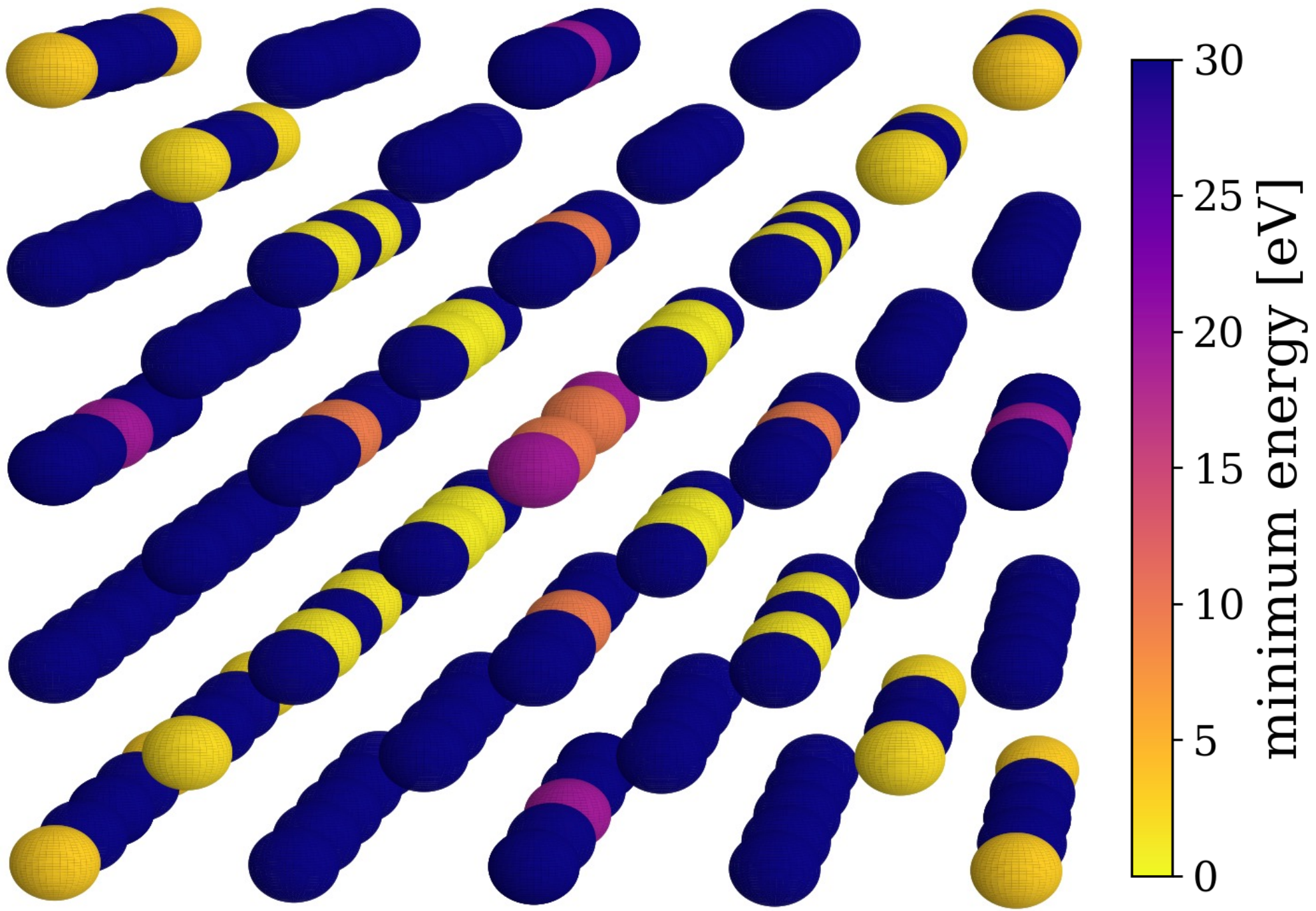}}
  	\end{tabular}
\end{ruledtabular}
\end{table}

Table~\ref{tab:min_E} shows that the contribution to diffusion in W and Fe from direct electron recoils is dominated by the events involving high-energy electrons colliding with atoms situated along the $\langle111\rangle$ crystallographic directions. In bcc metals there are $N_{\langle111\rangle}=8$ atoms nearest to a  vacancy. The electron collision contribution to the vacancy diffusion coefficient is proportional to the weighted  sum of frequencies $\nu_{\langle111\rangle}^{(k)}$ with which one of the atoms at a $k$-th position along a $\langle111\rangle$ atomic string causes a vacancy to perform a hop following an event of interaction with a high-energy electron
\begin{equation}\label{eq:D_el}
    D_{\textnormal{el}}= \frac{N_{\langle111\rangle}}{6}d^2\left(\nu_{\langle111\rangle}^{(1)}+\nu_{\langle111\rangle}^{(2)}+\ldots\right),
\end{equation}
where $d=a\sqrt{3}/2$ is the nearest neighbour distance between sites in bcc lattice. This provides a lower bound for $D_{\textnormal{el}}$ as it neglects the events where the struck atom is not situated along a $\langle111\rangle$ atomic string.

To evaluate $\nu_{\langle111\rangle}^{(k)}$, we note that a vacancy hop to a nearest lattice site involves the following sequence of events:
\begin{enumerate}
    \item an atom near a vacancy is struck by an electron with sufficiently high energy $E_{\textnormal{el}}$ to initiate a recoil in the eV energy range;
    \item the atom recoils in a specific direction with kinetic energy $E_{\textnormal{R}}$ transferred in the collision with the electron;
    \item one of the atoms in the vicinity of a vacancy, not necessarily the atom impacted by the electron, crosses the potential barrier and moves into a stable energy minimum position at the initially vacant lattice site.
\end{enumerate}

Recalling Eq.~\eqref{collision_rate_total}, the \emph{total} frequency of collisions between an atom and electrons depends on the integrated over the solid angle electron flux $\phi_{\textnormal{el}}(E)$, defined by (\ref{isotropic_flux}), and the total cross section of scattering of electrons by an atom $\sigma_{\textnormal{tot}}(E)$,
\begin{equation}
    \nu_{\textnormal{tot}} = \int\textnormal{d}E\phi_{\textnormal{el}}(E)\sigma_{\textnormal{tot}}(E).\label{nu_E_R}
\end{equation}
To account for points 1 and 2 above, we need the frequency \emph{distribution} of atomic recoils as a function of recoil energy $E_{\textnormal{R}}$, resulting from collisions with electrons over the entire spectrum of electron energies. The recoil energy-resolved frequency of impact events for an atom at a lattice site is
\begin{eqnarray}
    \nu (E_{\textnormal{R}})&=&\int \text{d}o \int \text{d}o' \int \text{d}E' {\text{d} \sigma ({\bf n}' \rightarrow {\bf n}) \over \text{d}o} \phi_{\textnormal{el}}({\bf n}', E')\nonumber \\
    &\times &\delta \left\{E_{\textnormal{R}}- {1\over 2}E_{\textnormal{R}}^{\textnormal{max}}(E')[1-\cos ({\bf n}\cdot {\bf n}')]\right\},\label{eq:frequency_recoils_E_Resolved}
\end{eqnarray}
where $E_{\textnormal{R}}^{\textnormal{max}}(E)$ is given by Eq.~\eqref{eq:recoil_energy_maximum}, and $\delta (x)$ is the Dirac delta function. This equation involves only the elastic cross section of scattering of electrons by an atom because inelastic scattering of high-energy electrons primarily contributes to energy losses but not to atomic recoils \cite{LandauQuantumMechanics}.
Integrating (\ref{eq:frequency_recoils_E_Resolved}) over $E_{\textnormal{R}}$, we recover expression (\ref{nu_E_R}). 

If the flux of electrons bombarding the atoms is isotropic, see Eq.~\eqref{isotropic_flux}, in Eq.~\eqref{eq:frequency_recoils_E_Resolved} we integrate over the directions of ${\bf n}'$ and arrive at
\begin{eqnarray}
    \nu (E_{\textnormal{R}})&=&2\pi \int \limits _0 ^{\pi}\text{d}\theta \sin \theta {\text{d} \sigma \over \text{d}\theta} \int \limits _0 ^{\infty} \text{d}E\, \phi_{\textnormal{el}}(E)\nonumber \\
    &\times &\delta \left\{E_{\textnormal{R}}- {1\over 2}E_{\textnormal{R}}^{\textnormal{max}}(E)[1-\cos \theta]\right\}.
    \label{eq:frequency_recoils_E_Resolved1}
\end{eqnarray}

Since the angle of scattering $\theta$ enters Eq.~\eqref{eq:frequency_recoils_E_Resolved1} only as an argument of $\cos \theta$, we change the variable of integration to $\xi =1-\cos \theta$ and write
\begin{eqnarray}
    \nu (E_{\textnormal{R}})&=&2\pi \left(\frac{Ze^2}{4\pi\varepsilon_0 m c^2}\right)^2\int \limits _0 ^{2}\text{d}\xi  \int \limits _0 ^{\infty} \text{d}E\, \phi_{\textnormal{el}}(E)\left(\frac{1-\beta^2}{\beta^4}\right)\nonumber \\
    &\times& \frac{1}{(\kappa +\xi)^2} \delta \left[E_{\textnormal{R}}- {\xi \over 2}E_{\textnormal{R}}^{\textnormal{max}}(E)\right],
    \label{eq:frequency_recoils_E_Resolved2}
\end{eqnarray}
where the cross section of scattering is given by the screened Rutherford expression (\ref{eq:screened_Rutherford}). Integrating over $\xi$, we find
\begin{equation}
\begin{aligned}
    \nu (E_{\textnormal{R}}) = 2\pi \left(\frac{Ze^2}{4\pi\varepsilon_0 m c^2}\right)^2\int \limits _0 ^{\infty} \text{d}E\, \phi_{\textnormal{el}}(E)\left(\frac{1-\beta^2}{\beta^4}\right) \\
    \times {2E^{\textnormal{max}}(E) \over\left[\kappa E_{\textnormal{R}}^{\textnormal{max}}(E)+2E_{\textnormal{R}} \right]^2 }\Theta \left[E_{\textnormal{R}}^{\textnormal{max}}(E)-E_{\textnormal{R}}\right],
    \label{eq:frequency_recoils_E_Resolved3}
\end{aligned}
\end{equation}
where $\Theta (x)$ is the Heaviside function, $\Theta (x)=1$ for $x>0$ and $\Theta (x)=0$ for $x<0$.

For a mono-energetic flux of electrons with kinetic energy ${\cal E}$, the expression for the flux has the form ${\phi_{\textnormal{el}}(E)=\Phi_0\delta (E-{\cal E})}$ and the distribution of the frequency of atomic recoils with respect to the energy of recoils is
\begin{eqnarray}
    \nu (E_{\textnormal{R}})&=&2\pi \left(\frac{Ze^2}{4\pi\varepsilon_0 m c^2}\right)^2 \Phi_{0}{(1+{\cal E}/mc^2)^2\over [(1+{\cal E}/mc^2)^2-1]^2}\nonumber \\
    &\times& {2E_{\textnormal{R}}^{\textnormal{max}}({\cal E}) \over\left[\kappa E_{\textnormal{R}}^{\textnormal{max}}({\cal E})+2E_{\textnormal{R}} \right]^2 },
    \label{eq:frequency_recoils_E_single}
\end{eqnarray}
for $E_{\textnormal{R}}<E_{\textnormal{R}}^{\textnormal{max}}({\cal E})$. There are no recoils with energies higher than $E_{\textnormal{R}}^{\textnormal{max}}({\cal E})=2{\cal E}({\cal E}+2mc^2)/Mc^2$.
\begin{figure}[t]
  \includegraphics[width=0.45\textwidth]{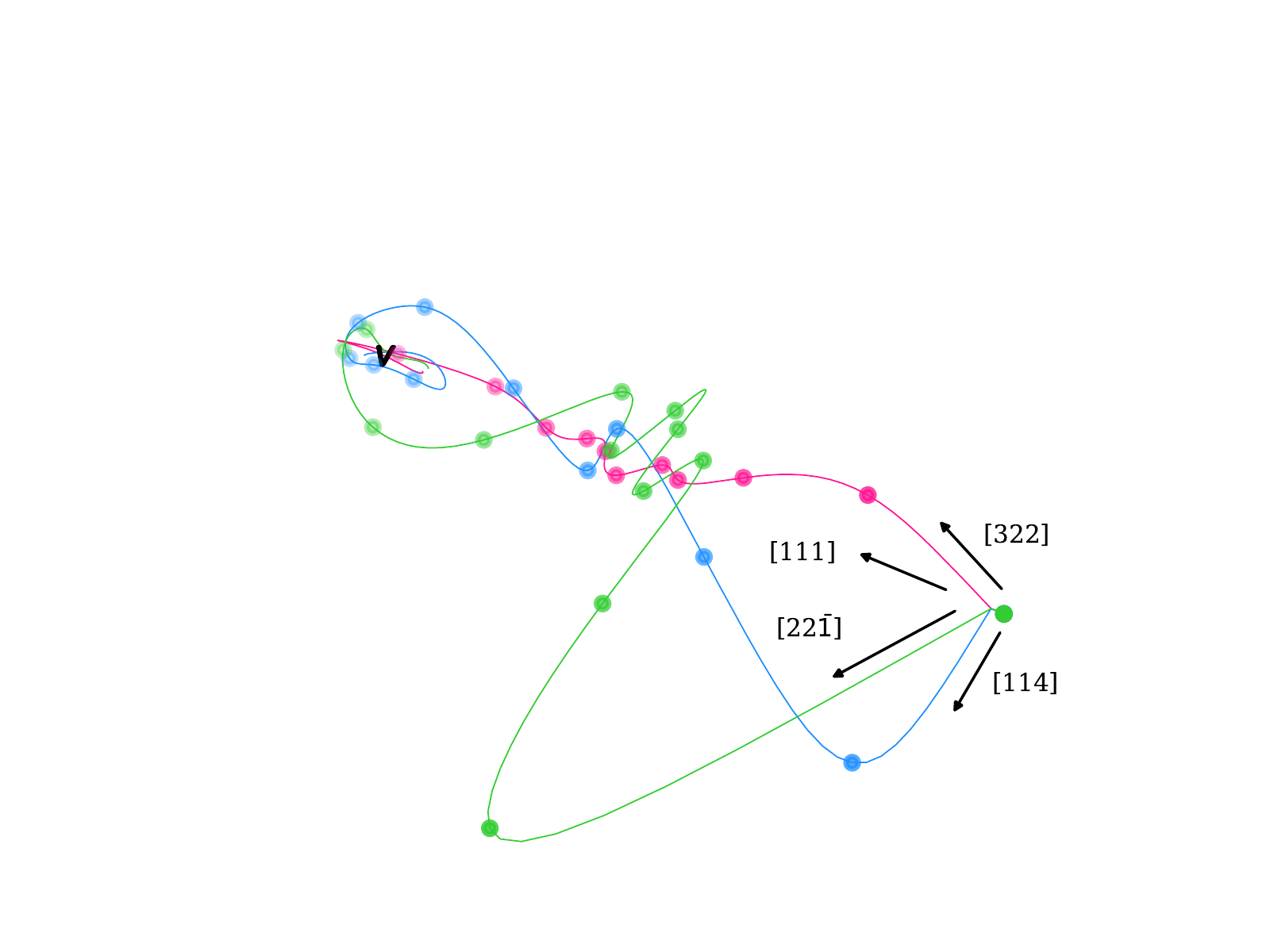}%
    \caption{Representative examples of  trajectories found in events involving collisions of electrons with atoms in a 1NN shell of a vacancy in W. The [322], $[1 1 4]$ and $[22\bar{1}]$ trajectories start at about 11$^\circ$, 35$^\circ$ and 55$^\circ$ off the $[111]$ crystallographic direction, respectively, and require recoil energies of 2, 5 and 9 eV to enable the atom to successfully complete a hop into a vacant site (marked as $V$).}
    \label{fig:trajectories}
\end{figure}

\begin{figure*}[t]
\subfloat[]{%
  \includegraphics[width=0.49\textwidth]{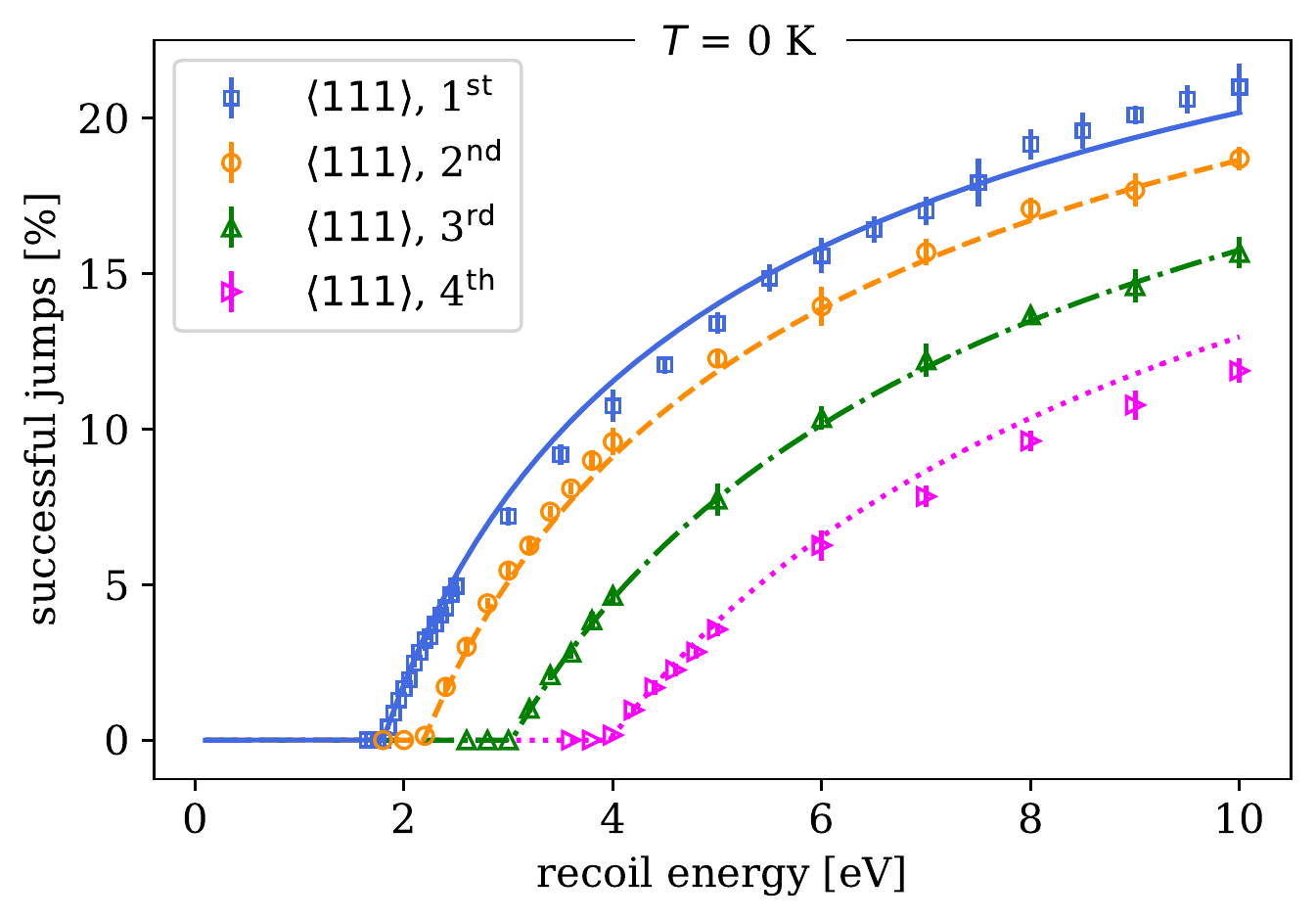}%
}\hfill
\subfloat[]{%
  \includegraphics[width=0.49\textwidth]{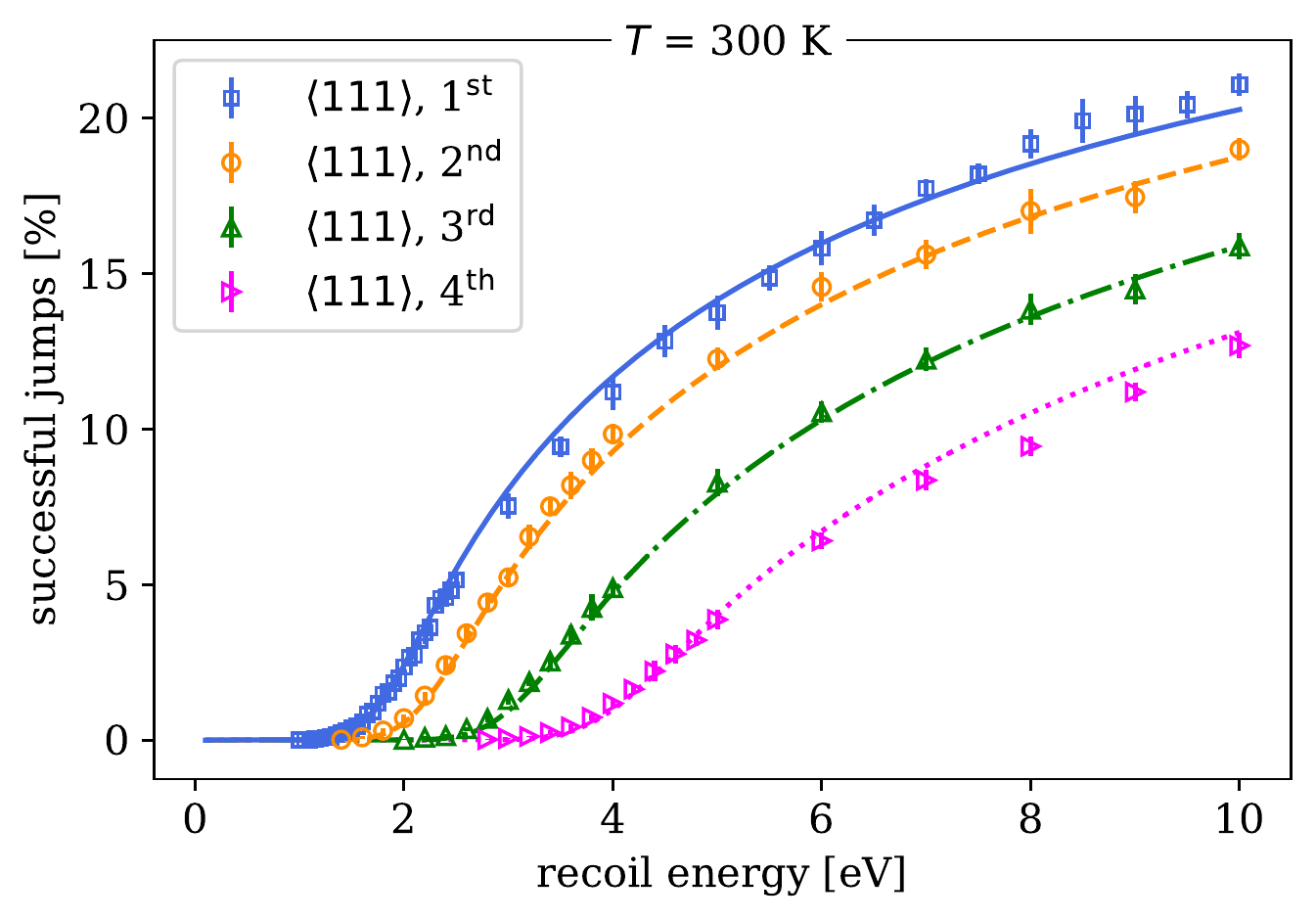}%
}\hfill
\subfloat[]{%
  \includegraphics[width=0.49\textwidth]{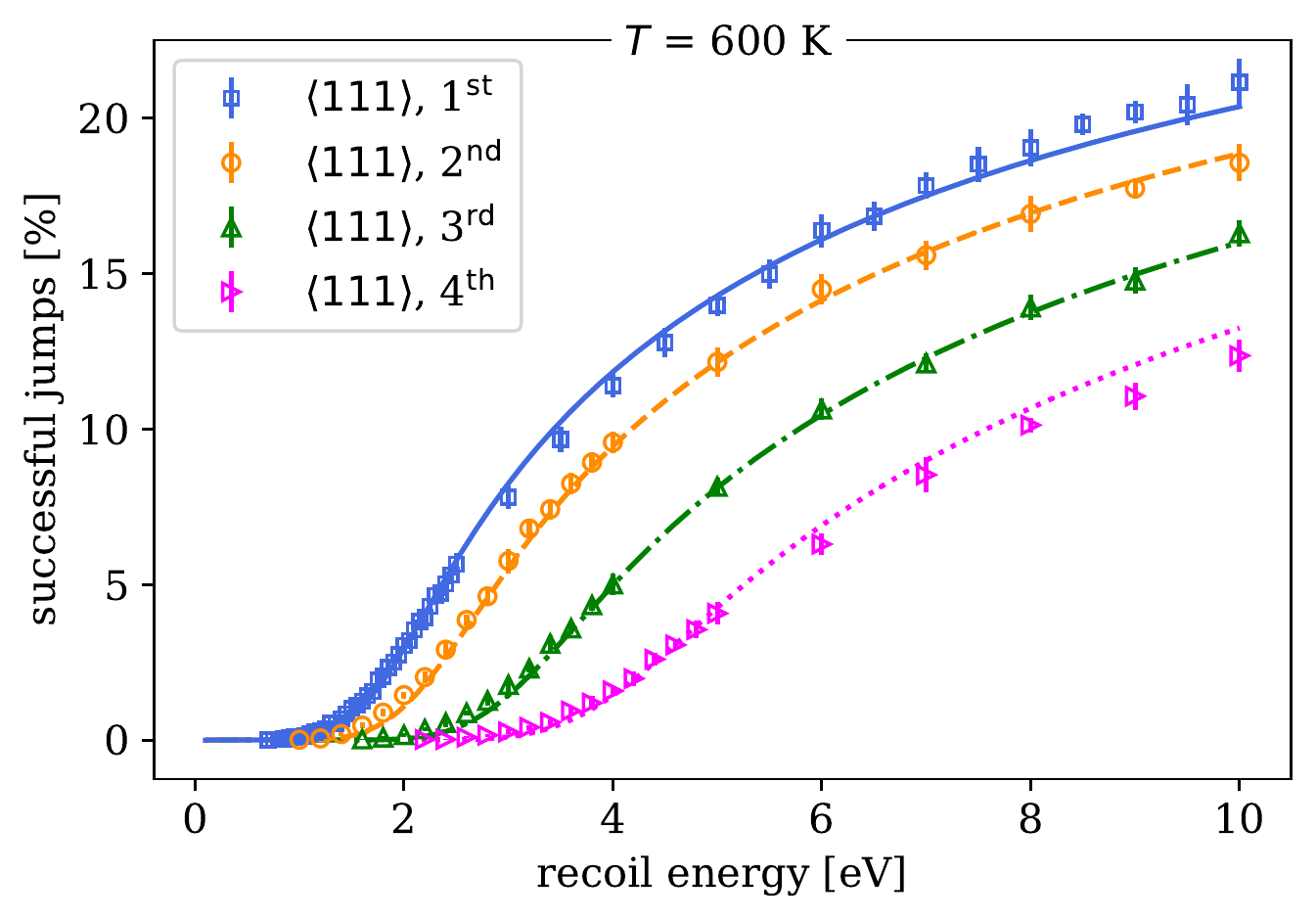}%
}\hfill
\subfloat[]{%
  \includegraphics[width=0.49\textwidth]{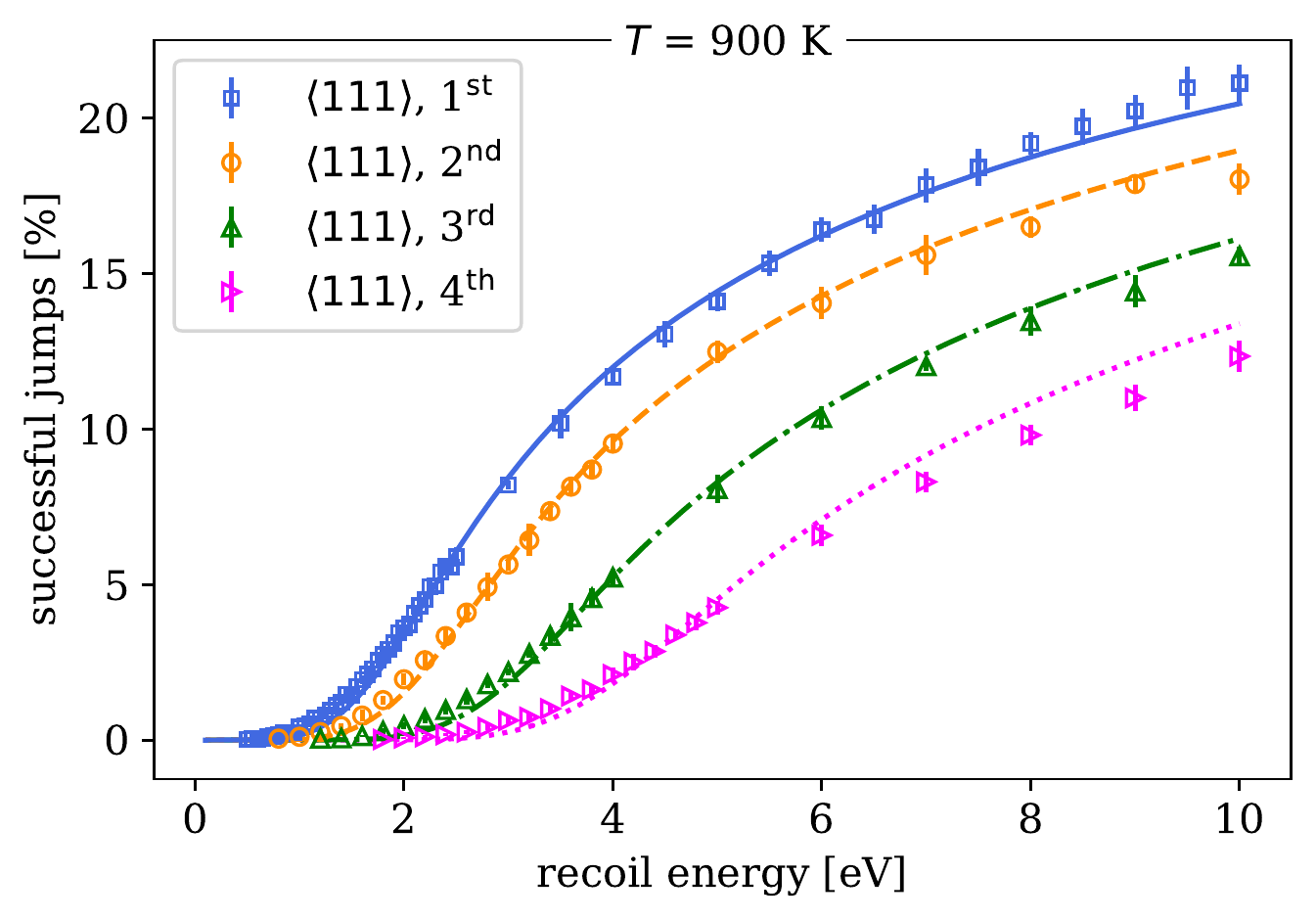}%
}\hfill
    \caption{Fraction of successful vacancy hops in W for 0~K (a), 300~K (b), 600~K (c), 900~K (d) if the kicked atom is one of the 1$^{\text{st}}$, 2$^{\text{nd}}$, 3$^{\text{rd}}$ or 4$^{\text{th}}$ atoms along the $\langle 111\rangle$ directions. Discrete data points, with error bars indicating one standard deviation, are derived from MD simulations (potential \cite{Mason2017}), whereas the fitting curves are given by Eq.~\eqref{eq:J_E_R}. Fitting all the data simultaneously yielded a single pair $\alpha=0.3504$, $k_{\textnormal{B}}T_0=2.038$~eV (with values of $E_{\textnormal{a}}^0$ in Eq.~\eqref{eq:E_a} from Table~\ref{tab:min_E}). Eq.~\eqref{eq:J_E_R} fits all the data points with $R^2=99.7$~\%.}
    \label{fig:jump_prob_W_fit} 
\end{figure*}

The same result can be obtained from the energy-differential cross sections, Eq.~\eqref{eq:dS_dEr_simplified} for the screened Rutherford interaction and Eq.~\eqref{eq:dS_dEr_simplified_Mott} for the Mott scattering, resulting in
\begin{equation}\label{eq:nu_E}
    \nu(E_{\textnormal{R}}) = \int \limits_0^\infty \text{d}E_{\textnormal{el}}\phi_{\textnormal{el}}(E_{\textnormal{el}})\frac{\text{d}\sigma}{\text{d} E_{\textnormal{R}}}.
\end{equation}
In the above equations, $\nu(E_{\textnormal{R}})$ is expressed in units s$^{-1}$eV$^{-1}$ whereas the flux of electrons is expressed in the  cm$^{-2}$s$^{-1}$eV$^{-1}$ units. 

We now address point 3 above by quantifying the probability of the vacancy hopping to a neighbouring lattice site if a collision with a high-energy electron transfers energy $E_{\textnormal{R}}$ to one of the neighbouring atoms. This probability, which in what follows we refer to as $J(E_{\textnormal{R}}, T)$, depends on the recoil energy and temperature. 

Since the spectrum of recoil energies given by Eq.~\eqref{eq:frequency_recoils_E_single} is broad, vacancy hopping events are expected to be dominated by the electron impacts generating recoil energies comparable or greater than the vacancy migration energy. This energy scale is many times the energy of thermal motion of atoms, and hence the statistics of transitions stimulated by electron impacts are expected to be different from the statistics of thermally activated events, detailed in Ref. \cite{Allnatt1993}.

The treatment of thermally activated transitions involves computing the rate of a many-body atomic system crossing a potential barrier associated with a reaction where an atom hops to a neighbouring vacant site \cite{Vineyard1957,Landauer1961,Allnatt1993}. For thermally activated processes, this rate is exponentially small, and the thermal motion of atoms at equilibrium positions is assumed to be unaffected by the occurrence of transitions. 

A typical recoil event stimulated by a high-energy electron collision with an atom has the energy well above the thermal energy but still below the threshold for the formation of a Frenkel pair. The motion of an atom receiving the recoil and its neighbours remains confined to the vicinity of equilibrium positions, although the amplitude of motion is significantly larger than the amplitude of thermal vibrations. The initial energy of the recoil dissipates by phonon radiative transfer. Occasionally, on the timescale of several Debye oscillations, this energy gets partially projected onto the trajectory of a many-body reaction involving a vacancy hopping to a neighbouring lattice site.

Providing an accurate quantitative estimate of $J(E_{\textnormal{R}},T)$ is difficult because of the atomic many-body character of recoil-stimulated vacancy hopping events. Trajectories of MD simulations shown in Fig.~\ref{fig:trajectories} suggest that the initial direction of an atomic recoil does not have a strong effect on the outcome of a sequence of local interactions between the atoms that eventually result in vacancy migration. Even the recoils at a large angle to the direction of a vacancy hop appear to contribute significantly to the vacancy hopping rate. 

At $T=0$ K, function $J(E_{\textnormal{R}}, 0)$ vanishes for all the recoil energies $E_{\textnormal{R}}<E_{\textnormal{a}}$, where $E_{\textnormal{a}}$ is the minimum recoil energy required to initiate a vacancy migration. At a finite temperature, $J(E_{\textnormal{R}}, T)$ is a monotonically increasing function of $E_{\textnormal{R}}$, likely saturating at high recoil energies $E_{\textnormal{R}}\gg E_{\textnormal{a}}$, provided that $E_{\textnormal{R}}$ is still well below the Frenkel pair threshold production energy. MD simulations performed in this study and in Ref. \cite{Satoh2017} suggest that at high recoil energies function $J(E_{\textnormal{R}}, T)$ approaches a limit close to 0.27. Notably, this value is higher than the $1/8=0.125$ fraction of the solid angle spanned by the directions from an atom in a corner of a cubic cell towards a vacancy in its centre.   

Temperature effects are expected to assist the recoil-stimulated vacancy diffusion. Thermal velocity adds to the velocity of an atom derived from a recoil event if both vectors are aligned towards a vacancy. Also, the effective free-energy migration barrier is lowered by the vibrations of atoms though the entropy effect \cite{Gilbert2013}. A treatment similar to a transition rate theory approach, see Appendix~\ref{app:prob} for detail, suggests a suitable functional form for $J(E_{\textnormal{R}}, T)$
\begin{align}\label{eq:J_E_R}
    J(E_{\textnormal{R}}, T)&=\alpha\Bigg(1+\frac{1}{2}\left(1-\sqrt{\frac{E_{\textnormal{a}}}{E_{\textnormal{R}}}}\right)\text{erf}\left[\frac{\sqrt{E_{\textnormal{R}}}-\sqrt{E_{\textnormal{a}}}}{\sqrt{k_{\textnormal{B}}T}}\right] \nonumber \\
    &-\frac{1}{2}\left(1+\sqrt{\frac{E_{\textnormal{a}}}{E_{\textnormal{R}}}}\right)\text{erf}\left[\frac{\sqrt{E_{\textnormal{R}}}+\sqrt{E_{\textnormal{a}}}}{\sqrt{k_{\textnormal{B}}T}}\right]  \\
    &+\sqrt{\frac{k_{\textnormal{B}}T}{\pi E_{\textnormal{R}}}}\text{sinh}\left[\frac{2\sqrt{E_{\textnormal{R}}E_{\textnormal{a}}}}{k_{\textnormal{B}}T}\right]\exp\left[-\frac{E_{\textnormal{a}}+E_{\textnormal{R}}}{k_{\textnormal{B}}T}\right]\Bigg),\nonumber
\end{align}
where $\alpha$ is a constant that can be determined from MD simulations. 
Eq.~\eqref{eq:J_E_R} was derived assuming that the activation energy is the same as in Eq.~\eqref{eq:th_D}. To account for the entropy term \cite{Gilbert2013}, we treat $E_{\textnormal{a}}$ as a weakly temperature dependent quantity and define
\begin{equation}\label{eq:E_a}
E_{\textnormal{a}} (T) = E_{\textnormal{a}}^0\left(1-\frac{T}{T_0}\right),   
\end{equation}
where the constants are material-dependent with ${T_0\sim 10^4}$~K. $E_{\textnormal{a}}^0$ is the minimum energy required for an atom to hop into a vacant lattice site and it can be derived from a single MD simulation for each of the atoms in a $\langle111\rangle$ atomic string, see Table~\ref{tab:min_E}. As for the choice of parameters $\alpha$ and $T_0$, we found that a single set of two constants is sufficient to fit all of the data derived from extensive MD simulations, involving randomly kicking of atoms in the vicinity of a vacancy, from one to four nearest neighbour distances in a $\langle111\rangle$ direction. The MD data and the fits based on analytical formula \eqref{eq:J_E_R} are shown in Fig.~\ref{fig:jump_prob_W_fit}.

For a given electron flux and the resulting distribution of recoil energies, the hopping success rate is given by the product of Eq. \eqref{eq:nu_E} and Eq.~\eqref{eq:J_E_R}. Since both the electron and recoil energies are given by continuous probability distributions, the total success rate $\nu_{\langle111\rangle}^{(k)}$ is obtained by integrating over all the electron and recoil energies, yielding
\begin{equation}\label{eq:rate_success_continuous}
    \nu_{\langle111\rangle}^{(k)}(T)=\int \limits_0^\infty \text{d}E_{\textnormal{el}} \int \limits_0^\infty \text{d}E_{\textnormal{R}} \phi_{\textnormal{el}}(E_{\textnormal{el}})\frac{\text{d}\sigma}{\text{d} E_{\textnormal{R}}} J_k(E_{\textnormal{R}}, T).
\end{equation}
Numerically, the electron flux is represented by a set of discrete values $\Phi_{\textnormal{el}}^{(i)}$ corresponding to discrete electron energies $E_i$, and thus the electron flux in the above equation can be written as 
\begin{equation}
    \phi_{\textnormal{el}}(E_{\textnormal{el}}) = \sum_{n=1}^N \Phi_{\textnormal{el}}^{(i)}\delta(E_{\textnormal{el}}-E_i).
\end{equation}
Substituting this in Eq.~\eqref{eq:rate_success_continuous} we arrive at 
\begin{equation}\label{eq:rate_success}
    \nu_{\langle111\rangle}^{(k)}(T)=\sum_{i=1}^N \Phi_{\textnormal{el}}^{(i)}\int \limits_0^\infty \frac{\text{d}\sigma}{\text{d} E_{\textnormal{R}}}J_k(E_{\textnormal{R}}, T)\text{d}E_{\textnormal{R}}.
\end{equation}
The $k$-index in $J_k(E_{\textnormal{R}}, T)$ indicates that the corresponding energy threshold refers to the $k$-th atom along a $\langle111\rangle$ direction. Inserting Eq.~\eqref{eq:rate_success} into Eq.~\eqref{eq:D_el} enables calculating $D_{\textnormal{el}}$, the high-energy electron recoil contribution to the diffusion of vacancies in a material exposed to neutron irradiation. 
The barriers entering the formulae, evaluated for pure Fe and W, are listed in Table ~\ref{tab:min_E}.

As it will be shown below, Eq.~\eqref{eq:rate_success} is weakly dependent on the temperature of the material. This enables deriving a simple analytical formula for the coefficient of driven diffusion. Taking the limit $\lim_{T\to0} J(E_{\textnormal{R}},T)$ in Eq.~\eqref{eq:J_E_R} we arrive at 
\begin{equation}\label{eq:J_T_indep}
    J(E_{\textnormal{R}}, 0)=\alpha\left(1-\sqrt{\frac{E_{\textnormal{a}}}{E_{\textnormal{R}}}}\right)\Theta[E_{\textnormal{R}}-E_{\textnormal{a}}],
\end{equation}
where Eq.~\eqref{eq:J_T_indep} sets a lower bound on $J(E_{\textnormal{R}},T)$. Setting $\kappa=0$ in Eq.~\eqref{eq:screened_Rutherford}, we evaluate the integral in Eq.~\eqref{eq:rate_success}, which in combination with Eq.~\eqref{eq:D_el} results in 
\begin{align}\label{eq:D_el_simplified}
    \widetilde{D}_{\textnormal{el}}=&\frac{\pi\alpha}{3}\left(\frac{Ze^2}{4\pi\varepsilon_0 m c^2}\right)^2 \Phi_{0}a^2\left(\frac{1-\beta^2}{\beta^4}\right)\times \nonumber\\ &\sum_{k=1}^{N}\Bigg[\frac{E_{\textnormal{R}}^{\textnormal{max}}}{E_{\textnormal{a}}^{(k)}}+2\sqrt{\frac{E_{\textnormal{a}}^{(k)}}{E_{\textnormal{R}}^{\textnormal{max}}}}-3\Bigg]
\end{align}
if $E_{\textnormal{R}}^{\textnormal{max}}>E_{\textnormal{a}}^{(k)}$, whereas $\widetilde{D}_{\textnormal{el}}=0$ otherwise.

Considering a 2000~keV mono-energetic beam of electrons typical of a TEM illumination, where ${\Phi_{\textnormal{el}}=2\times10^{21}}$~cm$^{-2}$s$^{-1}$ and $T=300$~K, from (\ref{eq:rate_success}) we find that ${D_{\textnormal{el}}=2.22\times 10^{-15}}$~cm$^{2}$/s if using the Mott scattering cross-section and ${1.52\times 10^{-15}}$~cm$^{2}$/s if using the Rutherford cross-section, taking the lowest 4 activation energies from potential \cite{Mason2017}. Eq.~\eqref{eq:D_el_simplified} gives the close value of ${1.47\times 10^{-15}}$~cm$^{2}$/s.

A qualitatively similar result can be derived from the treatment developed by Kiritani \cite{Kiritani1976}, who investigated the low temperature limit and included a contribution from high-energy recoils (HER). Kiritani's expression for the diffusion coefficient, derived in Appendix~\ref{app:high_E_rec}, is
\begin{align}\label{eq:D_el_HER}
    \widetilde{D}_{\textnormal{el}}^{\textnormal{HER}}=&\pi\left(\frac{Ze^2}{4\pi\varepsilon_0 m c^2}\right)^2 \Phi_{0}a^2\left(\frac{1-\beta^2}{\beta^4}\right)\times \nonumber\\ &\frac{E_{\textnormal{R}}^\textnormal{max}}{E_\textnormal{a}}\log\left(\frac{E_{\textnormal{R}}^\textnormal{max}}{E_\textnormal{a}}\right)
\end{align}
if $E_{\textnormal{R}}^{\textnormal{max}}>E_{\textnormal{a}}$, and zero otherwise. The numerical values estimated from formula (\ref{eq:D_el_HER}) are about an order of magnitude higher than those derived from (\ref{eq:D_el_simplified}), e.g. ${1.77\times 10^{-14}}$~cm$^{2}$/s for the above TEM case. From these estimates, at very low temperatures vacancies driven by electron impacts in a TEM are as mobile as they would be in pure tungsten at 680-720~K. This is consistent with experimental observations showing high mobility of vacancies in tungsten examined {\it in situ} in an electron microscope  \cite{Arakawa2020}.

Under the DEMO and HFR neutron irradiation conditions, the flux of electrons, integrated over the spectrum of energies, is lower, of order 1-$4\times10^{12}$~cm$^{-2}$s$^{-1}$, resulting in the estimated coefficients of driven diffusion of order 
$5\times 10^{-24}$ cm$^2$/s for the DEMO case and $2 \times 10^{-23}$ cm$^2$/s for the HFR case, based on equation (\ref{eq:D_el_simplified}). Equation (\ref{eq:D_el_HER}) gives estimates of $4\times 10^{-23}$ cm$^2$/s and $2 \times 10^{-22}$ cm$^2$/s for the DEMO and HFR cases. These values weakly depend on the choice of the activation energy barrier in formulae (\ref{eq:D_el_simplified}) and (\ref{eq:D_el_HER}).

The characteristic diffusion distance $\sqrt{6D_{\textnormal{el}}\Delta t}$ derived from the above values, assuming the timescale of operation of a reactor component of $10^7$ s,  is 5-10 \AA\ for the DEMO and 10-20 \AA\ for the HFR scenarios. While these spatial scales are comparable with the average distance between the defects in a heavily irradiated tungsten, where the vacancy content is close to 0.1~\%  \cite{Buswell1970,Mason2021,Hollingsworth2022}, the rate of diffusion stimulated by direct electron impacts under typical reactor conditions appears relatively low, and unlikely has an appreciable effect on microstructural evolution.

In the next section, by following a similar line of mathematical argument, we show that a far more pronounced effect of $n-\gamma -e$ scattering on diffusion of defects stems from the interaction of high-energy electrons with vacancy-impurity clusters in engineering materials.


\section{Vacancy migration in impure W, stimulated by high-energy electrons}

Evidence from high-voltage transmission electron microscope (TEM) observations show that high-energy electron irradiation is effective in stimulating vacancy migration \cite{Kiritani1976,Arakawa2020}, especially in heavy elements \cite{Kiritani1976}. The point highlighted by the TEM observations is the pivotal part played by impurities \cite{Satoh2008,Arakawa2020} that in engineering materials immobilise vacancies in the temperature range well above the onset temperature of migration of vacancies observed in high purity materials \cite{Ma2019}. For example, in pure tungsten vacancies are mobile above 350$^{\circ}$C \cite{Thompson1960}, whereas impurities immobilise vacancies in the entire temperature range extending to  900$^{\circ}$C \cite{Ferroni2015,Liu2014}.

In the preceding section we showed that the high-energy electrons generated by the $\gamma$-photons can stimulate vacancy migration by depositing a sufficient amount of energy to atoms near a vacancy. In a pure material, within  a typical operating temperature range of a tokamak reactor, the frequency of such electron-stimulated events is not high compared to the frequency of thermal hopping events. 

In pure W, where the activation energy for vacancy hopping is $\sim$1.66~eV \cite{Liu2014}, the Arrhenius law (\ref{eq:th_D}) predicts the hopping rate of 1~Hz at 615~K, in agreement with experimental observations \cite{Hu2016, Debelle2008}. 
The presence of carbon impurities has a dramatic effect on thermally activated migration of vacancies. DFT calculations show that in tungsten the activation energy for the dissociation of a vacancy-carbon impurity cluster, which is a rate-limiting stage for the thermally activated migration of vacancies, is 3.39 eV or 3.43~eV, depending on whether a vacancy is bound to one or two carbon impurities \cite{Liu2014}. The dissociation of vacancy-carbon impurity clusters controls the rate of release of mobile vacancies into the material, and this explains why the defect and dislocation microstructure of tungsten irradiated with ions at 500$^\circ$C remains stable  to nearly 900$^\circ$C \cite{Ferroni2015}. The shift of the onset of vacancy migration temperature from 350$^\circ$C to 900$^\circ$C implies the presence of a process with the activation energy of $\sim$3.2~eV, consistent with the predicted \cite{Liu2014} energy barrier for thermal dissociation of vacancy-impurity clusters.

Below 900$^\circ$C, the Arrhenius law (\ref{eq:th_D}) makes vacancies bound to impurities effectively immobile. At the same time, the efficiency of high-energy electron impacts in de-trapping vacancies from impurities is nearly the same as the efficiency of electron-driven events of vacancy hopping in a pure material. This is because formulae (\ref{eq:D_el_simplified}) and (\ref{eq:D_el_HER}) 
are barely sensitive to $E_\textnormal{a}$ in comparison with the Arrhenius formula (\ref{eq:th_D}). 
By setting the energy barrier in Eq.~\eqref{eq:rate_success} to $E_\textnormal{a}=3.2$~eV, we estimate the rate of dissociation of carbon-vacancy clusters by high-energy electron impacts. 

\begin{figure}[t]
  \includegraphics[width=0.45\textwidth]{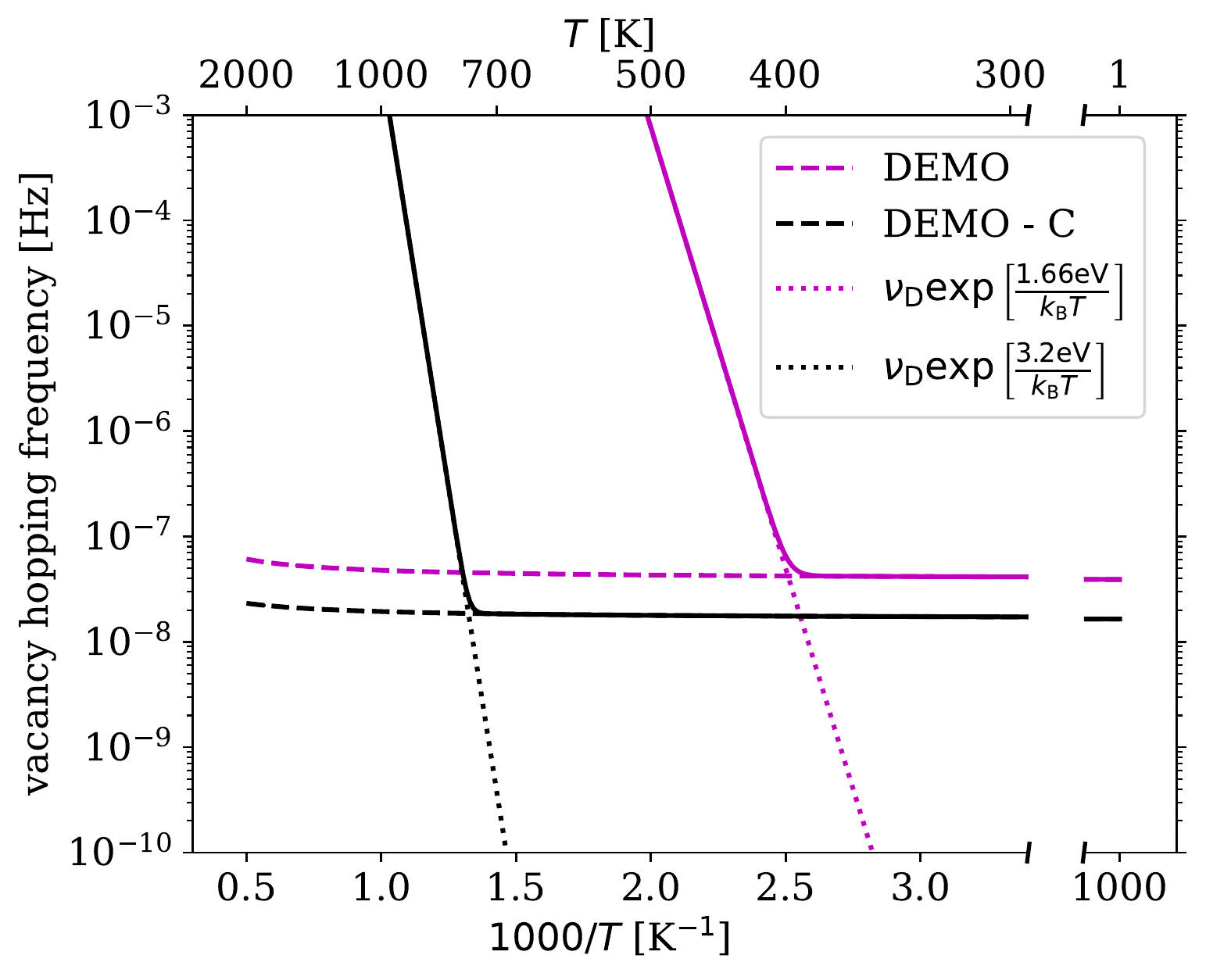}%
  \caption{Rates of vacancy hopping in pure and impure tungsten.  The Arrhenius steeply inclined dotted lines represent the rates of thermally activated hopping of vacancies; the nearly horizontal curves correspond to the rates of vacancy hopping driven by the high-energy electron impacts, evaluated assuming the DEMO spectrum of $\gamma$-photons.  Black lines correspond to the activation energy of vacancies pinned by impurities; the magenta lines refer to vacancies in ideal pure tungsten. Solid lines show the combined effect of thermally activated and athermal $\gamma$-photon stimulated processes, evaluated as the square root of the sum of the two hopping frequencies squared, using Eq.~2 from Ref. \cite{Arakawa2020}. Note that if an impurity-vacancy cluster dissociates, the mobility of a vacancy released in the process equals the mobility of a vacancy in a pure material at the same temperature, and is several orders of magnitude higher than the mobility of vacancies in impure tungsten.}\label{fig:freq}
\end{figure}

The rates of thermally activated vacancy hopping and the rates of vacancy hopping stimulated by electron impacts are compared in Fig.~\ref{fig:freq} for ideal pure tungsten and for industrially produced tungsten where vacancies are bound to carbon or other impurities, assuming the DEMO first wall $\gamma$-photon electron spectrum. The photons included in the generation of high-energy electrons are only those produced \emph{internally} by the neutrons in the bulk of the material. In a relatively thin tungsten layer involved in some of the current designs of fusion tritium breeding blankets, the \emph{external} photon flux generated by the plasma can make a substantial additional contribution to the spectrum of high-energy electrons bombarding the vacancy-impurity clusters. 

Fig.~\ref{fig:freq} highlights the entirely different temperature dependence of vacancy mobility in pure and impure tungsten. At high temperature, thermal hops always dominate but at lower temperatures, electron-collison-induced athermal vacancy hopping events take precedence. A fundamentally similar phenomenon was observed in cryogenic TEM observations of dynamic de-trapping of dislocation loops, where thermally activated dynamics was dominant above approximately 100~K but was replaced by the electron-collision-induced processes at lower temperatures where thermal vibrations are suppressed \cite{Arakawa2020}. The underlying physics of motion of loops observed at low temperature in TEM experiments \cite{Arakawa2020} is different \textemdash the migration of loops in pure crystalline lattice is driven by quantum fluctuations of atomic positions \cite{Arakawa2020} whereas in our case the hopping of vacancies is stimulated by the intense $\gamma$-photon radiation \textemdash but the manifestations of the effects are similar. Notably, the frequency of collision-driven vacancy hopping is almost insensitive to the activation energy, due to the power-law distribution of recoil energies \eqref{eq:dS_dEr_simplified_Mott} involving a large proportion of high-energy events.

Whereas vacancy hopping driven at every step by high-energy electron collisions is still a rare phenomenon, the key aspect of vacancy dynamics in an impure material is the rate-limiting part played by the vacancy-impurity cluster dissociation events.
In the temperature range from $\sim350^\circ$C to $\sim900^\circ$C, corresponding to the expected operating temperature range for tungsten components in a tokamak reactor, thermal diffusion of free vacancies is exceedingly fast. In this temperature range, once an electron collision triggers the dissociation of a vacancy-impurity cluster, the subsequent thermally activated diffusion of a vacancy occurs with low activation energy of $\sim 1.66$ eV and, as Fig.~\ref{fig:freq} shows, is many orders of magnitude faster than the diffusion of vacancies in the same material in the absence of background $\gamma$-radiation triggering the dissociation events. As a representative example, if we take the temperature of 500$^\circ$C in the above temperature interval, the diffusion coefficient of a vacancy increases as a result of a vacancy-impurity dissociation event from about $10^{-7}$~cm$^2$/s to about $10^{3}$~cm$^2$/s, i.e. by nearly ten orders of magnitude.

Although the argument above is relatively simplified, experimental observations confirming the occurrence of such events are abundant. For example, Fig. 11 from Ref. \cite{Yao2008} or Fig. 3 from Ref. \cite{Dudarev2010} show typical examples of observations of de-trapping events, where the movement of a mobile defect between the traps appears instantaneous on the experimental timescale, highlighting the dramatic difference between the mobilities of the freely moving and impurity-confined defects. When assessing the overall effect of vacancy-impurity dissociation events on microstructural evolution, one also needs to consider the re-capture events, involving a vacancy and an impurity and occurring due to their  attractive interaction \cite{Hudson2005}. We estimate that these events reduce the effective dissociation frequency by about an order of magnitude, but this is still a small correction to the dramatic increase of the vacancy diffusivity resulting from the electron-collision-triggered dissociation of vacancy-impurity clusters.

In addition to vacancy diffusion, electron recoils can stimulate other processes involving high activation energies. The unpinning of dislocations from Frank loops by ion-induced recoils was observed in Cu in experiments and in MD simulations \cite{Khiara2022}. In simulations, only the high PKA energies approaching 10~keV were considered. Electrons in a neutron irradiated material generate numerous but less energetic recoils that can readily driven processes involving high activation energies. Electron impacts can drive the motion of dislocation loops \cite{Dudarev2007}, observed in TEM experiments \cite{Arakawa2020}. Simulations also show that irradiated microstructures can undergo large-scale avalanche type re-organisation\textemdash triggered by a relatively small scale low-energy single Frenkel pair generation event \textemdash over length scales much greater than that of the perturbed region \cite{Derlet2020}. Further work is required to quantitatively assess the effect of $\sim$eV recoils, which statistically are far more frequent than the $\sim$keV neutron-initiated recoils producing radiation damage in the first place. 

We conclude that the effect $\gamma$-photons on microstructure of real materials is particularly relevant at intermediate temperatures, close to the expected operating temperature range of a tokamak reactor. Whereas the discussion in this section was limited to individual vacancies, the treatment of electron-stimulated motion of self-interstitial atoms and vacancy clusters is outlined in Appendix~\ref{app:SIA_V_cl}.

\section{General implications}\label{sec:general}
The accelerated diffusion of defects is one of the consequences of exposure to intense $\gamma$-photon fluxes resulting from the bombardment of materials by neutrons. We now give an overview of other implications of $\gamma$-photon exposure for the development of a fusion energy source. In steels and tungsten, the dominant part of the energy deposited by neutron irradiation does not give rise to the generation of crystal defects by atomic recoils, but rather to the production of $\gamma$-photons with energies approaching and exceeding 10 MeV. Partially, these $\gamma$-photons are absorbed in the material, generating a fluctuating athermal dynamic steady state population of high energy electrons. Some of the energetic $\gamma$-quanta escape from the materials back into the reactor environment, producing additional physical effects, different from those directly associated with neutron exposure \cite{Fischer2017,Fischer2019}. 

Tritium diffusion into reactor materials and its retention must be minimised to satisfy the radioactive safety requirements  \cite{Humphry2019}. $\gamma$-radiation enhances deuterium permeation through steels following exposure to the dose rate of $\gamma$-radiation above several Gy/s, equivalent to the power density deposition rate of several Watts per kg \cite{Fujita2018}. We can see from Table~\ref{tab:heating} that in DEMO, this rate is higher approaching kGy/s; the permeation enhancement would in this case be more pronounced, as confirmed by observations summarised in Ref. \cite{Fujita2018}. Deuterium diffusion and trapping are phenomena of central significance for ceramic breeder blanket materials, where ionizing radiation modifies the electronic structure of defects and therefore the corresponding trapping energies \cite{Gonzalez2015}.

Helium is among the transmutation products forming in structural fusion materials exposed to neutron irradiation. In plasma-facing materials, however, higher helium concentration originate from the trapping of \(\alpha\)-particles (i.e.,~\(^4\)He) produced by the fusion reaction and escaping from the plasma, and also via the decay of tritium (also escaping as un-burnt fuel from the plasma) producing \(^3\)He. For example, the equilibrium \(^3\)He concentration, which depends on the tritium retention in W, was estimated to be close to 600~appm assuming $\sim$1~\% of retained tritium \cite{Shimada2017, Markelj2020}. He concentration in fusion power plant materials is expected to be comparable to the concentration of vacancies generated by irradiation. Helium trapped by radiation defects immobilizes dislocations and vacancies \cite{Bonny2014, Markelj2020}, giving rise to radiation embrittlement and swelling. Vacancies in W strongly bind to He atoms. The binding energy of a single He atom to a vacancy is 5.4~eV and remains positive and as high as 3.8~eV even for a sixth He atom added to a He$_5$V cluster \cite{Bonny2014}. Thermal He de-trapping is therefore negligible at any operating temperature. However, electron collisions can kick He atoms out of a vacancy or kick W atoms into a vacancy, as we demonstrated above using molecular dynamics simulations. The physical origin of the effect is similar to the impact dissociation of carbon-vacancy impurity clusters considered in the previous section.

Compact spherical tokamak concepts are attractive from the perspective of reduced economic investment, but pose a challenge because of the reduced shielding of superconductors in the central column \cite{Humphry2019}. Tungsten carbide has been proposed as a candidate material for shielding; according to the above, this increases the intensity of $\gamma$-photon spectra in the plasma chamber. Monte Carlo neutron and $\gamma$-transport calculations for the central column found a nearly constant average energy of the $\gamma$-photons of 2~MeV throughout the shield and the coils due to (n,$\gamma$) reactions in tungsten \cite{Windsor2017}.

$\gamma$-photon exposure also affects functional insulators, polymers, and the rates of corrosion of structural materials by coolants, and can trigger plasma instabilities. 

The exposure of insulating materials to $\gamma$-photons gives rise to the deterioration of optical properties \cite{Pintsuk2022}. The phenomenon stems from the generation of electron-hole pairs and the subsequent decay of electronic excitations, accompanied by the formation of structural defects \cite{Lushchik1977,Lushchik1989}. The exposure of polymers to $\gamma$-photons is also considered to be mostly detrimental \cite{Phillips1988}.  

Irradiation is also a concern for the superconducting coils generating the magnetic field confining the plasma. It is generally found that the critical temperature $T_\textnormal{C}$ does not change significantly at small radiation exposure, but then it decreases with further exposure until irradiation eventually destroys the superconducting state. Fast neutron irradiation of fusion-relevant Nb$_3$Sn and yttrium barium copper oxide (YBCO) was found to reduce $T_\textnormal{C}$ by a few percent after exposure to the fluence of $10^{22}$~m$^{-2}$ \cite{Nishimura2012, Fischer2018}. Further exposure drastically impaired the performance once the fluence reached $10^{23-24}$~m$^{-2}$ \cite{Karkin1976, Nishimura2012}. $\gamma$-radiation in a compact spherical tokamak may remain high even if the neutron field is well screened \cite{Windsor2017}. There are contradictory data in literature showing both improvement and degradation of performance of $\gamma$-irradiated YBCO, with moderate effects up to $\sim500$~kGy of irradiation \cite{Leyva1995}. However, the maximum dose reached in the tests ($\lesssim1$~MGy) is far lower than the total $\gamma$ dose of 10-10,000~MGy expected in ITER between the vacuum vessel and the first wall \cite{Vayakis2008, Kashaykin2021}.  

Among the insulating materials we mention sapphire that, together with diamond, is one of the candidate materials for the transmission windows of electron cyclotron resonance plasma heaters. Its resistance to ionising radiation up to a dose rate of 0.5~MGy/s is confirmed \cite{Heidinger1998}, notably using somewhat less energetic X-rays ($\lesssim100$~keV) than the $\sim1$~MeV $\gamma$-photons that are to be expected in a fusion environment. Sapphire optical fibres are also a potential option for plasma diagnostics. Simultaneous neutron (3.5 dpa) and $\gamma$ irradiation at 95-298$^\circ$C showed acceptable transparency but increasing the irradiation temperature to 688$^\circ$C led to a major degradation of physical properties \cite{Petrie2022}.

The radiolysis of water, commonly used as a coolant in nuclear applications, is known to be a contributing factor accelerating the corrosion of both zirconium \cite{Burns1976} and ferritic \cite{Was2020} alloys. Here, the relative significance of $\gamma$-photon versus neutron or ion irradiation depends on the geometry of a reactor component \cite{Gilbert_2017}, highlighting the role of highly spatially resoved simulations of neutron and $\gamma$-photon fields in the context of an advanced reactor design. Of specific interest to fusion is the fact that in the current divertor designs, where the neutron field is nearly as high as in the first wall, the water coolant flows in copper pipes encapsulated in a tungsten armour, which poses concerns due the $\gamma$-generation in the latter.

Finally, we note an effect of intense neutron-induced $\gamma$-photon emission from the tungsten walls of a fusion power plant on the fusion plasma. Table \ref{tab:heating} illustrates a remarkable difference between the $\gamma$-photon emission from beryllium, chosen as the plasma-facing material for {ITER} \cite{Temmerman2021}, and tungsten selected as a plasma-facing and shielding material for a commercial fusion power plant \cite{Rieth2013}. The use of tungsten might give rise to unusual $\gamma$-emission-stimulated phenomena stemming from the seeding of a population of very high energy runaway electrons in the plasma \cite{Breizman2019,Martin-Solis2017,Vallhagen2020}, different from the effects of contamination or melting commonly explored in the context of plasma-facing materials technology.

\section{Conclusions}
Neutron irradiation deposits energy in materials through a variety of nuclear interactions; these include elastic and inelastic collisions with atomic nuclei and non-elastic interactions such as neutron capture or multiplication. The resulting nuclei are often left in internally excited states that emit $\gamma$-photons as they decay. The fraction of energy of neutrons converted into a flux of $\gamma$-photons strongly depends on the elemental content of the material. For example, in W the proportion of kinetic energy of neutrons converted into electromagnetic $\gamma$-radiation approaches 99~\%, which is higher than the light generation efficiency of LED devices, whereas in Be this fraction is less than ~1~\%, assuming exposure to identical neutron spectra.

It is important to assess the effect of generation of energetic $\gamma$-photons in W and medium-weight materials exposed to neutron irradiation. In a two-step process, neutrons generate $\gamma$-photons that in turn excite a steady-state population of high-energy electrons. Starting from an energy-resolved neutron spectrum, we compute the energy-resolved $\gamma$-photon and electron spectra. In W and  Fe, the DEMO fusion and HFR fission neutrons produce $\gamma$-photon and electron spectra that have energies in the range from hundreds of keV to several MeV. The high-energy electrons have sufficient energies to stimulate athermal migration of vacancies and other defects through relatively low energy atomic recoils, leading primarily to the dissociation of vacancy-impurity clusters. The frequency of vacancy hops stimulated by the interaction with electrons depends on their flux, on the energy-differential atomic cross section and on the nature of many-body atomic transitions triggered by the recoils. A functional form describing the latter is confirmed by molecular dynamics simulations. The results illustrate a Doppler effect lowering the threshold energy for vacancy migration. The diffusion coefficient for electron-stimulated diffusion in W is evaluated assuming DEMO, HFR and high-voltage TEM irradiation exposure. 

Our estimates suggest that the magnitude of the effect of $\gamma$-photons on the transport of defects may be profound, particularly in real engineering materials where vacancy mobility in the operating temperature range of a reactor is normally impeded by the presence of impurities. 
The high energy electrons generated by the $\gamma$-photons stimulate the dissociation of vacancy-impurity clusters, giving rise to a dramatic increase of the effective mobility of vacancies.
Experimental observations show that in the absence of $\gamma$-radiation, impurities pin vacancies and form clusters stable up to 900$^\circ$C, as the thermal hopping frequency depends exponentially on the migration barrier whereas electron-induced hopping shows barely any dependence at all. Atomic recoils, initiated by the high-energy electrons produced by the $\gamma$-quanta, have a broad energy spectrum extending into the several eV range even in tungsten, and hence the same mechanism likely triggers a variety of other reactions involving high activation energies.

Finally, in section \ref{sec:general} above we noted a broad range of general implications, resulting from the effect of neutron-induced $\gamma$-photon emission, for nuclear reactor materials, technology, and reactor operation.

\begin{acknowledgements}
The authors are grateful to M. Rieth, D. Terentyev,  S. Kalcheva, G. Pintsuk, D. R. Mason, M. Short, P. Helander, S. Chislett-McDonald, H. Campbell, and an anonymous reviewer of this paper, for stimulating discussions. This work has been carried out within the framework of the EUROfusion Consortium, funded by the European Union via the Euratom Research and Training Programme (Grant Agreement No 101052200 — EUROfusion) and from the EPSRC Energy Programme (grant number EP/W006839/1). To obtain further information on the data and models underlying the paper please contact PublicationsManager@ukaea.uk. Views and opinions expressed are however those of the authors only and do not necessarily reflect those of the European Union or the European Commission. Neither the European Union nor the European Commission can be held responsible for them. We gratefully acknowledge the provision of computing resources by the IRIS (STFC) Consortium. 
\end{acknowledgements}

\begin{appendix}
\section{The total cross section of Compton scattering} \label{app:kernel}
The kernel of Eq.~\eqref{kernel} is the energy-differential cross section involving only the Compton scattering of photons. This cross section describes scattering of a photon with the initial energy $E'$ into a state with any energy between $E'$ and $E'/(1+2E'/mc^2)$. Therefore, the total Compton scattering cross section in Eq.~\eqref{eq:sigma_total}, tabulated in Ref. \cite{Hubbell1975}, can also be found by integrating Eq.~\eqref{kernel} over all the energies $E$ after scattering
\begin{align}
    \tilde{\sigma}_{\textnormal{CS}}(E')=& \int  \text{d}E\int  \text{d}o' {\text{d}^2 \sigma ({\bf n}',E' \rightarrow {\bf n},E) \over \text{d}o' \text{d}E'} \\ \nonumber 
    =&   \int \text{d}E\,  K(E,E') .
\end{align}
By integrating Eq.~\eqref{kernel} over the interval of energies $E'/(1+2E'/mc^2) < E < E'$ we find that
\begin{align}\label{eq:integral_kernel}
    \tilde{\sigma}_{\textnormal{CS}}(\varepsilon)= &2\pi r_{\textnormal{c}}^2\bigg[\frac{2}{\varepsilon^2}+\frac{(\varepsilon+1)}{(2\varepsilon+1)^2} +\frac{\log(1+2\varepsilon)}{2\varepsilon} \\ \nonumber 
    &-\frac{2(\varepsilon+1)}{\varepsilon^3}\text{tanh}^{-1}\left(\frac{\varepsilon}{\varepsilon+1}\right)\bigg] ,
\end{align}
where $\varepsilon=E'/mc^2$.

\begin{figure}[t]
\includegraphics[width=0.45\textwidth]{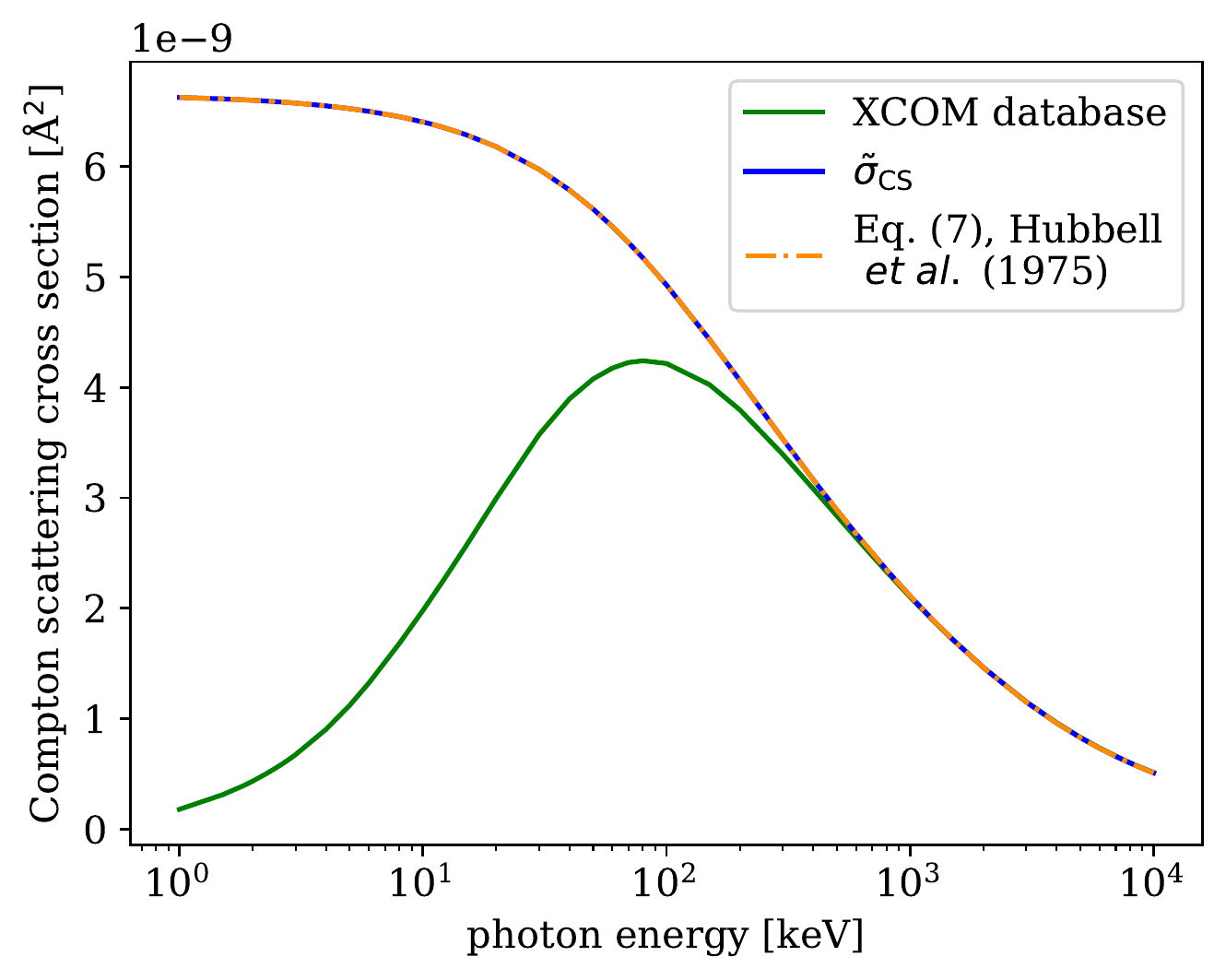}%
\caption{Total Compton cross section of scattering for a single electron, obtained by integrating the Klein-Nishina cross section (blue), is compared with the data available in the XCOM database for W \cite{Berger1987} (green) and analytical expression from Ref.~\cite{Hubbell1975}. The assumption that even electrons in the inner shells of atoms act as independent scattering centres is very accurate if the photon energy exceeds $\sim$300~keV.}
\label{fig:compare_sigma_CS}
\end{figure}

Comparison of Eq.~\eqref{eq:integral_kernel} and $\sigma_{\textnormal{CS}}$ from \cite{Berger1987} is given in Fig.~\ref{fig:compare_sigma_CS}. Eq.~(7) of Ref.~\cite{Hubbell1975} provides a different expression for Eq.~\eqref{eq:integral_kernel}, which is also included in the figure. We see that for energies of interest of $\sim$300~keV and above, the assumption that there are $Z/\Omega$ independent electrons, where $\Omega$ is the atomic volume, is very accurate. At 100 keV  Eq.~\eqref{eq:integral_kernel} overestimates the data from \cite{Berger1987} by only {16\,\%}. 

\section{Thermal broadening of recoil energy} \label{app:Doppler}
Consider a collision between a high-energy electron and an atom. If the latter is at rest, it recoils with velocity $(v_{x,\textnormal{R}}, v_{y,\textnormal{R}}, v_{z,\textnormal{R}})$ and kinetic energy ${E=E_{\textnormal{R}}=\frac{m}{2}(v_{x,\textnormal{R}}^2+v_{y,\textnormal{R}}^2+v_{z,\textnormal{R}}^2)}$. If the atom also had some initial thermal velocity $(v_{x,T}, v_{y,T}, v_{z,T})$, the kinetic energy $E$ after the collision is
\begin{equation}
    E = \frac{m}{2}v^2=E_T+E_{\textnormal{mix}}+E_{\textnormal{R}},
\end{equation}
namely the sum of the thermal energy ${E_T=m(v_{x,T}^2+v_{y,T}^2+v_{z,T}^2)/2}$, the athermal recoil energy $E_{\textnormal{R}}$, and a mixing term 
\begin{equation}
    E_{\textnormal{mix}} = m(v_{x,T}v_{x,\textnormal{R}}+v_{y,T}v_{y,\textnormal{R}}+v_{z,T}v_{z,\textnormal{R}}).
\end{equation}
At $T>0$, the atom may receive more or less energy than it would at absolute zero, enabling collisions below the threshold to provide sufficient energy to cause a transition. The distribution of energy $E$ depends on the distribution of the individual terms, one of which is a constant $E_{\textnormal{R}}$.

$E_T$ follows the Maxwell-Boltzmann distribution for a particle of mass $m$ at temperature $T$. $E_{\textnormal{mix}}$ is the sum of three independent terms, where each is a product of thermal velocity that obeys a 1D Maxwell-Boltzmann distribution, i.e. a Gaussian distribution, and a term that is uniformly distributed between two extrema $\pm\sqrt{2E_{\textnormal{R}}/m}$. It can be shown that $E_{\textnormal{mix}}$ follows a Gaussian distribution whose mean is zero and whose standard deviation is $\sigma=\sqrt{2k_{\textnormal{B}}TE_{\textnormal{R}}}$. For the simulations that we considered, $E_{\textnormal{R}}$ is of the order of 1-10 eV, whereas $k_{\textnormal{B}}T$ is of the order of 0.03-0.08 eV. Hence, the distribution of $E$ is primarily that of $E_{\textnormal{R}}+E_{\textnormal{mix}}$, i.e. Eq.~\eqref{eq:Gauss_kTEr}
\begin{equation}\label{eq:Gauss_kTEC}
    f_{\textnormal{R}}(E)=\frac{1}{\sqrt{4\pi k_{\textnormal{B}}TE_{\textnormal{R}}}}\exp \left[-\frac{(E-E_{\textnormal{R}})^2}{4 k_{\textnormal{B}}TE_{\textnormal{R}}}\right],
\end{equation}
plus a correction given by the thermal contribution $E_T$, which is not trivial to evaluate as $E_{\textnormal{mix}}$ and $E_T$ are not independent. However, this correction is of second-order if $k_{\textnormal{B}}T\ll E_{\textnormal{R}}$. We note that the effect of temperature on the kinetic energy after the collision does not depend on the thermal energy $k_{\textnormal{B}}T$, but rather on $\sqrt{(k_{\textnormal{B}}T)E_{\textnormal{R}}}$, and therefore can be much greater than what one might anticipate from some simple energy addition argument \cite{Boersch1967, Fujikawa2006}.

\section{Interatomic potential for Fe}\label{app:Fe_pot}
The energy landscape that an atom experiences when hopping into a vacancy, treated as a function of reaction coordinate, may have local minima. If this happens, the atom can get stuck in one of these minima. We found that this commonly occurs in Fe, but does not happen in W. The local minimum corresponds to a ``split vacancy'' defect, where the Fe atom finds itself in a stable position between two neighbouring lattice sites, which both can be identified as vacancies. For this to happen, the atom must arrive at the midpoint between the two lattice sites with a low kinetic energy. The depth of the local minimum depends on the potential. As this local minimum is an artifact of parameterization of the potential, we selected a potential where this minimum is as shallow as possible. To this end, we tested three different potentials, namely by Mendelev {\it et al.} \cite{Mendelev2003}, Malerba {\it et al.} \cite{Malerba2010}, and Gordon {\it et al.} \cite{Gordon2011}.   

Consider the simple case where one of the first nearest neighbours of the vacancy receives a kick precisely along the $\langle111\rangle$ direction towards the vacancy, at 0~K. If using the Mendelev potential, a split vacancy remains stable after kicks in the recoil energy range from 0.9-1.05~eV, while the hop is not successful if the recoil energy is 0.85~eV and successful if it is above 1.1~eV. Similarly, the Malerba potential predicts a split vacancy for the recoil energies between 1.05-1.15~eV, an unsuccessful hop at 1.0 eV and a full hop at 1.2~eV. On the other hand, with the Gordon potential an attempt is unsuccessful at 0~K if the recoil energy is 0.81~eV and is fully successful if it is 0.83~eV, with a much more narrowly-defined activation energy of 0.82~eV.

\begin{figure*}
\subfloat[]{%
  \includegraphics[width=0.25\textwidth]{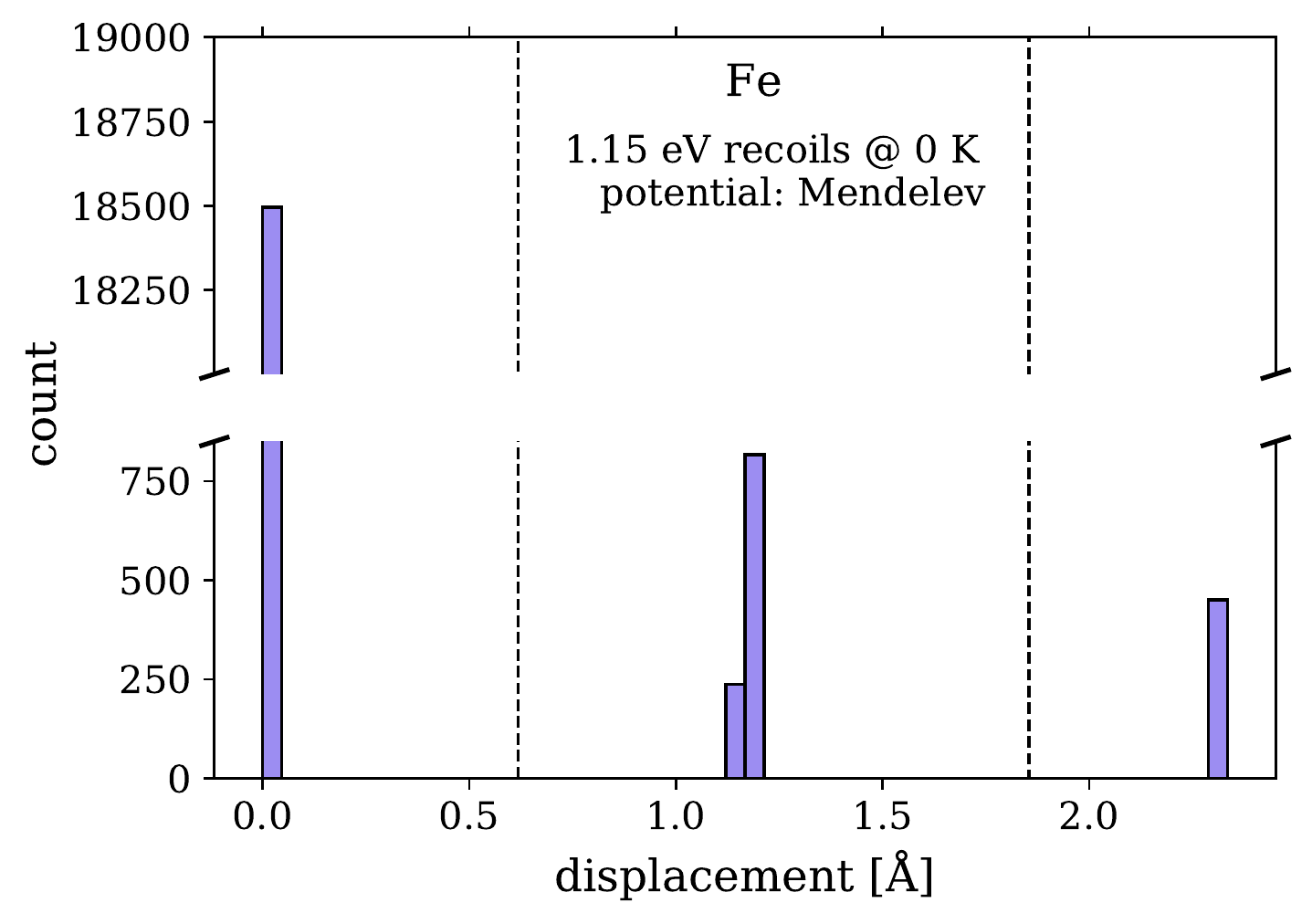}%
}\hfill
\subfloat[]{%
  \includegraphics[width=0.25\textwidth]{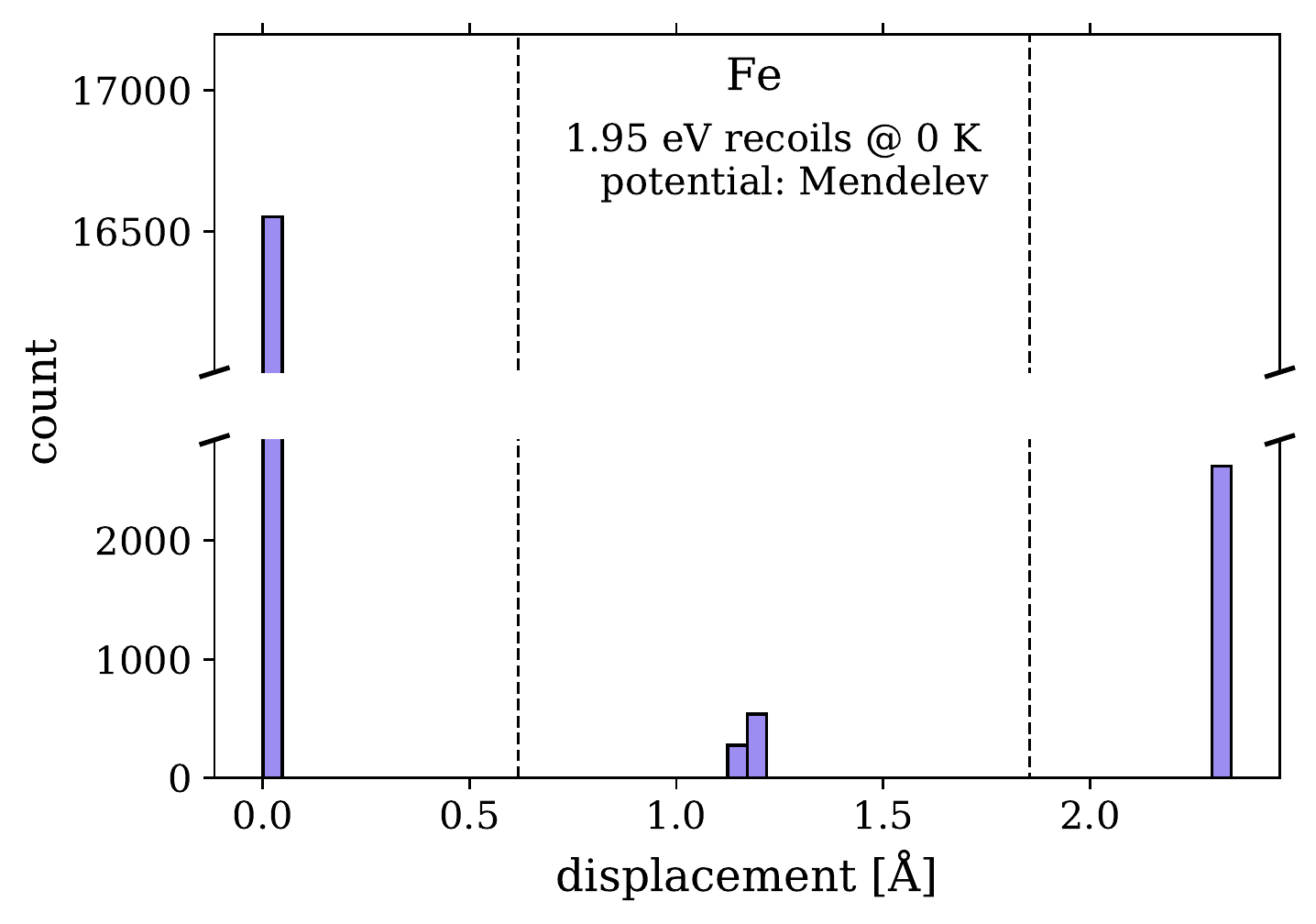}%
}\hfill
\subfloat[]{%
  \includegraphics[width=0.25\textwidth]{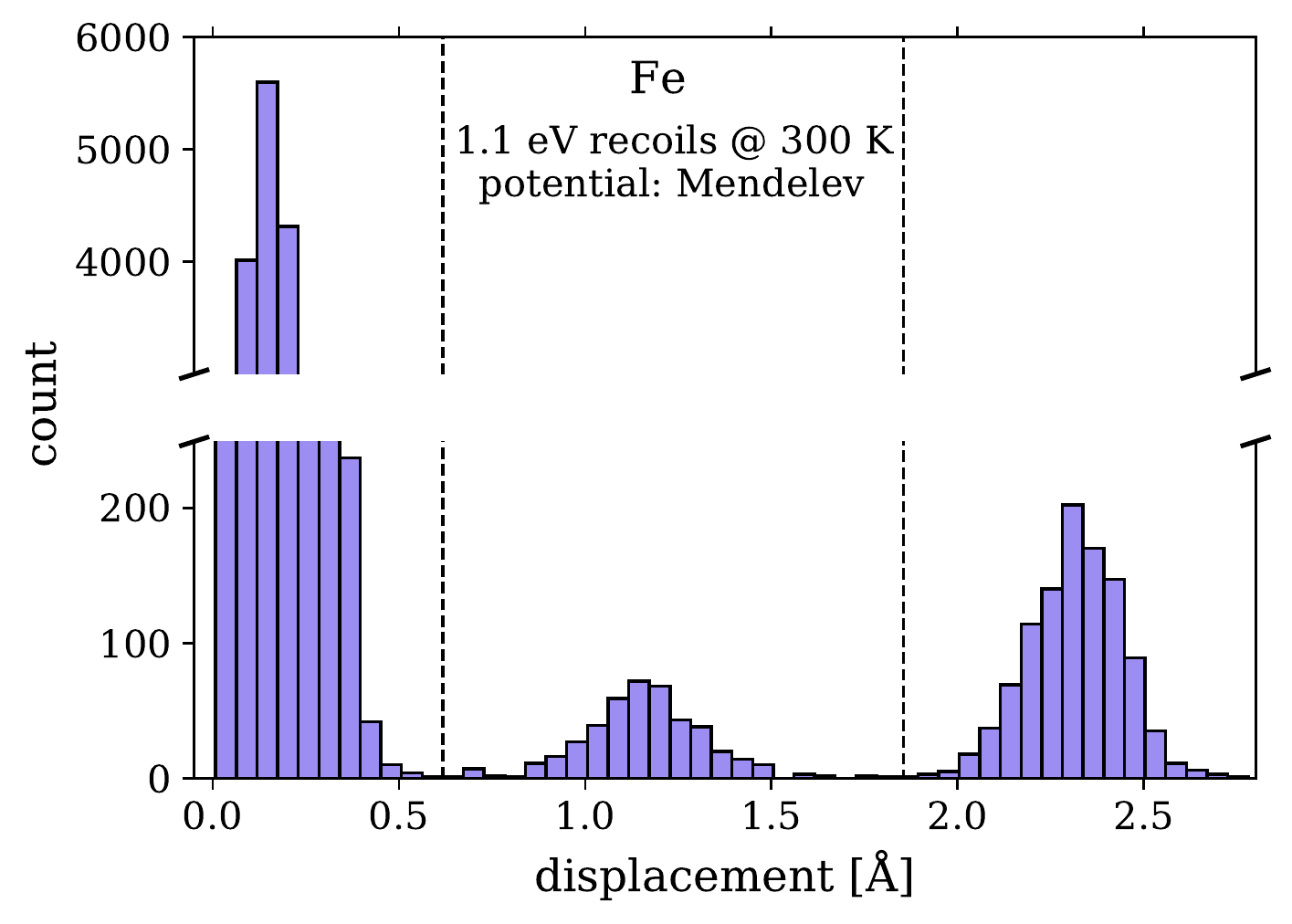}%
}\hfill
\subfloat[]{%
  \includegraphics[width=0.25\textwidth]{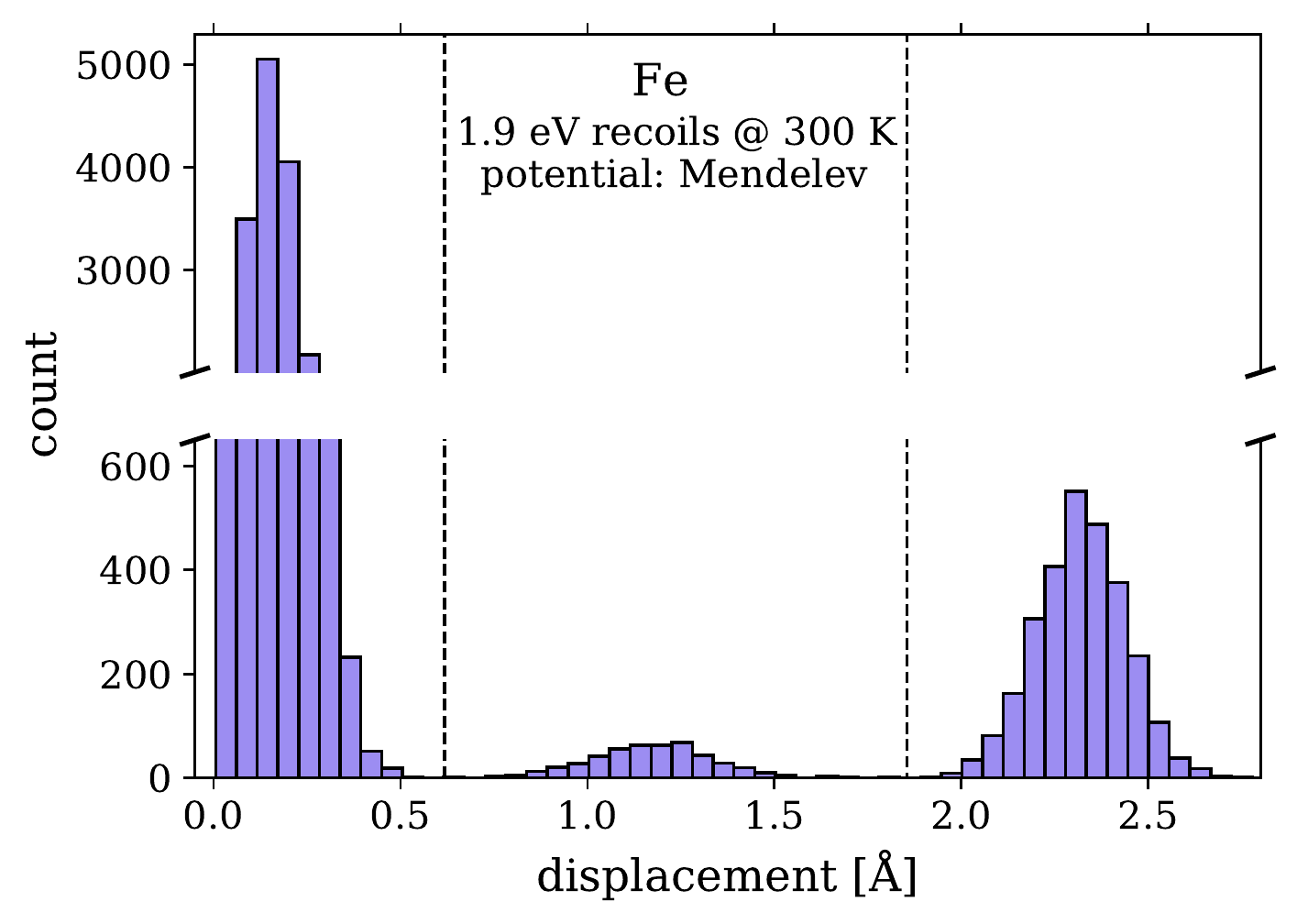}%
}\hfill

\subfloat[]{%
  \includegraphics[width=0.25\textwidth]{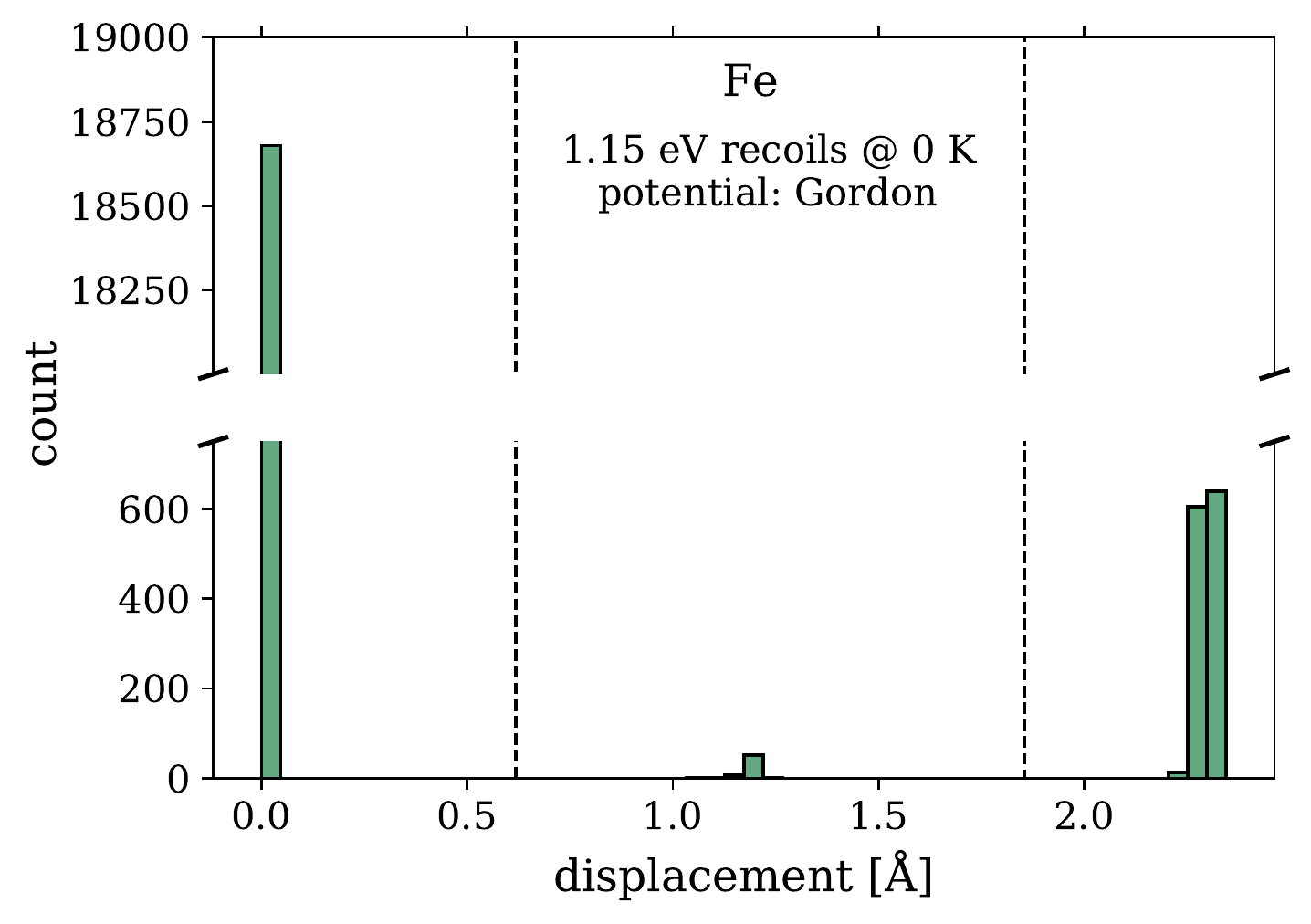}%
}\hfill
\subfloat[]{%
  \includegraphics[width=0.25\textwidth]{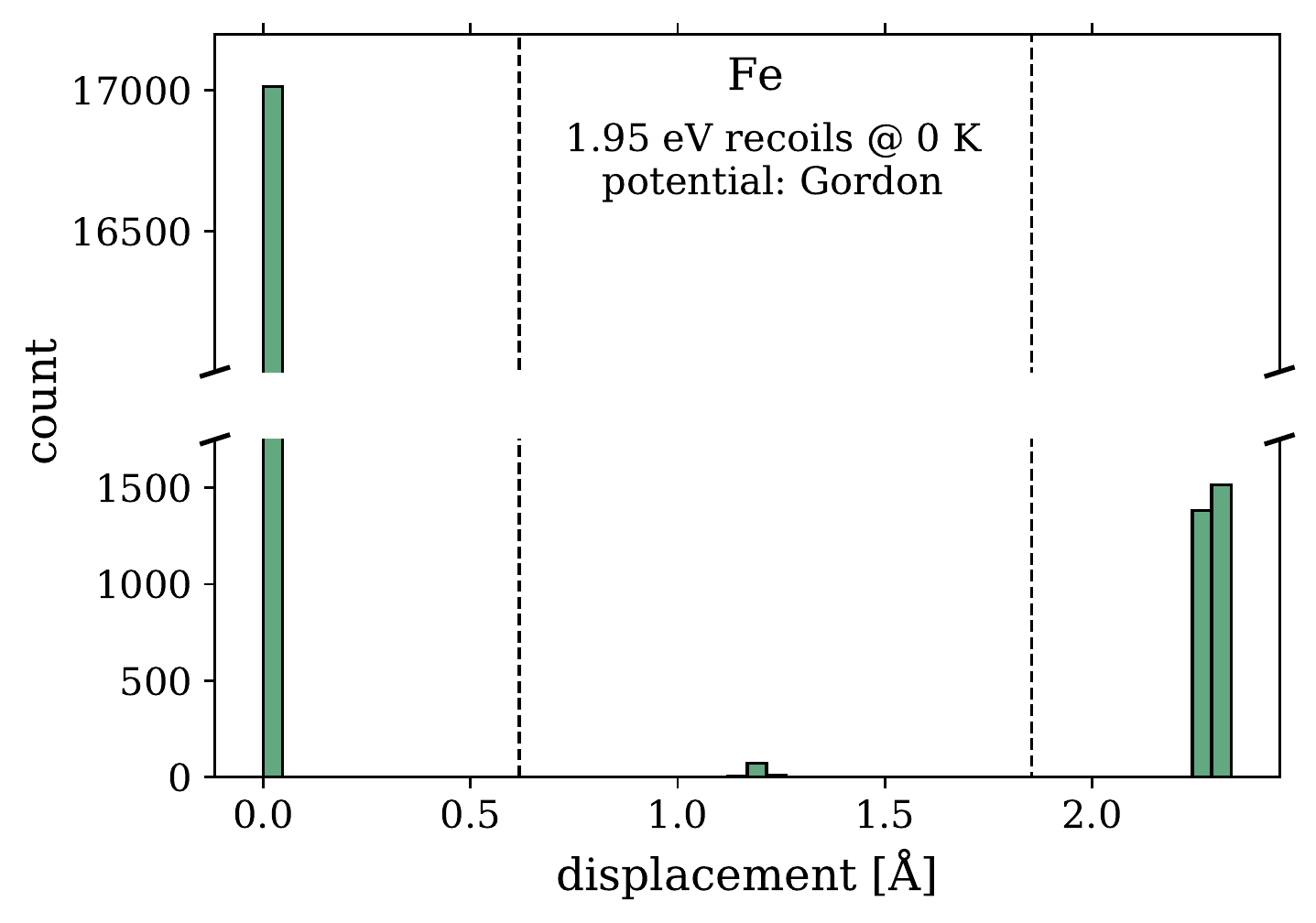}%
}\hfill
\subfloat[]{%
  \includegraphics[width=0.25\textwidth]{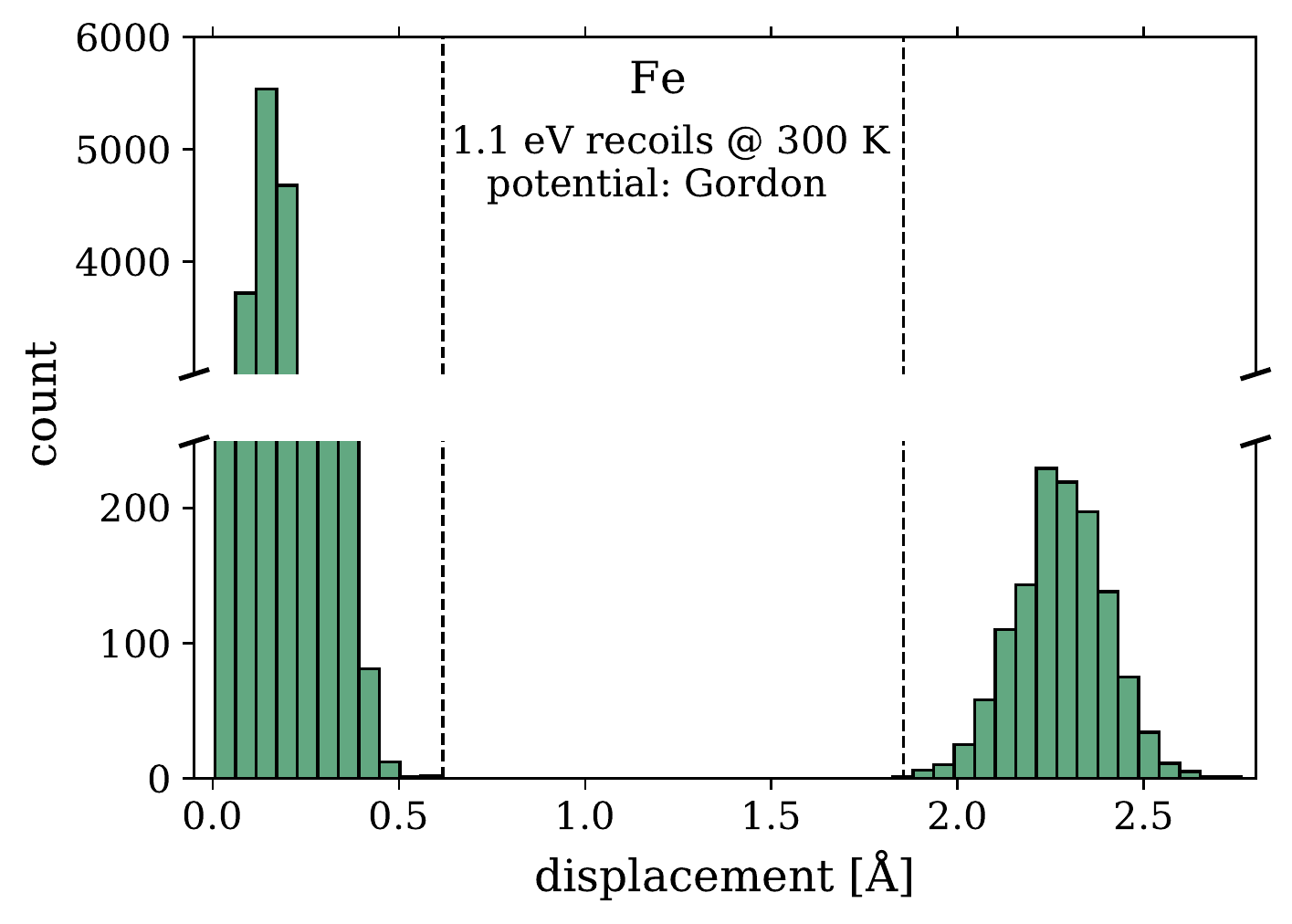}%
}\hfill
\subfloat[]{%
  \includegraphics[width=0.25\textwidth]{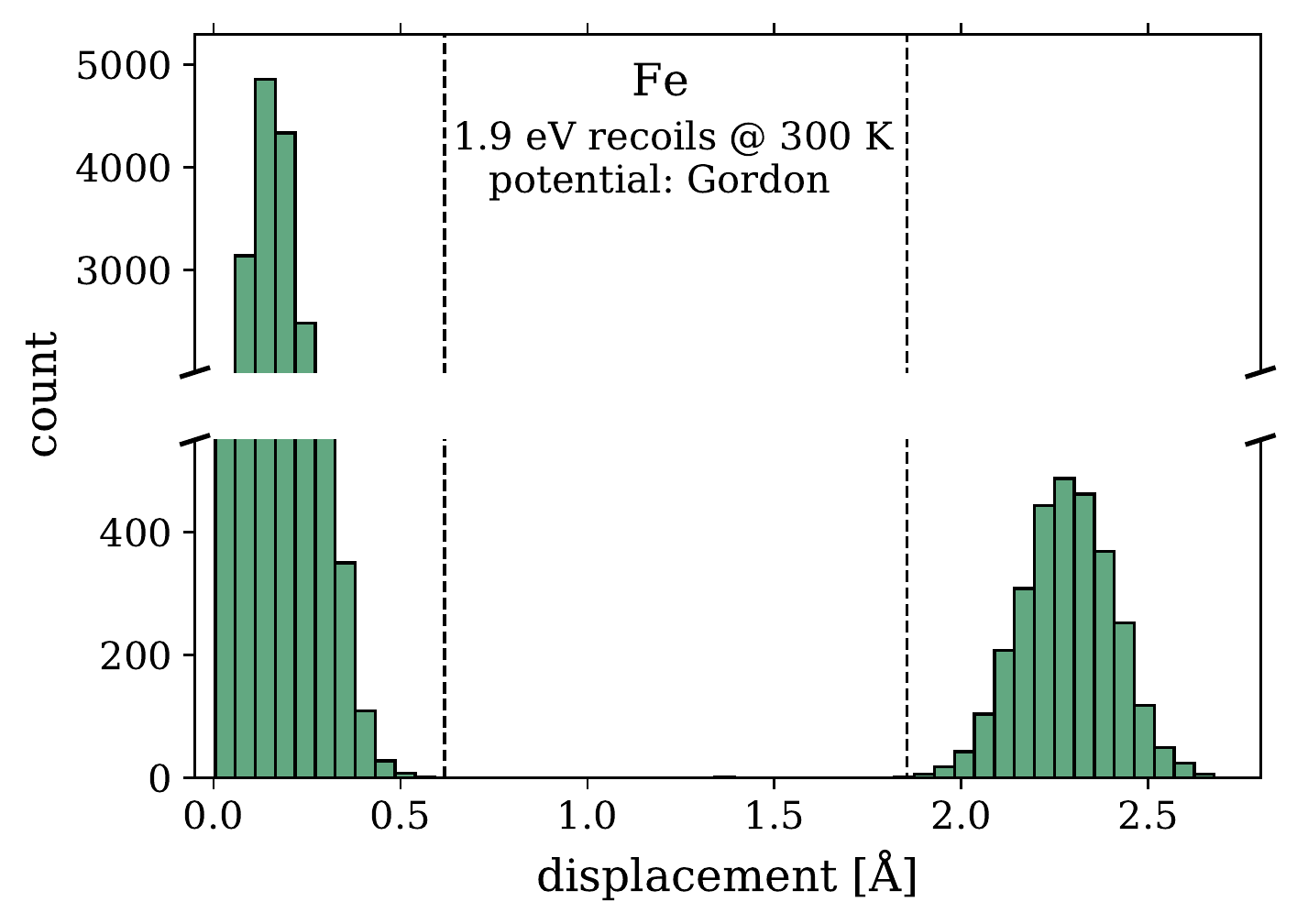}%
}\hfill
    \caption{Histograms showing distributions of  displacements of 1NN hops after recoils in random directions simulated at 0~K and at 300~K. Displacement is defined as the distance between the coordinate of the atom at the beginning and the end of each realisation. The most likely outcome is that the atom returns to the initial position. Successful hops form the distributions centred at $\approx2.48$~\AA. The instances of formation of ``split vacancy'' configurations are responsible for the distributions shown in the centre. The split vacancy configuration forms much more often with the Mendelev potential (a-d) than with the Gordon potential (e-h).}
    \label{fig:hist_d_Fe_2pot} 
\end{figure*}

The influence of the intermediate local minimum on the results is very strong. Fig.~\ref{fig:hist_d_Fe_2pot} shows histograms of the displacement experienced by the 1NN between the beginning and the end of 20,000 recoils in random directions with various energies and at different temperatures. The two expected outcomes for the displacement are either approximately null (if the attempt is unsuccessful), or approximately $\sqrt{3}a/2\approx2.48$~\AA\ (if the attempt is successful), with possibly a spread due to thermal vibrations. In the top row of Fig.~\ref{fig:hist_d_Fe_2pot}, produced using the potential by Mendelev {\it et al.}, one can clearly appreciate the presence of a third cluster of realisations, where the atom is stuck in the middle of the hop. This undesirable third outcome is substantially more likely than the full hop if the recoil energy is close to the vacancy migration energy. On the other hand, the same simulations repeated with the potential by Gordon {\it et al.} show that half-hops are very unlikely at 0~K and never occurr at 300~K.

\begin{figure*}
\subfloat[]{%
  \includegraphics[width=0.3\textwidth]{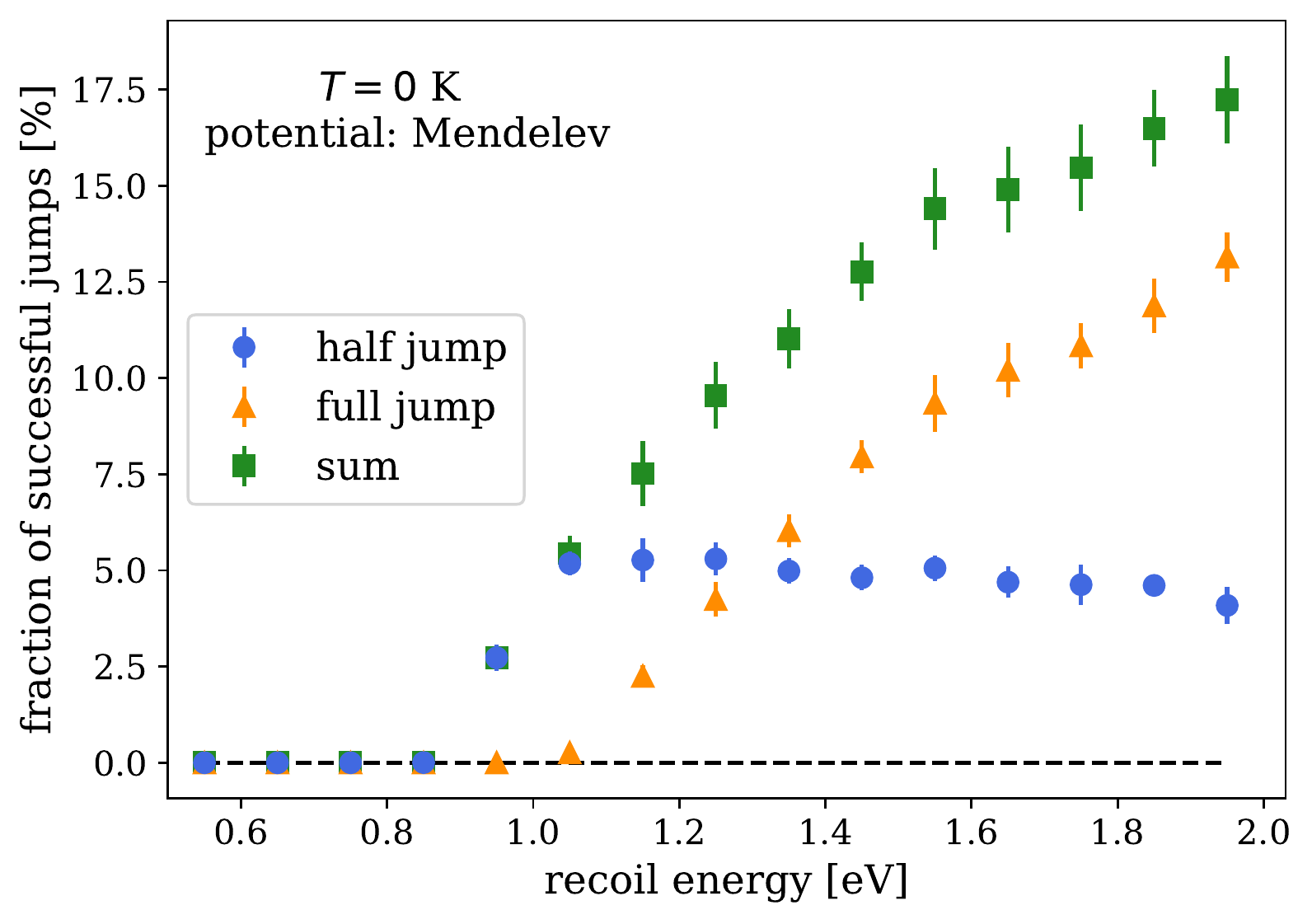}%
}\hfill
\subfloat[]{%
  \includegraphics[width=0.3\textwidth]{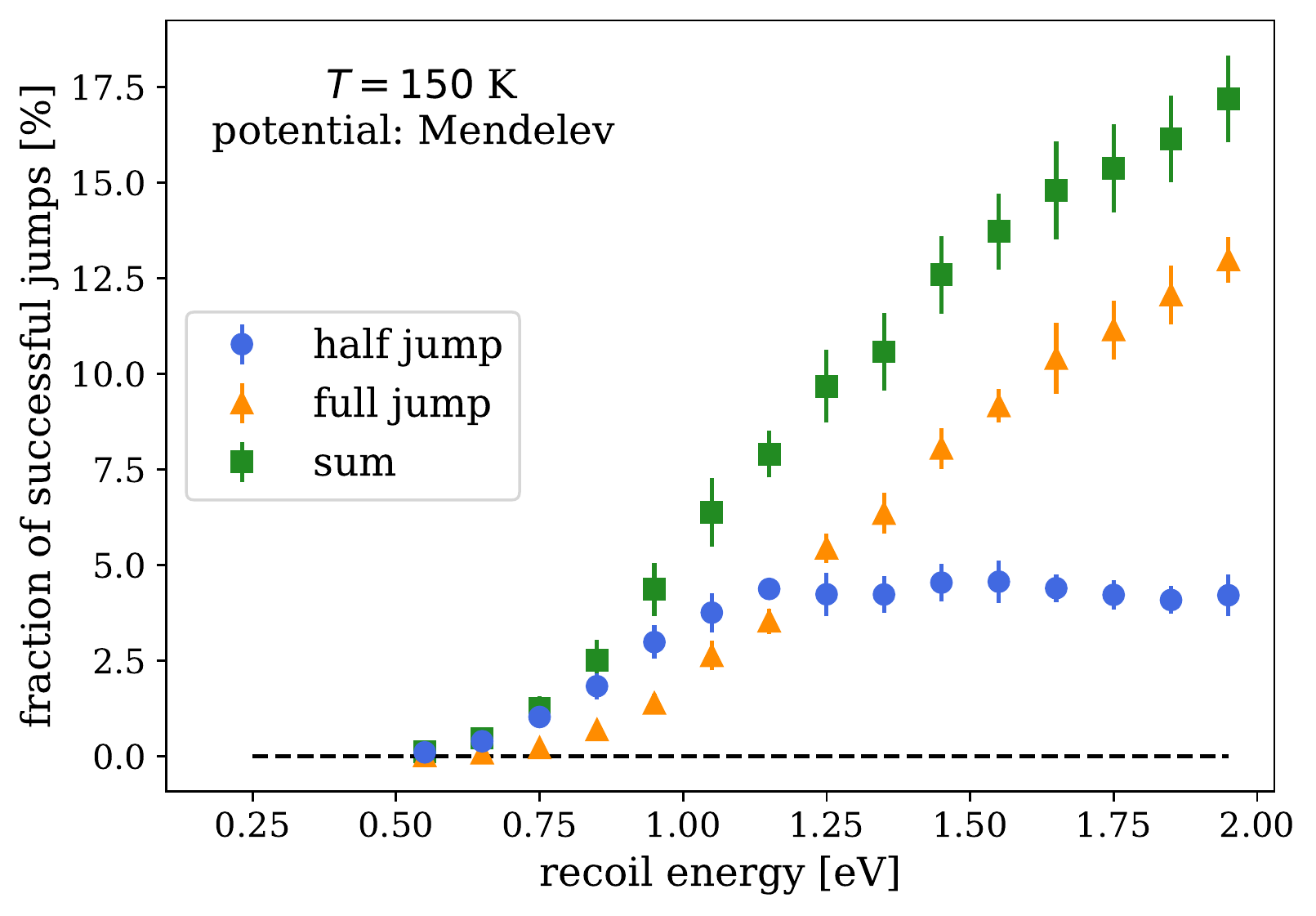}%
}\hfill
\subfloat[]{%
  \includegraphics[width=0.3\textwidth]{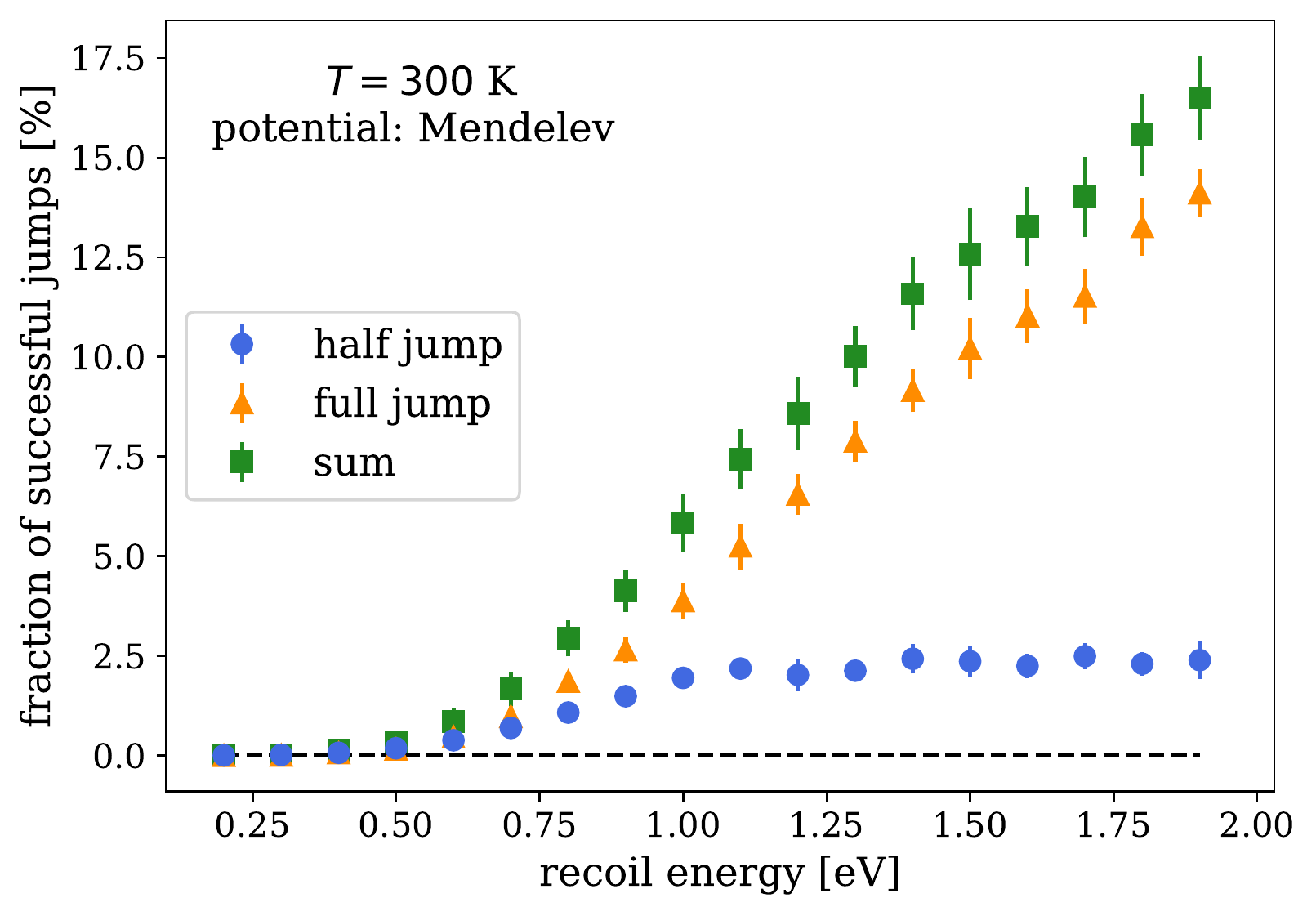}%
}\hfill

\subfloat[]{%
  \includegraphics[width=0.3\textwidth]{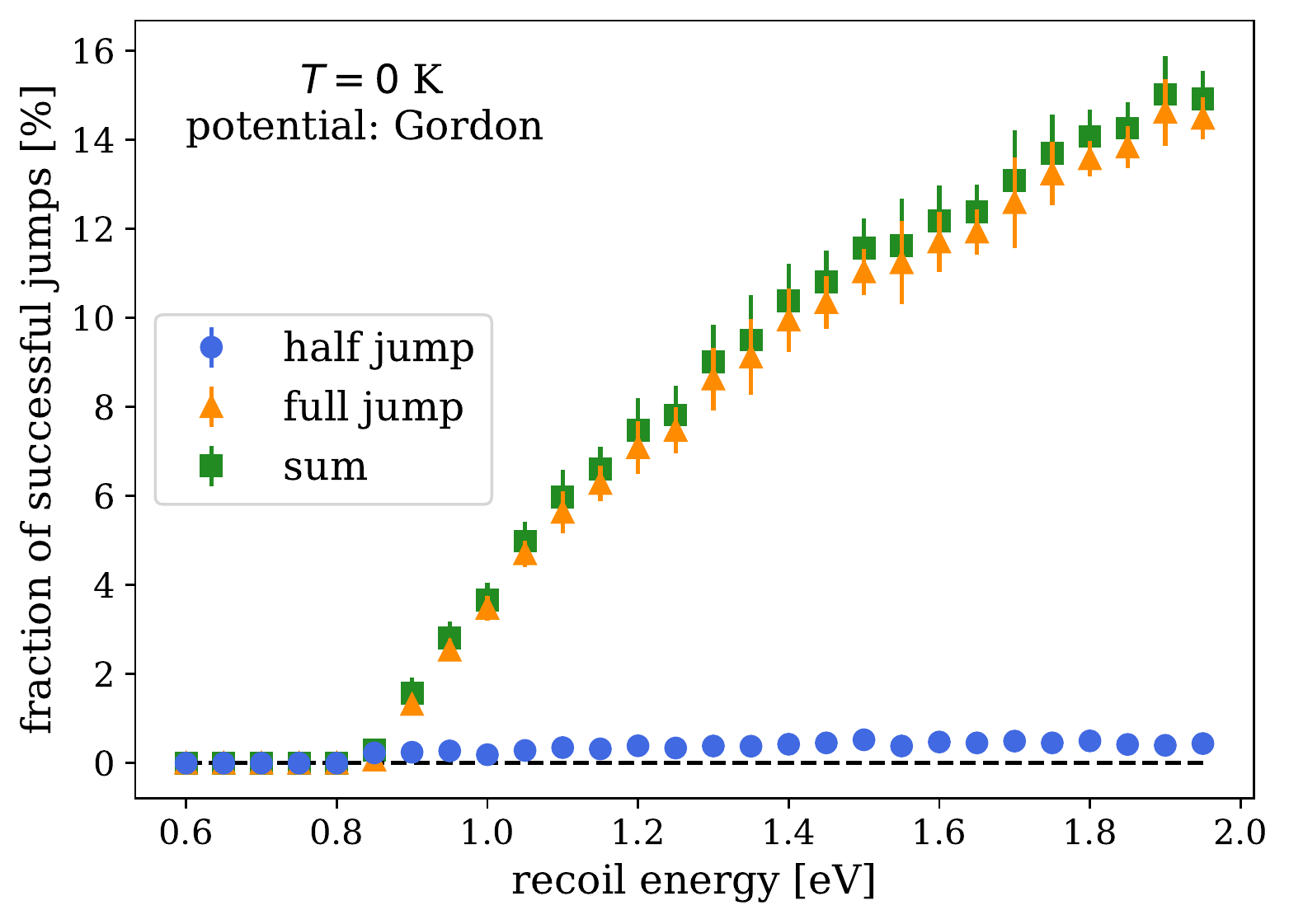}%
}\hfill
\subfloat[]{%
  \includegraphics[width=0.3\textwidth]{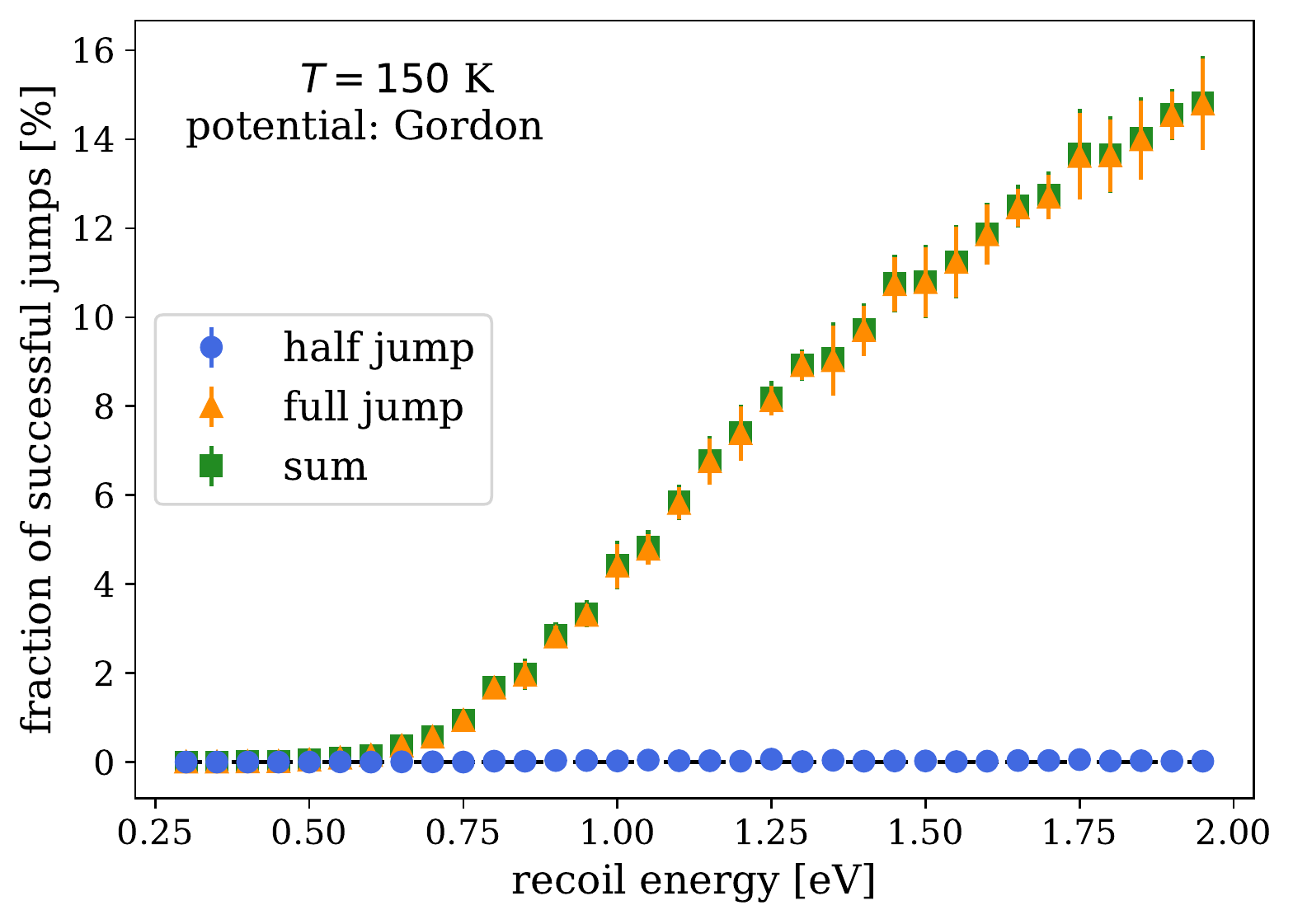}%
}\hfill
\subfloat[]{%
  \includegraphics[width=0.3\textwidth]{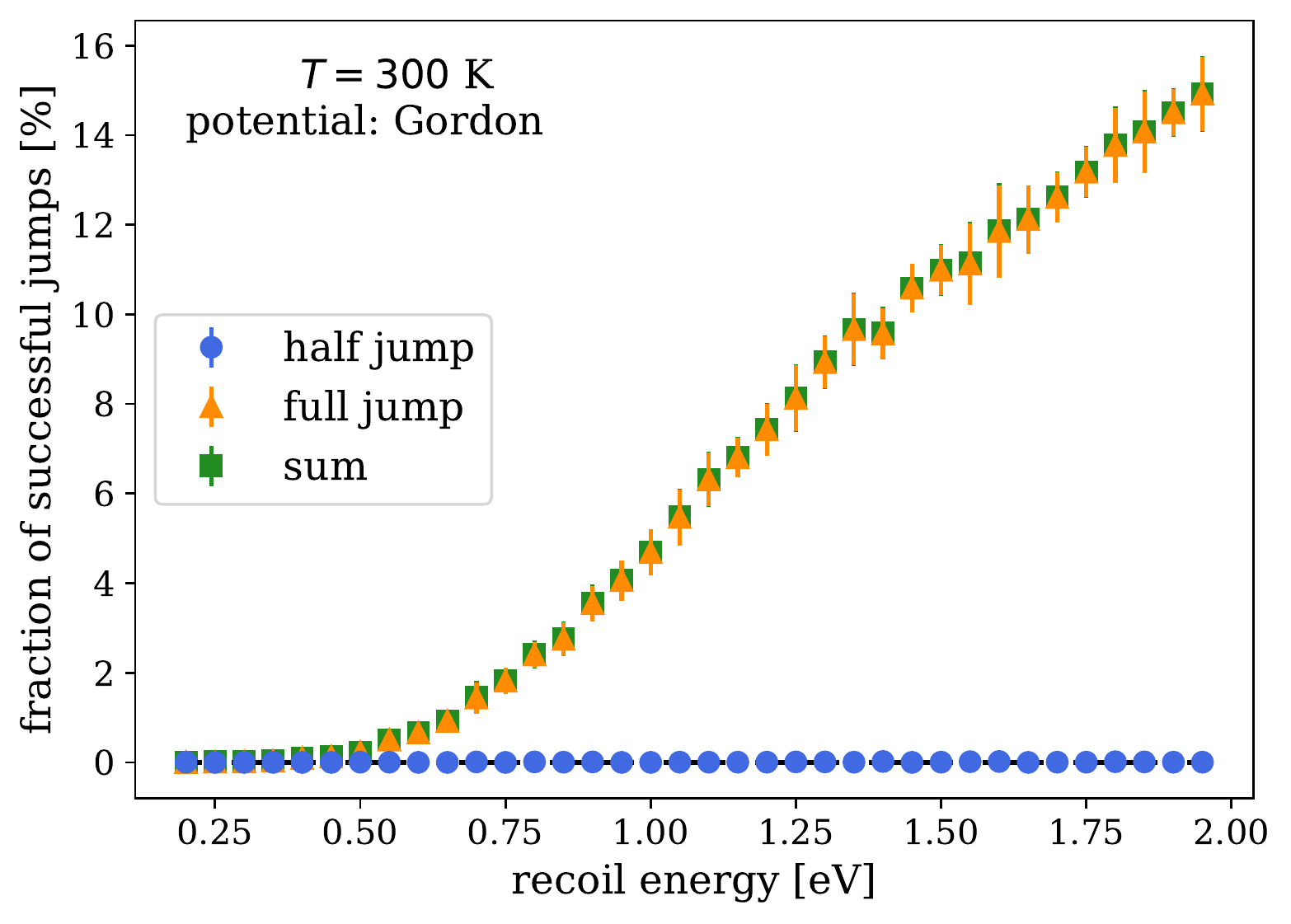}%
}\hfill
    \caption{Fraction of successful hops in Fe plotted for the Mendelev (a-c) and Gordon (d-f) potentials, similarly to Fig.~\ref{fig:jump_frequency}. For the former potential, the instances of half hop are comparable in the frequency of occurrence to those of complete hops, and are dominant for the recoil energies close to the top of the migration barrier. For the latter potential, the occurrence of half hops is very unlikely and only detectable at 0~K.}
    \label{fig:frac_Fe_2pot} 
\end{figure*}

Fig.~\ref{fig:frac_Fe_2pot} shows the effect of the choice of potential on the fraction of successful hops for the same potentials and for different temperatures. The presence of half hops fundamentally alters the results for the Mendelev potential, but has a negligible effect for the Gordon potential. For the results, and for Fig.~\ref{fig:jump_frequency}, the very few instances of half hops that were found at 0~K were equally divided between successful and unsuccessful outcomes.

\section{Probability function of successful hops: \texorpdfstring{Eq.~\eqref{eq:J_E_R}}{}} \label{app:prob}

In order to relate electron-induced recoils and vacancy diffusion, one has to quantify the probability that a collision of an electron with a random atom in a random direction leads to a vacancy hopping by one or more lattice spacing. We assume that this function ${J(E_{\textnormal{R}})}$ is a function of the recoil energy only. The electron flux $\phi_{\textnormal{el}}({\bf n},E_{\textnormal{el}})$ is assumed to be isotropic and thus solely a function of the electron energy because the $\gamma$-photons that are generated by nonelastic neutron reactions are emitted in a random direction and have long attenuation distance of the order of a cm before they generate high energy electrons that in turn are excited in a random directions. We present a semi-analytical model to predict $J(E_{\textnormal{R}})$, and verify the model by MD simulations, which are also used to parameterise the analytical result.

Assuming the vacancy as a quasi-particle moving in a one-dimensional potential energy landscape, we start from a one-dimensional Langevin equation for a particle of mass $m$ and friction coefficient $\gamma$,
\begin{equation}
    m\frac{\partial^2 x}{\partial t^2}=-\frac{\partial V(x)}{\partial x}-\gamma\frac{\partial x}{\partial t}+\eta(x, t).
\end{equation}
The particle moves in a potential that we assume for simplicity to be
\begin{equation}
    V(x)=E_{\textnormal{a}} \sin ^2 \left(\frac{\pi x}{\lambda}\right),
\end{equation}
with the periodicity of $\lambda=\frac{\sqrt{3}}{2}a$ in a bcc crystal. $\eta(x, t)$ is a stochastic force that satisfies the so-called white noise conditions
\begin{equation}
    \left< \eta(x, t)\right> = 0
\end{equation}
\begin{equation}
    \left< \eta(x, t)\eta(x', t')\right> = 2k_{\textnormal{B}} T \gamma \delta(x-x')\delta(t-t'),
\end{equation}
i.e. it has zero mean and is uncorrelated in space and time.

The particle initially is at $x_0$ and has velocity $v_0$. At $t=0$, it receives a recoil so that the velocity becomes
$$
v=v_0+v_{\textnormal{R}},
$$ 
where $v_{\textnormal{R}}=\sqrt{2E_{\textnormal{R}}/m}$ is defined at 0~K. We assume that the influence of stochastic force and friction are both negligible. The former gives rise to thermal fluctuations on the energy scale about 10-100 times lower than $E_{\textnormal{R}}$. The latter is weak on the time scale ${\sim1}$ ns of the transition. For a vacancy to migrate, the kinetic energy after the recoil must be above the barrier height
$$
\frac{1}{2}mv^2 > E_{\textnormal{a}}-V(x_0).
$$
If the temperature is low, we assume that ${E_{\textnormal{a}}-V(x_0)\approx E_{\textnormal{a}}}$, leading to the condition
\begin{equation}
    v^2>\frac{2E_{\textnormal{a}}}{m}.
\end{equation}
To find the likelihood of this condition being met, we need to know the probability distribution function (PDF) for $v^2$. In a 1D problem, the thermal part is a Gaussian distribution $f_{v_0}(v)$ with zero mean and the standard deviation of $\sigma=k_{\textnormal{B}} T/m$. Contribution from collisions $v_{\textnormal{R}}$ is drawn from the uniform distribution
$$
f_{v_{\textnormal{R}}}(v)=
\begin{cases}
\displaystyle \frac{1}{2v_{\textnormal{R}}}, & |v| < v_{\textnormal{R}}\\
0, & \text{otherwise,}
\end{cases}
$$ 
as this generates recoils with random orientations in 3D. The PDF for the sum of the two is given by their convolution
\begin{align}\label{eq:pdf_v}
    f_v(v)&=\int \limits_{-\infty}^{\infty}\text{d}v'\, f_{v_0}(v-v')f_{v_{\textnormal{R}}}(v') \nonumber \\
    &=\frac{1}{4v_{\textnormal{R}}}\left[\text{erf}\left(\frac{v+v_{\textnormal{R}}}{\sqrt{2}\sigma}\right)-\text{erf}\left(\frac{v-v_{\textnormal{R}}}{\sqrt{2}\sigma}\right)\right].
\end{align}
The PDF of $v^2$ follows from Eq. \eqref{eq:pdf_v} through a change of variable $g(y)=y^2$ with $y=v^2$. From
\begin{equation}
    f_y(y)=f_v\left(g^{-1}(y)\right)\bigg|\frac{\text{d}\left(g^{-1}(y)\right)}{\text{d}y}\bigg|,
\end{equation}
we find
\begin{equation}
    f_{v^2}(v^2)=\frac{1}{4v_{\textnormal{R}}\sqrt{v^2}}\left[\text{erf}\left(\frac{\sqrt{v^2}+v_{\textnormal{R}}}{\sqrt{2}\sigma}\right)-\text{erf}\left(\frac{\sqrt{v^2}-v_{\textnormal{R}}}{\sqrt{2}\sigma}\right)\right].
\end{equation}
The probability of a successful hop $p(v^2>\frac{2E_{\textnormal{a}}}{m})$ follows as
\begin{align}
    &\qquad p\left(v^2>\frac{2E_{\textnormal{a}}}{m}\right)=\frac{1}{2}\int \limits_{2E_{\textnormal{a}}/m}^\infty \text{d}s\, f_{v^2}(s) \nonumber \\ & =\frac{1}{2}+\frac{1}{4}\left(1-\sqrt{\frac{E_{\textnormal{a}}}{E_{\textnormal{R}}}}\right)\text{erf}\left[\frac{\sqrt{E_{\textnormal{R}}}-\sqrt{E_{\textnormal{a}}}}{\sqrt{k_{\textnormal{B}}T}}\right] \nonumber \\
    &-\frac{1}{4}\left(1+\sqrt{\frac{E_{\textnormal{a}}}{E_{\textnormal{R}}}}\right)\text{erf}\left[\frac{\sqrt{E_{\textnormal{R}}}+\sqrt{E_{\textnormal{a}}}}{\sqrt{k_{\textnormal{B}}T}}\right] \nonumber \\
    &+\sqrt{\frac{k_{\textnormal{B}}T}{4\pi E_{\textnormal{R}}}}\text{sinh}\left[\frac{2\sqrt{E_{\textnormal{R}}E_{\textnormal{a}}}}{k_{\textnormal{B}}T}\right]\exp\left[-\frac{E_{\textnormal{a}}-E_{\textnormal{R}}}{k_{\textnormal{B}}T}\right],
\end{align}
where the factor of $1/2$ in front of the integral indicates that the kick can be given either towards or away from the vacancy. We assume that the 1D approximation adequately describes a reaction along a one-dimensional reaction pathway, and that the 3D atomic dynamics alters the result only though an extra scaling factor that we call $\alpha$. This leads to the functional form $J(E_{\textnormal{R}}, T)$ for the transition probability given by Eq. \eqref{eq:J_E_R}.

\section{Vacancy hops stimulated by high-energy recoils}\label{app:high_E_rec}
Eq.~\eqref{eq:rate_success} gives the vacancy hopping frequency for a given electron spectrum. This is an appropriate expression if the recoil energy does not exceed significantly the vacancy migration barrier, and applies to direct collisions of electrons with the nearest neighbour atoms of a vacancy or the atoms along the $\langle111\rangle$ close packed directions in bcc crystal structure. If the energy delivered by electrons is many times greater than the vacancy migration barrier, an alternative estimate was given by Kiritani \cite{Kiritani1976}: since the recoil energy gets distributed to atoms near the struck one, one of the nearest neighbours of a vacancy might gain the excess energy greater than the barrier $E_\textnormal{a}$. Kiritani proposed that the vacancy jump frequency per atom that can contribute to the process (e.g. 8 atoms if the electron flux is angular isotropic and 4 if it is unidirectional in a bcc metal) in the high-energy recoil (HER) regime is
\begin{equation}\label{eq:nu_HER}
\nu_{\textnormal{HER}}=\Phi_{\textnormal{el}}\int\limits_{E_\textnormal{a}}^{E_\textnormal{R}^\textnormal{max}}\frac{E_\textnormal{R}}{E_\textnormal{a}}\frac{\text{d}\sigma}{\text{d} E_{\textnormal{R}}}\text{d}E_{\textnormal{R}}.
\end{equation}
Under a 2~MV beam and flux $\Phi_{\textnormal{el}}=10^{19}$~cm$^{-2}$s$^{-1}$ vacancies are expected to hop approximately 0.1 to 1 times per second, with higher hopping frequency for higher atomic number \cite{Kiritani1976}. This agrees with the estimate given by Arakawa \emph{et al.} \cite{Arakawa2020}.

If the energy recoil cross section is given by Eq.~\eqref{eq:dS_dEr_simplified}, calculating the integral in Eq.~\eqref{eq:nu_HER} is straightforward and
\begin{equation}\label{eq:nu_HER_solved}
\nu_{\textnormal{HER}}=\pi\Phi_{\textnormal{el}}\left(\frac{Ze^2}{4\pi\varepsilon_0 m c^2}\right)^2 \left(\frac{1-\beta^2}{\beta^4}\right)\frac{E_{\textnormal{R}}^\textnormal{max}}{E_\textnormal{a}}\log\left(\frac{E_{\textnormal{R}}^\textnormal{max}}{E_\textnormal{a}}\right),
\end{equation}
if $E_{\textnormal{R}}^\textnormal{max}>E_\textnormal{a}$, and $\nu_{\textnormal{HER}}=0$ otherwise. The presence of parameter $\kappa$ in Eq.~\eqref{eq:dS_dEr_simplified} is negligible at very high electron energies (e.g. $\kappa=3\times10^{-5}$ for 2~MeV electrons).

\begin{figure}[t]
\includegraphics[width=0.45\textwidth]{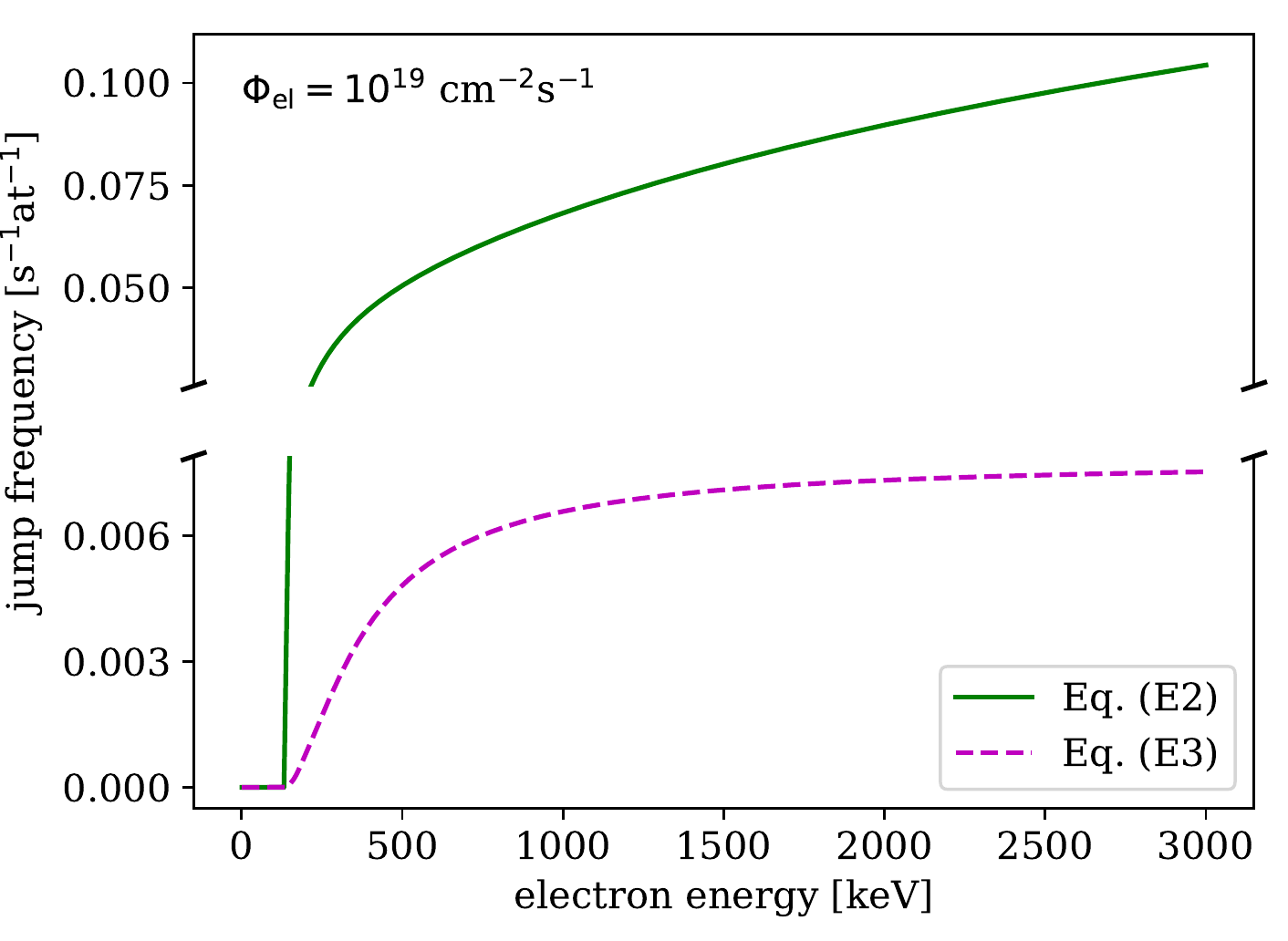}%
\caption{Vacancy hopping frequency due to an electron flux in W per atom that can contribute to a transition. As opposed to Eq.~\eqref{eq:nu_vac_111}, which shows saturation above $E_\textnormal{el}\sim1$~MeV, Eq.~\eqref{eq:nu_HER_solved} does not saturate. The values derived from the phenomenological model of Kiritani \cite{Kiritani1976} expressed by Eq.~\eqref{eq:nu_HER_solved} are about 10 to 15 higher than those evaluated using the theory developed in Section~\ref{sec:D_el}.}
\label{fig:nu_vac}
\end{figure}

Integration of Eq.~\eqref{eq:rate_success} in the limit $T\rightarrow0$, using Eq.~\eqref{eq:J_T_indep}, gives the vacancy hopping frequency stimulated by only the direct collision with one of the atoms situated along the $\langle111\rangle$ directions, which is
\begin{align}\label{eq:nu_vac_111}
    \nu_{\langle111\rangle}=&\frac{\pi\alpha}{3}\Phi_{\textnormal{el}}\left(\frac{Ze^2}{4\pi\varepsilon_0 m c^2}\right)^2 \left(\frac{1-\beta^2}{\beta^4}\right)\times \nonumber\\ &\sum_{k=1}^{4}\Bigg[\frac{E_{\textnormal{R}}^{\textnormal{max}}}{E_{\textnormal{a}}^{(k)}}+2\sqrt{\frac{E_{\textnormal{a}}^{(k)}}{E_{\textnormal{R}}^{\textnormal{max}}}}-3\Bigg]
\end{align}
if $E_{\textnormal{R}}^\textnormal{max}>E_\textnormal{a}$, and $\nu_{\langle111\rangle}=0$ otherwise. Fig.~\ref{fig:nu_vac} shows how the vacancy hopping frequency varies for incident electron energies up to 3~MeV; the values computed using  Eq.~\ref{eq:nu_HER_solved} are at least one order of magnitude higher than those computed using Eq.~\ref{eq:nu_vac_111}. In both cases, the electron kinetic energy enters the expressions via $\beta=v/c$, where $v$ is the electron velocity, and via the maximum target recoil energy $E_{\textnormal{R}}^{\textnormal{max}}$.

\section{Electron collision stimulated migration of self-interstitial atom and vacancy clusters} \label{app:SIA_V_cl}
While vacancies in W have a high migration barrier of about 1.7~eV, self-interstitial atoms have a very low migration barrier of about 0.01~eV \cite{Swinburne2017} and exhibit non-Arrhenius diffusion \cite{Dudarev2008}; they are mobile at cryogenic temperatures and, as observed in atomistic simulations, are able to migrate athermally, driven solely by elastic interactions \cite{Derlet2020}. Therefore, the electron-stimulated motion of self-interstitial atom defects is not expected to play a significant part.

In the main text, we considered only single vacancies, although vacancy clusters may also be present in the material. Under low temperature proton irradiation, mainly single vacancies are produced \cite{Ogorodnikova2019}. Moreover, \emph{ab initio} calculations show that di-vacancies are unstable in W \cite{Becquart2007}. Regardless of these considerations, divacancy, trivacancy and 4-vacancy clusters do not have markedly different migration energies in comparison with to single vacancies. The migration energies can be slightly higher or lower but on average the absolute value of the deviation from the single-vacancy migration energy does not exceed 10~\% \cite{Mason2017}, including the unusually low migration barrier of 1.15~eV found for a trivacancy configuration. Since the above analysis shows that the electron-stimulated recoil distribution is not strongly varying across this range of a fraction of an eV, the hopping rates are not expected to be vastly different if vacancies were arranged in small clusters, with the possible exception of trivacancy clusters. A second relevant point is that in the presence of an intense electron flux driving single-vacancy diffusion, enhanced clustering of vacancies may be expected. The binding energy of small vacancy clusters is significantly smaller than their migration energy \cite{Mason2017}, and the barrier for dissolving them is of the order of vacancy migration energy. Therefore, electron collisions should break the clusters apart at a faster rate than the enhanced vacancy mobility may help generate. An in-depth discussion about the two points about cluster diffusion and formation is beyond the scope of this work, and requires further analysis.

\end{appendix}

\bibliography{ref_short}

\end{document}